\documentclass[11pt,preprint]{aastex}
\usepackage{epsfig}
\newcommand       \Angstrom     {\,{\rm \AA}} 
\newcommand       \AU           {\,{\rm AU}}          

\newcommand       \cm           {\,{\rm cm}}
\newcommand       \mm           {\,{\rm mm}}
\newcommand       \erg          {\,{\rm erg}}
\newcommand       \eV           {\,{\rm eV}}
\newcommand	  \g		{\,{\rm g}}
\newcommand       \K            {\,{\rm K}}
\newcommand	  \pc		{\,{\rm pc}}
\newcommand	  \s		{\,{\rm s}}

\newcommand	  \yr		{\,{\rm yr}}

\newcommand	  \Myr		{\,{\rm Myr}}

\newcommand	  \Urad		{U}

\newcommand       \simlt        {\lesssim}
\newcommand       \simgt        {\gtrsim}
\newcommand       \gtsim        {\gtrsim}
\newcommand       \ltsim        {\lesssim}

\newcommand       \mum          {\,{\rm \mu m}}

\newcommand	  \Teff	        {T_{\rm eff}}

\newcommand	  \hra	        {{\rm HR\,4796A}}

\newcommand	  \hda	        {{\rm HD\,141569A}}
\newcommand	  \hdb	        {{\rm HD\,141569B}}
\newcommand	  \hdc	        {{\rm HD\,141569C}}
\newcommand	  \amin	        {a_{\rm min}}
\newcommand	  \amax	        {a_{\rm max}}
\newcommand	  \rin	        {r_{\rm in}}
\newcommand	  \rout	        {r_{\rm out}}
\newcommand	  \msil         {m_{\rm sil}} 
\newcommand	  \mcarb        {m_{\rm carb}} 
\newcommand	  \mice         {m_{\rm ice}} 
\newcommand	  \mpah         {m_{\rm PAH}} 
\newcommand	  \md           {m_{\rm d}}
\newcommand	  \rp           {r_{\rm p}}

\newcommand	  \rhosil       {\rho_{\rm sil}}
\newcommand	  \rhocarb      {\rho_{\rm carb}}
\newcommand	  \rhoice       {\rho_{\rm ice}}
\newcommand	  \Pice         {P^{\prime}}
\newcommand	  \sigmap       {\sigma_{\rm p}}
\newcommand	  \sigmar       {\sigma(r)}
\newcommand       \fion         {\phi_{\rm ion}}  
\newcommand       \apahmin      {a^{\rm PAH}_{\rm min}}
\newcommand       \fsil         {f_{\rm sil}}
\newcommand       \fcarb        {f_{\rm carb}}   
\newcommand       \fice         {f_{\rm ice}}  
\newcommand       \fsilp        {f_{\rm sil}^{\prime}}
\newcommand       \fcarbp       {f_{\rm carb}^{\prime}}   
\newcommand       \esil         {\epsilon_{\rm sil}}
\newcommand       \ecarb        {\epsilon_{\rm carb}}
\newcommand       \eice         {\epsilon_{\rm ice}}
\newcommand       \esilp        {\epsilon_{\rm sil}^{\prime}}
\newcommand       \ecarbp       {\epsilon_{\rm carb}^{\prime}}
\newcommand       \eeff         {\epsilon_{\rm eff}} 
\newcommand       \Rstar        {R_{\star}} 
\newcommand       \Fstar        {F^{\star}_\lambda}
\newcommand       \Mstar        {M_{\star}} 
\newcommand       \ummp         {u^{\rm ISRF}_\lambda} 
\newcommand       \lnT          {{\rm ln}T} 
\newcommand       \msun         {m_\odot}
\newcommand       \nud          {\nu_{\rm d}}  
\newcommand       \Ed           {E_{\rm d}} 
\newcommand       \Yion         {Y_{\rm ion}} 
\newcommand       \EIP          {E_{\rm IP}} 
\newcommand       \kir          {k_{\rm IR}} 
 
\newcommand       \kdes         {k_{\rm des}} 
\newcommand       \kdhy         {k_{\rm dhy}} 
\newcommand       \kion         {k_{\rm ion}} 
\newcommand	  \mH		{m_{\rm H}}

\newcommand	  \dof		{{\rm d.o.f.}}
\newcommand	  \Npar	        {N_{\rm par}}
\newcommand       \krec         {k_{\rm rec}}
\newcommand       \me           {m_{\rm e}}
\newcommand       \ve           {v_{\rm e}}
\newcommand       \nelc         {n_{\rm e}}
\newcommand       \Tg           {T_{\rm gas}}

\newcommand       \molH         {\rm H_{2}}
\newcommand       \molHi        {\rm H_{2}^{+}}
\newcommand       \triHi        {\rm H_{3}^{+}}
\newcommand       \nmolH        {n_{\rm H_{2}}}
\newcommand       \ntriHi       {n_{\rm H_{3}^{+}}}
\newcommand       \nco          {n_{\rm CO}}
\newcommand       \rcr          {\varsigma_{\rm CR}}
\newcommand       \ke           {k_{\rm e}}
\newcommand       \kco          {k_{\rm CO}}
\newcommand       \NmolH        {N_{\rm H_{2}}}
\newcommand       \NtriHi       {N_{\rm H_{3}^{+}}}
\newcommand       \Nco          {N_{\rm CO}}
\newcommand       \taustar      {\tau_\star}
\newcommand       \sigmaH       {\sigma_{\rm H_2}}
\newcommand       \mHtot        {m_{\rm H}^{\rm tot}}
\newcommand       \mearth       {\,{m_\oplus}}
\newcommand       \fhn          {f_{\rm hn}}
\newcommand       \kabs         {\kappa_{\rm abs}}
\newcommand       \taudes       {\tau_{\rm des}}
\newcommand       \mpahdot      {\dot{m}_{\rm PAH}}
\newcommand       \taurp        {\tau_{\rm RP}}
\newcommand       \mrpdot       {\dot{m}_{\rm RP}}
\newcommand       \taupr        {\tau_{\rm PR}}
\newcommand       \tausubl        {\tau_{\rm subl}}
\newcommand{\figwidth}{6.0in}

\countdef\decade=200
\advance\decade by \year
\countdef\hours=201
\advance\hours by \time
\divide\hours by 60
\countdef\mins=202
\advance\mins by \hours
\multiply\mins by 60
\multiply\hours by 100
\countdef\miltime=203
\advance\miltime by \hours
\advance\miltime by \time
\advance\miltime by -\mins

\shorttitle{The HD\,141569A Dust Disk}

\begin{document}

\title{
%------------- enable for labelling preprint ---------------------------
% \vspace*{-2.0em}
%  {\normalsize\rm submitted to {\it The Astrophysical Journal}}\\
% \vspace*{-2.0em}
%  \vspace*{1.0em}
 \vspace*{-2.0em}
  {\normalsize\rm {\it The Astrophysical Journal},
   vol.\,594, 987--1010}\\
  \vspace*{1.0em}
%-----------------------------------------------------------------------
Modeling the Infrared Emission from the HD\,141569A Disk 
%\\{\small DRAFT: \today ~~}
	 }

\author{Aigen Li and J.I. Lunine}
\affil{Theoretical Astrophysics Program,
       Lunar and Planetary Laboratory and Steward Observatory,
       %Departments of Astronomy and Planetary Sciences,
       University of Arizona, Tucson, AZ 85721;\\
        {\sf agli@lpl.arizona.edu, jlunine@lpl.arizona.edu}}

\begin{abstract}
We model the infrared (IR) emission from the double-ring disk  
of $\hda$, using a porous dust model that was previously shown
successful in reproducing the spectral energy distributions
(SEDs) of the disks around $\beta$ Pictoris and $\hra$.
The dust consists of either unaltered or highly processed 
interstellar materials and vacuum with a volume fraction of 
$\sim 90\%$. Together with a population of polycyclic aromatic 
hydrocarbon (PAH) molecules of which a large fraction is charged,
the porous dust model made of coagulated but otherwise unaltered 
protostellar interstellar grains provides an excellent fit to 
the entire SED from the mid-IR to millimeter wavelengths including 
the PAH mid-IR emission features.
The nondetection of the 21$\mum$ crystalline silicate feature
predicted from the porous dust model composed of highly-processed 
grains (mainly crystalline silicate dust) imposes an upper limit 
of $\sim 10\%$ on the mass fraction of crystalline silicates.
Spectroscopic and broadband photometric predictions
are made for SIRTF observations which will provide 
further tests of the applicability of the porous dust model
to circumstellar disks.
\end{abstract}


\section{Introduction\label{sec:intro}}
Since dust disks around young stars, depending on their age,
are the source material or the remnants of newly-formed planets,
the physical, chemical, and dynamical properties of circumstellar 
disks and their constituent grains are crucial in understanding 
the formation process of planetary systems.

Over the past 2 decades, impressive evidence has been assembled
for the commonality of circumstellar dust disks around 
pre-main-sequence stars (T Tauri stars and Herbig Ae/Be stars), 
main sequence (MS) stars, post-MS stars (red giants), and 
a white dwarf (see Backman \& Paresce 1993, Habing et al.\ 2001,
Zuckerman 2001, Lagrange, Backman, \& Artymowicz 2001 for reviews).
To date, at least 15\% of the A--K type stars are found to have
dust disks around them (see Lagrange, Backman, \& Artymowicz 2001).
The majority of these disks are spatially extensive, 
with their dust quantities and intensities of scattered-starlight 
and of dust thermal emission gradually decreasing outward from 
the central stars. However, so far there exist four exceptions:
(1) $\epsilon$ Eridani -- a $\sim 800\Myr$ old 
(Henry et al.\ 1996) K2V MS star -- has a nearly face-on dust ring 
with a mass and radius ($r \sim 60\AU$) similar to the solar system 
Kuiper Belt and is thought to be an analog to
the young solar system (Greaves et al.\ 1998); 
(2) Formahault ($\alpha$ PsA) -- a $\sim 200\Myr$ old 
(Barrado y Navascu\'{e}s et al.\ 1997) A3V MS star --
exhibits a ``doughnut''-like dust annulus with a radial distance 
of $\sim 150\AU$ from the star and a width of $\sim 60\AU$ 
(Holland et al.\ 1998, 2003); 
(3) $\hra$ -- a $\sim 8\pm3\Myr$ old A0V star (see Jura et al.\ 1998)
-- displays a ring-like disk peaking at $\sim 70\AU$ 
from the central star and abruptly truncated both inward and outward 
with a width of $\simlt 17\AU$ (Schneider et al.\ 1999; 
Koerner et al.\ 1998; Jayawardhana et al.\ 1998; Telesco et al.\ 2000);
(4) $\hda$ -- a nearby (distance to the Earth $d\approx 99\pc$;
van den Ancker, de Winter, \& Tjin A Djie 1998) 
$\sim 5\pm3\Myr$ old (Weinberger et al.\ 2000) pre-MS Herbig Ae/Be 
star with a spectral type of B9.5V, an effective temperature 
of $\Teff\approx 10000\K$ and a surface gravity of $\lg g\approx 4.0$
(Dunkin, Barlow, \& Ryan 1997).

The presence of dust around $\hda$ was first reported by
Jaschek, Jaschek, \& Egret (1986), de Grijp, Miley, \& Lub (1987), 
and Walker \& Wolstencroft (1988) based on the IRAS 
({\it Infrared Astronomical Satellite}) detection 
of infrared (IR) excess over expected from its photosphere.
It has recently been imaged in scattered light and in
thermal IR emission. With the discovery of its large
and morphologically complicated debris disk (see below), 
$\hda$ has recently aroused considerable interest.

The disk around $\hda$ appears to have two ring-like structures:
an outer ring at $\sim 325\AU$ from the star with a width of 
$\sim 150\AU$; and an inner ring at $\sim 185\AU$.
The outer ring was reported independently by Augereau et al.\ (1999a)
and by Weinberger et al.\ (1999), respectively based on
the 1.6$\mum$ and 1.1$\mum$ imaging of the scattered-light. 
They both used the {\it Near-IR Camera and Multiobject Spectrometer} 
(NICMOS) on the {\it Hubble Space Telescope} (HST).
The inner ring was discovered by Weinberger et al.\ (1999) 
at 1.1$\mum$ using the same instrument.
Weinberger et al.\ (1999) also reported the detection of
a gap or ``dip'' (a region of depleted material) at
$\sim 250\AU$ from the star with a width of $\sim 60\AU$.
They interpreted the gap as a sign of one or more planets 
orbiting within the disk (the planets do not have to be in 
the gap), but other explanations such as dust migration also 
exist (e.g. see Takeuchi \& Artymowicz 2001).

This ring-gap-ring structure was later confirmed by
optical coronagraphic observations made by Mouillet et al.\ (2001) 
at an effective wavelength of 5850$\Angstrom$ 
using the HST {\it Space Telescope Imaging Spectrograph} (STIS).
With a higher spatial resolution and a higher signal-to-noise ratio
than the NICMOS/HST data, Mouillet et al.\ (2001) found that 
the inner ring, with a FWHM (full width of half maximum) 
of $\sim 50\AU$, actually peaks at $\sim 200\pm 5\AU$. 
Strong brightness asymmetry was also reported by 
Mouillet et al.\ (2001). 
This asymmetry, interpreted as a result of the gravitational 
perturbation caused by putative massive planets within the disk,
by the two stellar companions $\hdb$ and $\hdc$, or both types 
of objects (Mouillet et al.\ 2001), was confirmed by 
the ground-based 2.2$\mum$ near-IR imaging observation 
at the Palomar 200 inch telescope using 
the adaptive optics system (Boccaletti et al.\ 2003).

In addition to the two annuli, the existence of a third dust 
component -- a population of warm, mid-IR emitting dust within
$\sim 100\AU$ from the star -- was recently reported by 
Fisher et al.\ (2000).
It was shown that the 10.8 and 18.2$\mum$ emission
observed with the University of Florida {\it Observatory Spectrometer 
and Camera for the Infrared} (OSCIR) on the Keck II telescope
extend out to $\sim 100\AU$ from the star (Fisher et al.\ 2000).  
The existence of such an extended, warm source close to the star,
originally suggested by Augereau et al.\ (1998) to account for
the disk IR emission at wavelengths $\lambda \simlt 10\mum$,
was later confirmed by the 12.5, 17.9, and  20.8$\mum$ imaging 
using the JPL {\it Mid-InfraRed Large-well Imager} (MIRLIN) 
on the Keck II telescope (Marsh et al.\ 2002). 

More recently, the HRC ({\it High Resolution Channel}) coronagraphic 
B, V, and I images of the $\hda$ disk obtained with 
the HST {\it Advanced Camera for Surveys} (ACS) revealed 
a tightly-wound spiral structure with two arms, one of which 
stretches outward to reach the nearby companion stars
(Clampin et al.\ 2003).

In a previous paper (Li \& Lunine 2003), we have shown that
the IR emission from the $\hra$ disk can be well fit by
a model invoking highly porous cometary-type dust.
The porous cometary dust model was originally proposed for 
the $\beta$ Pictoris disk and was shown successful in reproducing 
its spectral energy distribution from the near-IR to millimeter 
wavelengths including the 10$\mum$ amorphous and the 11.3$\mum$ 
crystalline silicate features (Li \& Greenberg 1998). 
In this paper we will model the observed IR and millimeter 
photometric and mid-IR spectroscopic signatures of $\hda$.
The objectives of this paper are three-fold: 
(1) We wish to infer the physical and chemical
properties of the dust in the $\hra$ disk and its
relationship to the formation of planets.
(2) We wish to know how widely applicable is the porous dust model;
if it can be shown that the porous dust model is also valid for 
many other disks at different evolutionary stages
and with different geometrical structures, it will be 
a valuable guide for interpreting future data sets. 
This is of particular importance given that the incoming 
{\it Space Infrared Telescope Facilities} (SIRTF) 
will accumulate a rich set of IR photometric
and spectroscopic data for a large number of dust disks;
we would hope that the porous dust model presented in this and 
related papers could serve as a starting point in analyzing these data.  
(3) We wish to make near- and mid-IR spectral and broadband
photometry predictions for the $\hda$ disk; these predictions
can be compared to future SIRTF observations in order to 
test further the validity of the porous dust model.   

This paper is organized as follows: 
we first summarize in \S\ref{sec:fluxobs} the photometric and 
spectroscopic data available for the $\hda$ disk
that are relevant in modeling the IR emission of the disk.
We then give an overview discussion in \S\ref{sec:model} of
the proposed dust model with emphasis on dust composition 
and morphology from the point of view of protoplanetary disk 
evolution, followed by detailed discussions of the PAH
component in \S\ref{sec:pah} and the porous dust component
in \S\ref{sec:pd}. The IR emission modeling technique is
presented in \S\ref{sec:iremmethod}.  
In \S\ref{sec:ionrec} the PAH photoelectric emission rates 
and the electron recombination rates are calculated, using
the methods described in \S\ref{sec:photophys} and \S\ref{sec:ne}. 
The key model parameters are discussed and specified in \S\ref{sec:para}.
Model results are presented in \S\ref{sec:results}.
In \S\ref{sec:discussion} we discuss 
(1) previous studies of the SED modeling of the $\hda$ disk, 
(2) comparison of the $\hda$ disk with the $\hra$ disk, 
(3) exploration of parameter space (\S\ref{sec:robust}), 
(4) PAH photodestruction (\S\ref{sec:pahdes}), and 
(5) dust loss through radiation expulsion and Poynting-Robertson drag
    (\S\ref{sec:rppr}).
In \S\ref{sec:sirtf} we calculate the dust IR flux densities
integrated over the SIRTF/IRAC and MIPS bands predicted for
our best-fitting models. Spectroscopic predictions are also
made for SIRTF/IRS observations.  
The major conclusions are summarized in \S\ref{sec:conclusion}.

\section{Available Photometric and Spectroscopic Data\label{sec:fluxobs}}
To facilitate modeling of the dust properties of the $\hda$ disk,
we have assembled the available optical, IR, and millimeter photometric 
data and mid-IR spectra. 
These include 
(1) the BV optical photometry (Sylvester et al.\ 1996);
(2) the JHKLL$^{\prime}$M near-IR photometry obtained by 
Sylvester et al.\ (1996) using the single-channel bolometer UKT9 
at the {\it United Kingdom Infrared Telescope} (UKIRT);
(3) the 12, 25, 60, and 100$\mum$ IRAS photometry 
(Sylvester et al.\ 1996);
(4) the 10.8 and 18.2$\mum$ OSCIR/KeckII imaging 
by Fisher et al.\ (2000);
%using the University of Florida {\it Observatory Spectrometer 
%and Camera for the Infrared} (OSCIR) on the Keck II telescope;
(5) the 12.5, 17.9, and 20.8$\mum$ MIRLIN/KeckII imaging 
by Marsh et al.\ (2002); 
%using JPL's {\it Mid-InfraRed Large-well Imager} (MIRLIN)
%on the Keck II telescope;
(6) the 7.5--13.5$\mum$ (spectral resolution $\Delta\lambda=0.17\mum$)
and 15.8--23.9$\mum$ ($\Delta\lambda=0.27\mum$) spectroscopy obtained
by Sylvester et al.\ (1996) using the UKIRT {\it Cooled Grating 
Spectrometer 3} (CGS3);  
(7) the 1350$\mum$ SCUBA 
({\it Submillimetre Common-User Bolometer Array}) 
measurement by Sylvester, Dunkin, \& Barlow (2001) 
using the {\it James Clerk Maxwell Telescope} (JCMT). 
Taking $E(B-V)=0.095$ (Weinberger et al.\ 1999),
we have corrected the interstellar reddening assuming a $R_V=3.1$ 
(total-to-selective extinction) interstellar extinction law 
(see Table 6 in Li \& Draine 2001).
%for the line of sight toward $\hda$.

\begin{figure}[h]
\begin{center}
\epsfig{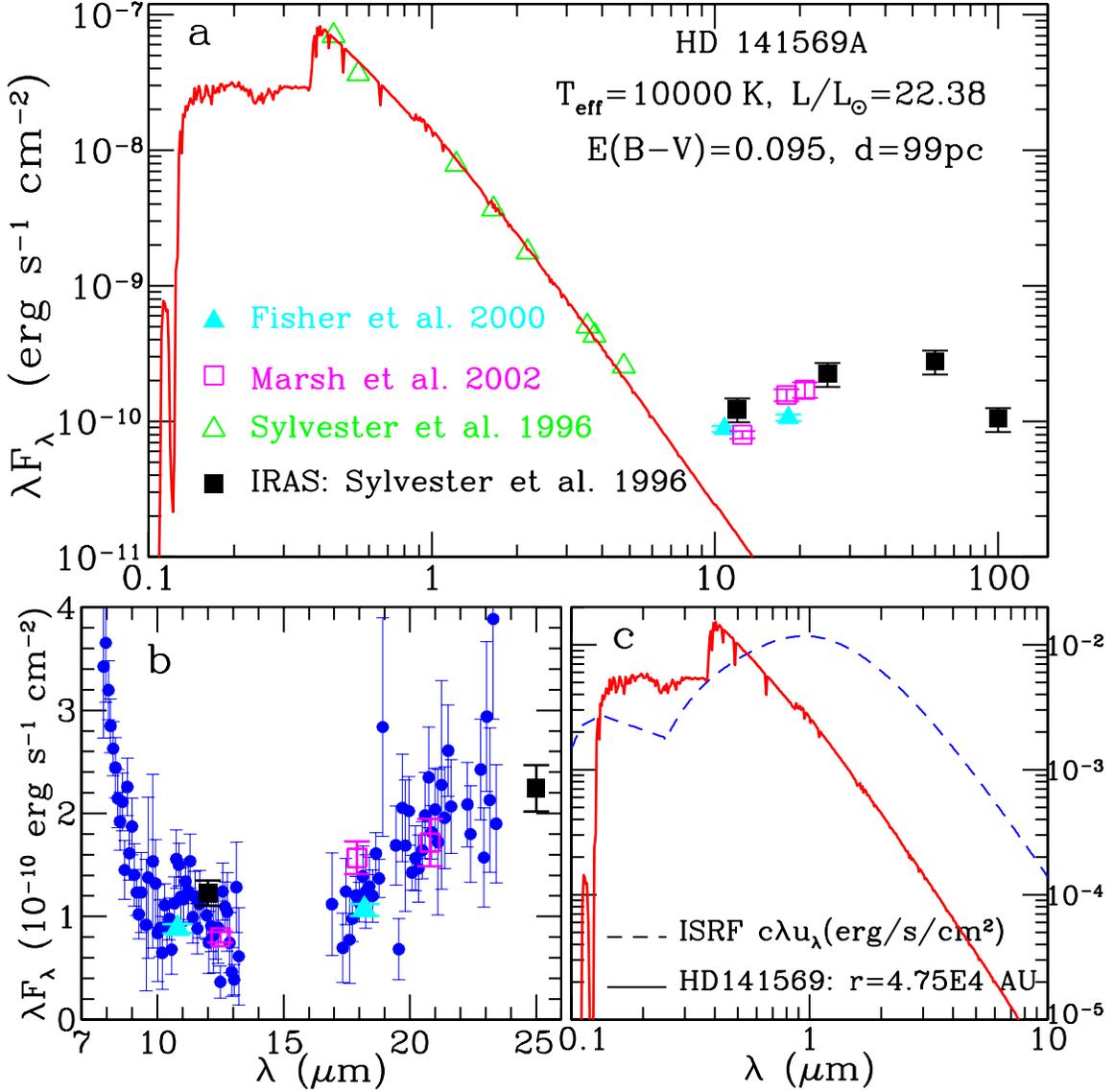}
\end{center}\vspace*{-1em}
\caption{
        \label{fig:starobs}
        \footnotesize
        ({\bf a}): Spectral energy distribution of
        the $\hda$ system (see \S\ref{sec:fluxobs}
        and Table \ref{tab:fluxobs}). 
        Red solid line: the $\Teff=10000\K$
        and $\lg g=4.0$ Kurucz model atmospheric spectrum;
        green open triangles: optical and near-IR photometry
        (Sylvester et al.\ 1996);
        cyan filled triangles: the 10.8 and 18.2$\mum$ OSCIR/KeckII
        photometry (Fisher et al.\ 2000); 
        black filled squares: the 12, 25, 60 and 100$\mum$ IRAS 
        photometry (Sylvester et al.\ 1996);
        magenta open squares: the 12.5, 17.9, and 20.8$\mum$ 
        MIRLIN/KeckII photometry (Marsh et al.\ 2002).
        ({\bf b}): The CGS3 mid-IR spectrum 
        (blue filled circles; Sylvester et al.\ 1996)
        as well as the OSCIR, MIRLIN, and IRAS photometry.
        All observational data in ({\bf a}) and ({\bf b})
        have been dereddened, but the stellar photospheric 
        contributions to the IR photometric and spectroscopic
        bands have not been subtracted.
        ({\bf c}): The $\hda$ radiation field at a distance 
        of $r=4.75\times 10^4\AU$ from the star (red solid line) 
        which equals the 912$\Angstrom$ -- 1$\mum$ general 
        interstellar radiation field (blue dashed line; 
        Mathis et al.\ 1983; see \S\ref{sec:pah}).
        }
\end{figure}

Table \ref{tab:fluxobs} summarizes the photometric data listed above,
the instruments/telescopes and their beam sizes through which
the observational data were obtained,
and the stellar atmospheric and dust disk contributions 
to each photometric band.
In Figure \ref{fig:starobs}a we plot the reddening-corrected 
optical, IR, and millimeter photometric results as well as the 
$\Teff = 10000\K$ and $\lg g=4.0$ Kurucz model atmospheric spectrum. 
Figure \ref{fig:starobs}b illustrates the CGS3 mid-IR spectrum
which clearly exhibits the 11.3$\mum$ out-of-plane C-H bending
feature and the long-wavelength wing of the 7.7$\mum$ C-C stretching
feature (see Li \& Draine 2001a for the PAH band assignment).
For the sake of comparison, the 10.8, 18.2$\mum$ OSCIR/KeckII data, 
the 12, 25$\mum$ IRAS data, 
and the 12.5, 17.9, 20.8$\mum$ MIRLIN/KeckII data,
are also plotted in Figure \ref{fig:starobs}b.
It is seen that these mid-IR data closely 
agree with one another.\footnote{%
 The 12$\mum$ IRAS data is about 12\% higher than the band-averaged
 CGS3 spectrum. We have therefore reduced the former by a factor
 of 1.12 to bring the 12$\mum$ IRAS data into agreement with
 the CGS3 spectrum. 
 }
The largest discrepancy ($\approx 30\%$) occurs for 
the 17.9$\mum$ MIRLIN/KeckII (Marsh et al.\ 2002)
data and the 18.2$\mum$ OSCIR/KeckII data (Fisher et al.\ 2000).
This discrepancy cannot be explained in terms of the difference
in the aperture sizes of these two instruments. 
As a matter of fact, the former (which has a higher 17.9$\mum$
flux density) was obtained using a smaller aperture.
The CGS3 mid-IR spectrum, obtained with a 5.5$^{\prime\prime}$
beam, appears to be intermediate between the MIRLIN and the OSCIR data.
Therefore, we ascribe this discrepancy to observational uncertainties.
Both data points will be included in our detailed dust modeling. 

\begin{table}[h,t]
\begin{center}
{\tiny
%{\scriptsize
%{\large
\caption[]{Existing optical, IR and millimeter photometric 
data for the HD\,141569A system (see \S\ref{sec:fluxobs}).\tablenotemark{a}\label{tab:fluxobs}}
\begin{tabular}{lcccccccccl}
\hline \hline
	$\lambda$ 
        %&Observed Flux Density\tablenotemark{b}
        &$F^{\rm obs}_\nu$\tablenotemark{b}
        %&Photospheric\tablenotemark{c}
        &$F^{\rm star}_\nu$\tablenotemark{c}
        %&Dust Disk\tablenotemark{d}
        &$F^{\rm dust}_\nu$\tablenotemark{d}
        %&Observed Flux Density\tablenotemark{b}
        &$F^{\rm obs}_\lambda$\tablenotemark{b}
        %&Photospheric\tablenotemark{c}
        &$F^{\rm star}_\lambda$\tablenotemark{c}
        %&Dust Disk\tablenotemark{d}
        &$F^{\rm dust}_\lambda$\tablenotemark{d}
        &Telescope/
	&Beam/Aperture %Size
        &Reference\\
        ($\mu$m)
        &(Jy)\tablenotemark{e}
        &(Jy)
        &(Jy)
        &(${\rm erg}\s^{-1}\cm^{-2}\mum^{-1}$)
        &(${\rm erg}\s^{-1}\cm^{-2}\mum^{-1}$)
        &(${\rm erg}\s^{-1}\cm^{-2}\mum^{-1}$)
        &Instrument
      	& Size %arcsec
        &\\%reference
\hline
0.45 (B)& 10.50
        &10.50
        & - %jnu-star
        &$1.56\times 10^{-7}$
        &$1.56\times 10^{-7}$
        &- %jnu-star
        &- %instrument
        &- %beam size
        & Sylvester et al.\ 1996\\
0.55 (V)& 6.70
        &6.70
        &- %jnu-star
        &$6.64\times 10^{-8}$
        &$6.64\times 10^{-8}$ 
        &- %jnu-star
        &- %instrument
        &- %beam size
        & Sylvester et al.\ 1996\\
1.22 (J)& 3.19
        &3.19
        &-
        &$6.43\times 10^{-9}$
        &$6.43\times 10^{-9}$ 
        &- %jnu-star
        &UKIRT%instrument 
        & 7.8$^{\prime\prime}$
        & Sylvester et al.\ 1996\\
1.65 (H)& 2.02
        &2.02
        &-
        &$2.23\times 10^{-9}$
        &$2.23\times 10^{-9}$
        &- %jnu-star
        &UKIRT%instrument 
        & 7.8$^{\prime\prime}$
        & Sylvester et al.\ 1996\\
2.18 (K)& 1.29
        &1.29
        &-
        &$8.14\times 10^{-10}$
        &$8.14\times 10^{-10}$
        &-
        &UKIRT%instrument  
        & 7.8$^{\prime\prime}$
        & Sylvester et al.\ 1996\\
3.55 (L)& 0.59
        &0.58
        &0.018
        &$1.41\times 10^{-10}$
        &$1.37\times 10^{-10}$
        &$4.29\times 10^{-12}$
        &UKIRT%instrument 
        & 7.8$^{\prime\prime}$
        & Sylvester et al.\ 1996\\
3.76 (L$^\prime$)& 0.54
        &0.50
        &0.040
        &$1.14\times 10^{-10}$
        &$1.06\times 10^{-10}$
        &$8.54\times 10^{-12}$
        &UKIRT%instrument 
        & 7.8$^{\prime\prime}$
        & Sylvester et al.\ 1996\\
4.77 (M)& 0.40
        &0.33
        &0.070
        &$5.33\times 10^{-11}$
        &$4.41\times 10^{-11}$
        &$9.25\times 10^{-12}$
        &UKIRT%instrument 
        & 5.0$^{\prime\prime}$
        & Sylvester et al.\ 1996\\
\hline
12      &0.49
        &0.058
        &0.43
        &$1.02\times 10^{-11}$
        &$1.21\times 10^{-12}$
        &$9.02\times 10^{-12}$
        &IRAS%instrument
        &$1^{\prime}$
        & Sylvester et al.\ 1996\\
25      &1.87
        &0.014
        &1.86
        &$8.98\times 10^{-12}$
        &$6.64\times 10^{-14}$
        &$8.91\times 10^{-12}$
        &IRAS%instrument
        &$1^{\prime}$
        & Sylvester et al.\ 1996\\
60      &5.54
        &0.0024
        &5.54
        &$4.62\times 10^{-12}$
        &$2.04\times 10^{-15}$
        &$4.61\times 10^{-12}$
        &IRAS%instrument
        &$1^{\prime}$
        & Sylvester et al.\ 1996\\
100     &3.48
        &0.0009
        &3.48
        &$1.04\times 10^{-12}$
        &$2.65\times 10^{-16}$
        &$1.04\times 10^{-12}$
        &IRAS%instrument
        &$1^{\prime}$
        & Sylvester et al.\ 1996\\
\hline
10.8    &0.318$\pm0.016$
        &0.071
        &0.25
        &$8.18\times 10^{-12}$
        &$1.83\times 10^{-12}$
        &$6.34\times 10^{-12}$
        &OSCIR/KeckII%instrument
        &$7.9^{\prime\prime}\times7.9^{\prime\prime}$
        & Fisher et al.\ 2000\\
18.2    &0.646$\pm 0.035$
        &0.026
        &0.62
        &$5.85\times 10^{-12}$
        &$2.34\times 10^{-13}$
        &$5.62\times 10^{-12}$
        &OSCIR/KeckII%instrument
        &$7.9^{\prime\prime}\times7.9^{\prime\prime}$
        & Fisher et al.\ 2000\\
\hline
12.5\tablenotemark{f}    &0.333$\pm0.022$
        &0.054
        &0.28
        &$6.39\times 10^{-12}$
        &$1.03\times 10^{-12}$
        &$5.36\times 10^{-12}$
        &MIRLIN/KeckII%instrument
        &5$^{\prime\prime}$
        & Marsh et al.\ 2002\\
17.9\tablenotemark{f}    &0.936$\pm 0.094$
        &0.027
        &0.91
        &$8.76\times 10^{-12}$
        &$2.50\times 10^{-13}$
        &$8.51\times 10^{-12}$
        &MIRLIN/KeckII%instrument
        &5$^{\prime\prime}$
        & Marsh et al.\ 2002\\
20.8\tablenotemark{f}    &1.19$\pm 0.16$
        &0.020
        &1.17
        &$8.25\times 10^{-12}$
        &$1.38\times 10^{-13}$
        &$8.11\times 10^{-12}$
        &MIRLIN/KeckII%instrument
        &5$^{\prime\prime}$
        & Marsh et al.\ 2002\\
\hline
1350    &$0.0054\pm0.001$
        &0
        &0.0054
        &$8.89\times 10^{-18}$
        &$8.03\times 10^{-21}$
        &$8.88\times 10^{-18}$
        &SCUBA%instrument
        &22$^{\prime\prime}$ 
        & Sylvester et al.\ 2001\\
\hline
\end{tabular}
\tablenotetext{a}{In the SED modeling presented in later sections, 
                  we will include both the CGS3 mid-IR spectrum 
                  and the IRAS, OSCIR/KeckII, MIRLIN/KeckII, 
                  and SCUBA photometry.}
\tablenotetext{b}{Total flux density observed from the Earth.}
\tablenotetext{c}{Stellar photospheric contribution to the
                  observed flux density.}
\tablenotetext{d}{Dust IR excess from the disk.}
\tablenotetext{e}{$\left(\frac{F_\nu}{\rm Jy}\right)\equiv \left(\frac{F_\lambda}{{\rm erg}\s^{-1}\cm^{-2}\mum^{-1}}\right) \left(\frac{\lambda}{\mu{\rm m}}\right)^2 \left(\frac{10^{9}}{3}\right)$.}
\tablenotetext{f}{The MIRLIN/KeckII instrument has a field of view
                  of $18^{\prime\prime}\times 18^{\prime\prime}$.
                  But the flux given in this table was obtained
                  by an integration over a $5^{\prime\prime}$ diameter 
                  circular aperture (see Marsh et al.\ 2002).}
}
\end{center}
\end{table}

\section{Modeling the Dust IR Emission}
\subsection{Dust Model\label{sec:model}}
Interstellar grains undergo coagulational growth
in cold, dense molecular clouds and protostellar 
and protoplanetary dust disks. This process creates
dust with fluffy, inhomogeneous structures and fairly 
quickly leads to the formation of planetesimals of 
cometary to asteroidal sizes.

Based on its short lifetime due to radiative expulsion,
the dust in the $\hda$ disk is thought to be of
secondary debris origin (Weinberger et al.\ 1999, 2000;
Fisher et al.\ 2000; Marsh et al.\ 2002); i.e, the dust is 
generated by the breaking down of large bodies like 
planetesimals, comets, and asteroids, rather than 
direct remnant left over from the star formation process.
Since planetesimals and comets are formed through coagulation
of porous dust aggregates (Cameron 1975, 1995;
Weidenschilling \& Cuzzi 1993; Greenberg \& Li 1999;
Beckwith, Henning, \& Nakagawa 2000),
the dust generated by collisions of planetesimals should 
resemble the original dust from which they are built up.

Following Li \& Lunine (2003), we will model the $\hda$ dust
as porous aggregates of either processed or unprocessed 
interstellar materials. We will consider two dust types 
-- each represents an extreme case -- 
the ``cold-coagulation'' dust model and the ``hot-nebula'' dust model. 
While the former assumes that
the dust aggregate is formed through cold-coagulation of unaltered 
protostellar interstellar grains, the latter goes to the other 
extreme by assuming that the constituent grains in the dust
aggregate are highly-processed in the protostellar nebula
where the originally amorphous silicate grains all are annealed 
and the carbon grains all are destroyed by oxidization
and are converted into CO (see Li \& Lunine 2003 for details).

A porous dust grain is characterized by 
(1) its fluffiness or porosity $P$ 
    (the fractional volume of vacuum); 
(2) its spherical radius $a$ 
    (the radius of the sphere encompassing the entire aggregate;
     we assume all grains are spherical in shape); 
(3) composition -- the ``cold-coagulation'' dust model
    assumes the dust to be composed of amorphous silicate and 
    carbonaceous materials (and H$_2$O-dominated ices in 
    regions colder than $\sim 110-120\K$);
    the ``hot-nebula'' dust model assumes 
    the dust to be composed of only crystalline silicates 
    (and ices in cold regions).
As discussed in Li \& Lunine (2003), 
for the ``cold-coagulation'' dust model,
the mixing mass ratios for the silicate, carbon and ice constituent
grains are approximately derived, by assuming cosmic abundance, 
to be $\mcarb/\msil \approx 0.7$ and $\mice/(\msil+\mcarb) \approx 0.8$
where $\msil$, $\mcarb$, and $\mice$ are respectively 
the total mass of the silicate, carbon, and ice subgrains.  
For the ``hot-nebula'' dust model, we will take 
$\mcarb/\msil = 0$ and $\mice/\msil = 1/3$ (see Li \& Lunine 2003).
We adopt a mass density of $\rhosil=3.5\g\cm^{-3}$,
$\rhocarb=1.8\g\cm^{-3}$, and $\rhoice=1.2\g\cm^{-3}$ 
respectively for the silicate, carbonaceous, 
and ice materials.\footnote{%
  For pure H$_2$O ice we have $\rhoice \approx 1.0\g\cm^{-3}$.   
  Since interstellar ice is a ``dirty'' mixture of
  H$_2$O, NH$_3$, CO, CO$_2$, CH$_3$OH, CH$_4$ and 
  many other molecules (see Appendix A in Li \& Lunine 2003), 
  we adopt $\rhoice=1.2\g\cm^{-3}$ for ``dirty ice'' material
  (Greenberg 1968).
  }

In addition to the porous dust component discussed above,
the presence of PAH molecules in the $\hda$ disk is
clearly indicated by the red wing of the 7.7$\mum$
feature and the 11.3$\mum$ feature seen in the CGS3 mid-IR
spectrum (Sylvester et al.\ 1996; see Fig.\,\ref{fig:starobs}b).  
Therefore, we will consider two dust populations in our models: 
the porous dust component (\S\ref{sec:pah})
and the PAH component (\S\ref{sec:pd}).

\subsubsection{Polycyclic Aromatic Hydrocarbon Dust\label{sec:pah}}
PAHs are excited by ultraviolet (UV), visible, and to a lesser 
extent by long wavelength photons (Li \& Draine 2002a). 
Following Li \& Draine (2002a), we define $\Urad(r)$ as 
the $912\Angstrom$ -- $1\mum$ starlight intensity incident
on the $\hda$ disk at a distance of $r$ from the central star 
relative to the value for the Mathis, Mezger, \& Panagia 
(1983; hereafter MMP) solar neighborhood interstellar 
radiation field\footnote{%
 For radiation fields with the same 912$\Angstrom$ -- 1$\mum$
 intensity, the  6--13.6$\eV$ starlight intensity of 
 the $\Teff=10000\K$ Kurucz radiation field is about 
 1.27 times that of the MMP field.
 }
\begin{equation}\label{eq:uisrf}
\Urad(r)\equiv \frac{\left(\Rstar/2r\right)^2
\int^{1\mum}_{912\Angstrom} \Fstar d\lambda}
{\int^{1\mum}_{912\Angstrom} c\ummp d\lambda}
\approx \left(\frac{4.75\times 10^4\AU}{r}\right)^2
\end{equation}
where $R_\star (\approx 7.32\times 10^{-3}\AU)$ is the $\hda$ 
stellar radius; $\Fstar$ is the flux per unit wavelength 
(${\rm erg}\s^{-1}\cm^{-2}\mum^{-1}$)
at the top of the illuminating star's atmosphere
which is approximated by the Kurucz model 
atmospheric spectrum for B9.5\,V stars 
($\Teff = 10000\K$ and $\lg g=4.0$; Kurucz 1979);
$c$ is the speed of light; $\ummp$ is the energy density 
of the MMP interstellar radiation field.
In Figure \ref{fig:starobs}c we plot the starlight spectrum 
at $r=4.75\times 10^4\AU$ which corresponds to $\Urad=1$.
Since the $\hda$ disk extends to a distance of only 
$\sim 500\AU$, the dust in this disk receives far more
{\it intense} irradiation than the dust in the diffuse 
interstellar medium: $\Urad > 9000$!

Illuminated by starlight with an intensity of $\Urad$,
the IR emissivity per {\bf gram} dust
(${\rm erg}\s^{-1}\mum^{-1}\g^{-1}$) for a mixture of 
neutral and ionized PAHs with a size distribution of
$dn_{\rm PAH}/da$ is 
\begin{eqnarray}
\nonumber
j^{\rm PAH}_{\lambda}(\Urad) &=& \int^{\infty}_{\apahmin}
da \frac{dn_{\rm PAH}}{da}
   \int^{\infty}_{0} dT\ 4\pi B_{\lambda}(T)\\
\nonumber
&&\times \Big\{ \fion(\Urad,a) C_{\rm abs}^{\rm PAH^{+}}(a,\lambda)
\left(\frac{dP^{\rm PAH^{+}}}{dT}\right)
      +
[1-\fion(\Urad,a)] C_{\rm abs}^{\rm PAH^{0}}(a,\lambda)
\left(\frac{dP^{\rm PAH^{0}}}{dT}\right)\Big\}\\
&&\times \Big\{ \int^{\infty}_{\apahmin}
da \frac{dn_{\rm PAH}}{da} \frac{4\pi}{3} a^3 \rho_{\rm PAH}\Big\}^{-1}
\label{eq:j_pah}
\end{eqnarray}
where $a$ is the spherical radius of a PAH molecule;\footnote{%
  Small PAHs (with $\ltsim 100$ carbon atoms) 
  are expected to be planar (see Appendix A in Draine \& Li 2001).
  The term ``PAH radius'' used in this paper
  refers to the radius $a$ of a spherical grain 
  with the same carbon density as graphite ($2.24\g\cm^{-3}$) 
  and containing the same number of carbon atoms $N_{\rm C}$: 
  $a \equiv 1.288 N_{\rm C}^{1/3}\Angstrom$
  [or $N_{\rm C}\equiv 0.468(a/{\rm \AA})^3$]. 
  }
$\apahmin$ is the lower-cutoff PAH size;
$C_{\rm abs}^{\rm PAH^{0}}$ and $C_{\rm abs}^{\rm PAH^{+}}$
are the absorption cross sections for neutral and ionized 
PAH molecules, respectively (see Li \& Draine 2001a);
$B_{\lambda}(T)$ is the Planck function at temperature $T$;
$dP^{\rm PAH^{0}}(\Urad,a,T)$ and $dP^{\rm PAH^{+}}(\Urad,a,T)$,
are, respectively, the probabilities that the vibrational temperature 
will be in $[T,T+dT]$ for neutral and charged PAHs
illuminated by starlight intensity $\Urad$;
$\fion(\Urad,a)$ is the probability of finding a PAH molecule 
of radius $a$ in a non-zero charge state;
$\rho_{\rm PAH}$ is the mass density of PAHs which is taken to 
be that of graphite ($\approx 2.24\g\cm^{-3}$).

Based on the ``thermal-discrete'' method (Draine \& Li 2001),  
the temperature distribution functions $dP/d\lnT$
have been computed by Li \& Draine (2002a) for neutral 
and charged PAHs and ultrasmall silicate grains of 
a range of sizes $3.5\Angstrom \simlt a \simlt 350\Angstrom$
illuminated by stars with a range of effective
temperatures $3000\K \simlt \Teff \simlt 30000\K$
and with a range of starlight intensities 
$0.1 \simlt \Urad \simlt 10^5$.
In this work we take the Li \& Draine (2002a) $dP/d\lnT$ 
results for $\Teff=10000\K$. 

We adopt a log-normal size distribution for the PAHs in the 
$\hda$ disk, characterized by two parameters: $a_{0}$ 
and $\sigma$; $a_{0}$ and $\sigma$ respectively determine 
the peak location and the width of the log-normal distribution:
\begin{equation}\label{eq:dnda_pah}
dn_{\rm PAH}/da = 
\frac{1}{\sqrt{\pi/2}\,\sigma
        \left\{ 1 - {\rm erf}\left[
        \ln\left(a^{\rm PAH}_{\rm min}/a_0\right)/\sqrt{2}\sigma
        \right] \right\}}     
        \frac{1}{a} \exp \left\{ - \frac{1}{2} 
       \left[ \frac{\ln (a/a_{0})}{\sigma} \right]^2 \right\},
       ~~~~~~~ {\rm for}\  a > \apahmin ~~;
\end{equation}
where ${\rm erf}(y)\equiv \int^{y}_{0}\exp\left(-x^2\right)\,dx$
is the error function.
The log-normal functional form for the PAH size distribution was 
shown successful in modeling the PAH mid-IR emission spectra 
observed for the Milky Way diffuse interstellar medium 
(Li \& Draine 2001a), reflection nebulae (Li \& Draine 2002a), 
and the Small Magellanic Cloud (Li \& Draine 2002b).

Lacking a priori knowledge of the balance between 
the photodestruction and collisional production
of PAHs as a function of PAH size in the $\hda$ disk, 
we adopt a lower-cutoff of $\apahmin\equiv3.5\Angstrom$ 
(corresponding to $N_{\rm C}\approx 20$ for PAHs) 
which is the minimum survival size for PAHs in the diffuse
interstellar medium (see Li \& Draine 2001a).
It will be seen in \S\ref{sec:pahdes} that PAHs smaller
than $\sim 4.6\Angstrom$ will be photolytically unstable 
in the inner $r<100\AU$ region around the star during 
the lifetime of the disk. 
PAHs are also subject to radiative expulsion. 
Continuous replenishment of this material,
presumably by sublimation of the icy mantles 
(in which interstellar PAHs have condensed during 
the dense molecular cloud phase) 
coated on the individual subgrains of large porous dust
produced by collisions of large parent bodies,
is required to maintain the PAH disk.
Therefore, it is reasonable to assume that small PAH molecules 
with $a< 4.6\Angstrom$ are also continuously replenished
so that there exists a stable distribution of small PAHs 
in the disk. Since the dust (including PAHs) in the disk 
originates from the interstellar medium (ISM), 
it is thus reasonable to adopt the interstellar lower-cutoff 
$\apahmin=3.5\Angstrom$ for the PAHs in the disk. 
But we will also see in \S\ref{sec:pahdes} that models with
$\apahmin=4.6\Angstrom$ are also capable of closely reproducing
the observed SED including the PAH emission features.

\subsubsection{Porous Dust\label{sec:pd}}
Large porous dust in the $\hda$ disk will attain 
a steady-state temperature determined by
balancing absorption and emission,
\begin{equation}
\left(\frac{R_\star}{2r}\right)^2 
\int^{\infty}_{0}C^{\rm PD}_{\rm abs}(a,\lambda) 
\Fstar d\lambda
= \int^{\infty}_{0}C^{\rm PD}_{\rm abs}(a,\lambda) 
4\pi B_\lambda\left(T\left[a,r\right]\right)d\lambda
\end{equation}
where $C^{\rm PD}_{\rm abs}(a,\lambda)$ is the absorption cross 
section of porous dust of spherical radius $a$ at wavelength $\lambda$;
$T(a,r)$ is the equilibrium temperature of dust 
of size $a$ at a radial distance of $r$ from the star.

The IR emissivity (${\rm erg}\s^{-1}\mum^{-1}$) for porous 
dust with a size distribution of $dn_{\rm PD}/da$
located at a distance of $r$ from the central star is
\begin{equation}\label{eq:j_pd}
j^{\rm PD}_{\lambda}(r) = \int^{\amax}_{\amin}da 
\frac{dn_{\rm PD}}{da} 4\pi B_{\lambda}(T[a,r])
C_{\rm abs}^{\rm PD}(a,\lambda) ~~~.
\end{equation}

We assume a power-law dust size distribution for the porous dust
\begin{eqnarray}\label{eq:dnda_pd}
\nonumber
\frac{dn_{\rm PD}}{da} &=&
\frac{1}{\ln\left[\amax/\amin\right]}a^{-1} ~, ~~\alpha= 1 ~;
\\ 
&=& \frac{(1-\alpha)}
{\amax^{1-\alpha}-\amin^{1-\alpha}}a^{-\alpha} ~, ~~\alpha\neq 1 ~;
\end{eqnarray}
where $a$ is the spherical radius, $\amin$ is the lower-cutoff, 
$\amax$ is the upper-cutoff, and $\alpha$ is the power-law index.
We take $\amin=1\mum$,\footnote{%
 A fluffy grain of $a=1\mum$ with a porosity of $P=0.90$
 consists of $\sim 100$ constituent individual (interstellar) particles
 which have a typical size of $a\sim 0.1\mum$ (see Li \& Greenberg 1997).
 Models with a smaller $\amin$ ($=0.1\mum$) and 
 a larger $\amin$ ($=10\mum$) will be discussed in \S\ref{sec:robust}.
 \label{fnt:pd} 
 }
and $\amax=1\cm$ (this is not a critical parameter since grains 
larger than $\sim 100\mum$ emit like blackbodies
and their IR emission spectra are size-insensitive;
see Figure \ref{fig:Td}).

We use Mie theory to calculate the absorption cross sections
for porous dust. The fluffy heterogeneous dust aggregate 
is represented by an equivalent homogeneous sphere with 
an effective dielectric function $\eeff$. 
Dielectric functions for the constituent dust materials are taken from 
(1) Draine \& Lee (1984) for amorphous silicate dust;
(2) Li \& Draine (2001b) for crystalline silicate dust;
(3) Li \& Greenberg (1997) for carbonaceous dust;
(4) Li \& Greenberg (1998) for H$_2$O-dominated ice.
Let $\esil$, $\ecarb$, and $\eice$ 
be the complex dielectric functions of silicate 
(either amorphous or crystalline), carbonaceous, 
and ice dust, respectively.
Let $\fsil$, $\fcarb$, and $\fice$ respectively be 
the volume fraction of silicate, carbonaceous, and 
ice dust in an aggregate, 
which can be obtained from the mass mixing ratio
$\mcarb/\msil$ and $\mice/\left(\msil+\mcarb\right)$ 
and the fluffiness $P$ ($\equiv 1-\fsil-\fcarb-\fice$)
of the aggregate (see \S\ref{sec:model}).
We first employ the Maxwell-Garnett effective medium theory
(Bohren \& Huffman 1983) to calculate the average dielectric 
functions for the ice-coated silicate subgrains $\esilp$
and the ice-coated carbonaceous subgrains $\ecarbp$
\begin{equation}\label{eq:mgemtsil}
\esilp = \eice \frac{
\left(1+2\fsilp\right)\esil + 2\left(1-\fsilp\right)\eice}
{\left(1-\fsilp\right)\esil + \left(2+\fsilp\right)\eice} ~~,
~~~~ \fsilp = \fsil/\left(\fsil+\xi\fice\right) ~~;
\end{equation}
\begin{equation}\label{eq:mgemtcarb}
\ecarbp = \eice \frac{
\left(1+2\fcarbp\right)\ecarb + 2\left(1-\fcarbp\right)\eice}
{\left(1-\fcarbp\right)\ecarb + \left(2+\fcarbp\right)\eice} ~~,
~~~~ \fcarbp = \fcarb/\left[\fcarb+\left(1-\xi\right)\fice\right] ~~;
\end{equation}
where $\xi$ is the fraction of ice condensed on the silicate cores.
For the ``cold-coagulation'' dust model, we assume $\xi=0.5$
(i.e. in an dust aggregate, the available ice equally condenses 
on the silicate subgrains and on the carbonaceous subgrains); 
for the ``hot-nebula'' dust model, we assume $\xi=1$ 
since we assume that all carbon dust has been destroyed 
by oxidation (see Li \& Lunine 2003).
We then use the Bruggman effective medium theory 
(Bohren \& Huffman 1983) to calculate the mean dielectric 
functions $\eeff$ for the porous heterogeneous dust aggregate
\begin{equation}\label{eq:brgemt}
\left(\fsil+\xi\fice\right) \frac{\esilp-\eeff}{\esilp+2\eeff}
+\left[\fcarb+\left(1-\xi\right)\fice\right] 
\frac{\ecarbp-\eeff}{\ecarbp+2\eeff}
+\left(1-\fsil-\fcarb-\fice\right)\frac{1-\eeff}{1+2\eeff}=0 ~~;
\end{equation}
where the third term accounts for the contribution from 
the vacuum ($\epsilon=1$). 
By definition, Eqs.(\ref{eq:mgemtsil}-\ref{eq:brgemt}) are valid
for both the ``cold-coagulation'' dust model ($\xi=0.5$)
and the ``hot-nebula'' dust model ($\xi=1$)
either with ($\fice>0$) or without ($\fice=0$) ice mantles. 

\subsubsection{IR Emission Spectrum}\label{sec:iremmethod}
For a given dust size distribution and a given disk structure 
(i.e., the dust spatial density distribution), the emergent IR 
emission spectrum can be obtained by integrating over the dust 
size range, and over the entire disk. 
Assuming an optically thin (see \S\ref{sec:results})
and radially symmetric disk, 
the flux density (${\rm erg}\s^{-1}\cm^{-2}\mum^{-1}$)
received at the Earth is
\begin{equation}\label{eq:Flambda}
F_\lambda = \frac{1}{4\pi d^2}
\left[ \int_{\rin}^{\rout}
j^{\rm PD}_{\lambda}(r) \sigmar\,2\pi\,rdr
+ j^{\rm PAH}_\lambda(\Urad) m_{\rm PAH}(\Urad)\right]
\end{equation}
where $d\approx 99\pc$ is the distance from the star to the Earth;
$\sigmar$ is the surface density for the porous dust component;
$\rin$ and $\rout$ are respectively the inner and outer
boundaries of the disk; $m_{\rm PAH}(\Urad)$ is the total mass
of the PAH dust required to account for the observed 
PAH IR emission bands {\it if} the PAH molecules in the $\hda$
disk are illuminated by starlight of an intensity of $\Urad$
times that of the 912$\Angstrom$ -- 1$\mum$ MMP interstellar 
radiation field. We will justify the optical thin treatment
in \S\ref{sec:results} (also see Table \ref{tab:para}). 
The $\hda$ disk is actually asymmetric 
(Mouillet et al.\ 2001; Boccaletti et al.\ 2003; Clampin et al.\ 2003),
therefore, our model, based on an assumption of a radially 
symmetric disk, is somewhat simplified. But this is not expected
to affect the conclusion of this paper since the asymmetric dust 
spatial distribution which would account for the observed 
brightness asymmetry, with some fine tuning, would also be able 
to reproduce the observed SED as can be seen from Eq.(\ref{eq:Flambda}).

Due to their single-photon
heating nature, the IR emission spectral shape of PAHs 
is independent of the starlight intensity $\Urad$;
the absolute emissivity level simply scales with $\Urad$
(Draine \& Li 2001; see Fig.\,13 in Li \& Draine 2001a).
Namely, $j^{\rm PAH}_{\lambda}(\Urad)/\Urad$ remains identical
over the entire dust disk, {\it provided} that the charging 
condition $\Urad\sqrt{T_{\rm gas}}/n_e$ for PAHs does not 
vary in the disk.\footnote{%
  The PAH ionization fraction $\fion$ depends on $\Urad/n_e$ 
  and $T_{\rm gas}$ where $n_e$ is the electron density in the
  disk and $T_{\rm gas}$ is the gas temperature
  (Weingartner \& Draine 2001). 
  }
For computing the PAH component, it is therefore not necessary 
to integrate over the disk; instead, it is sufficient just to
calculate the PAH IR emission spectrum for a single radiation
intensity $\Urad$ (i.e., a single region $r$ in the disk).
The required PAH mass is inversely proportional to $\Urad$.

The total mass for the porous dust component is
\begin{equation}\label{eq:pdmass}
m_{\rm PD} = \int_{\rin}^{\rout} \sigmar\,2\pi\,rdr
\int^{\amax}_{\amin}da 
\frac{dn_{\rm PD}}{da} \frac{4\pi}{3} a^3 \rho_{\rm PD}
\end{equation}
where $\rho_{\rm PD}$ is the mass density of the porous dust 
component (see Appendix B in Li \& Lunine 2003).

The vertical optical depth is
\begin{equation}\label{eq:opdp}
\tau_\lambda(r) = \int^{\amax}_{\amin}da 
\frac{dn_{\rm PD}}{da} \sigmar C^{\rm PD}_{\rm ext}(a,\lambda) 
\end{equation}
where $C^{\rm PD}_{\rm ext}(a,\lambda)$ is the extinction cross 
section of the porous dust of spherical radius $a$ 
at wavelength $\lambda$. The optical depth from the PAH component 
is negligible (see \S\ref{sec:results}).

The dust spatial distribution is well constrained by 
the near-IR imaging of scattered starlight (Weinberger et al.\ 1999;
Augereau et al.\ 1999a) and mid-IR imaging of dust thermal
emission (Marsh et al.\ 2002).
We adopt the following analytical formula to approximate
the dust distribution which was derived from the scattered light
(see Fig.\,3 of Weinberger et al.\ 1999) and dust thermal 
emission (see Fig.\,4 of Marsh et al.\ 2002)\footnote{%
 We assume that the dust vertical distribution is 
 determined by the vertical hydrostatic equilibrium
 (see Eqs.[\ref{eq:nH}-\ref{eq:sclhgt}] in \S\ref{sec:ne}).
 }
\begin{eqnarray}%\label{eq:dndr}
\sigmar/\sigmap &=&
0.223 \exp\left\{-4\ln2\left[\left(r/{\rm AU}-55\right)/70\right]^2\right\}
~, ~~ \rin < r < 105\AU ~;
\label{eq:dndr1}
\\ 
&=& \exp\left\{-4\ln2\left[\left(r/{\rm AU}-203\right)/90\right]^2\right\}
~, ~~ 105\AU < r < 235\AU ~;
\label{eq:dndr2}
\\
&=& 0.914\ 140^2/\left[\left(r/{\rm AU}-310\right)^2+140^2\right]
~, ~~ 235\AU < r < \rout ~;
\label{eq:dndr3}
\end{eqnarray}
where $\sigmap$ is the mid-plane surface density at $r=\rp\equiv 203\AU$.
We note that, since we use this dust spatial distribution,
it is expected that our models are also able to reproduce 
the imaging observations.
Eq.(\ref{eq:dndr1}) represents the 12.5, 17.9, 20.8$\mum$ 
mid-IR emitting dust component (Marsh et al.\ 2002);
Eq.(\ref{eq:dndr2}) and Eq.(\ref{eq:dndr3}) describe the two
annular structures peaking at $r\approx 200\AU, 325\AU$, respectively
(Augereau et al.\ 1999a; Weinberger et al.\ 1999; Mouillet et al.\ 2001).
The inner boundary $\rin$ is set at where grains are heated to 
$\gtsim 1500\K$. Therefore, $\rin$ is a function of dust size.
For micron-sized grains, the inner boundary is roughly at $\rin=0.2\AU$. 
The outer boundary is taken to be $\rout=500\AU$ which is 
expected from the disk truncation caused by the tidal effects 
of $\hdb$, a companion star of $\hda$ 
(Weinberger et al.\ 2000).\footnote{%
 It was shown by Artymowicz \& Lubow (1994) that circumstellar disks 
 will be truncated by the tidal effects of a companion star in circular 
 orbit at approximately 0.9 of the average Roche lobe radius.
 For the $\hda$ ($m_\star \approx 2.3\msun$; van den Ancker et al.\ 1998)
 and $\hdb$ ($m_\star \approx 0.5\msun$; Weinberger et al.\ 2000)
 system with a mass ratio of $q\approx 4.6$, the Roche lobe radius is 
 $R_{\rm cr} \approx 0.49\,A\,q^{2/3}/
 \left[0.6\,q^{2/3}+\ln\left(1+q^{1/3}\right)\right]
 \approx 0.51\,A$, 
 where $A$ ($\approx 990\AU$ [Weinberger et al.\ 2000];
 $\approx 1065\AU$ [Boccaletti et al.\ 2003]) is the physical 
 separation between $\hda$ and $\hdb$. 
 However, if $\hdb$ is out of the $\hda$ disk plane,
 the physical separation $A$ must be much larger than 990$\AU$.
 As will be discussed in \S\ref{sec:robust},
 models with a larger $\rout$ value also provide 
 good fits to the observed SED (see Figure \ref{fig:coldvarpara}).
 \label{fnt:rout}
 }
This is also consistent with the near-IR scattered light 
observations reported by Weinberger et al.\ (1999) and 
Augereau et al.\ (1999a) which show that the $\hda$ disk
extends to a radial distance of about 500$\AU$.

The mean starlight intensity $\langle \Urad \rangle$,
averaged over the different portions of the disk,
and weighted by the surface density of the porous dust
population, is
\begin{eqnarray}\label{eq:umean}
\nonumber
\langle \Urad \rangle & = & \frac{
\int_{\rin}^{\rout} \Urad(r) \sigmar\,2\pi\,rdr}
{\int_{\rin}^{\rout} \sigmar\,2\pi\,rdr}
\\
\nonumber
& = & 1.24 \times 10^6 ~, ~~ \rin = 1\AU < r < 105\AU ~;
\\ 
\nonumber
& = & 2.76 \times 10^4 ~, ~~ 105 < r < \rout = 500\AU ~;
\\
& = & 4.03 \times 10^4 ~, ~~ \rin = 1\AU < r < \rout = 500\AU ~.
\end{eqnarray}

PAHs smaller than $\sim 4.6\Angstrom$ will be photodestroyed
in a time scale shorter than the age of the $\hda$ system
(see \S\ref{sec:pahdes}). PAHs will also be efficiently
removed from the disk by radiation pressure (see \S\ref{sec:rppr}).
Therefore, the PAH component in the $\hda$ disk must be 
continuously replenished, most likely through the evaporation 
of the icy mantles\footnote{%
 PAHs are abundant and widespread in the interstellar medium 
 (see L\'eger \& Puget 1984, Allamandola, Tielens, \& Barker 1985, 
  Li \& Draine 2001a). It is expected that the icy mantles of the
  grains in dust disks, formed from condensation of volatile
  molecular species in the dense cloud phase,
  contain a substantial amount of PAHs since the freeze-out
  of PAHs onto the icy mantles are also expected to occur 
  in the dense cloud phase and during the aggregation process.
  }
of grains produced by the breakup of larger bodies. 
The evaporation of icy mantles takes place at $r<100\AU$
for micron-sized dust and at $r<30\AU$
for mm-sized dust (see Figure \ref{fig:Td}). 
Therefore, it is reasonable to assume that 
the free-flying PAH molecules, responsible for
the observed PAH IR emission bands, are mainly distributed in 
the inner $r<100\AU$ region. If we assume that the spatial
distribution of PAHs follows that of the porous dust at
$r<100\AU$, the mean starlight intensity to which the PAHs
in the $\hda$ disk are exposed is 
$\langle \Urad \rangle \approx 1.24\times 10^6$
(see Eq.[\ref{eq:umean}]).

\subsubsection{PAH Ionization}\label{sec:ionrec}
PAHs acquire charge through photoelectric emission
and collisions with electrons and ions. Ideally, one
can calculate the steady-state charge distribution for
a PAH molecule of radius $a$ from the balance between 
the electron capture rates and the photoelectron emission
rates plus the ion capture rates 
(e.g. see Weingartner \& Draine 2001).
Since little is known regarding the electron density $n_e$ and
its distribution in the $\hda$ disk, we will not carry out
detailed calculations of the PAH charge distribution;
instead, we will just calculate and compare 
the photoionization rates ($\kion$; see \S\ref{sec:photophys}) 
and the electron recombination rates 
($\krec$; see \S\ref{sec:photophys} and \S\ref{sec:ne})
for PAHs as a function of size and as a function of
distance from the central star.

\begin{figure}[h]
\begin{center}
\epsfig{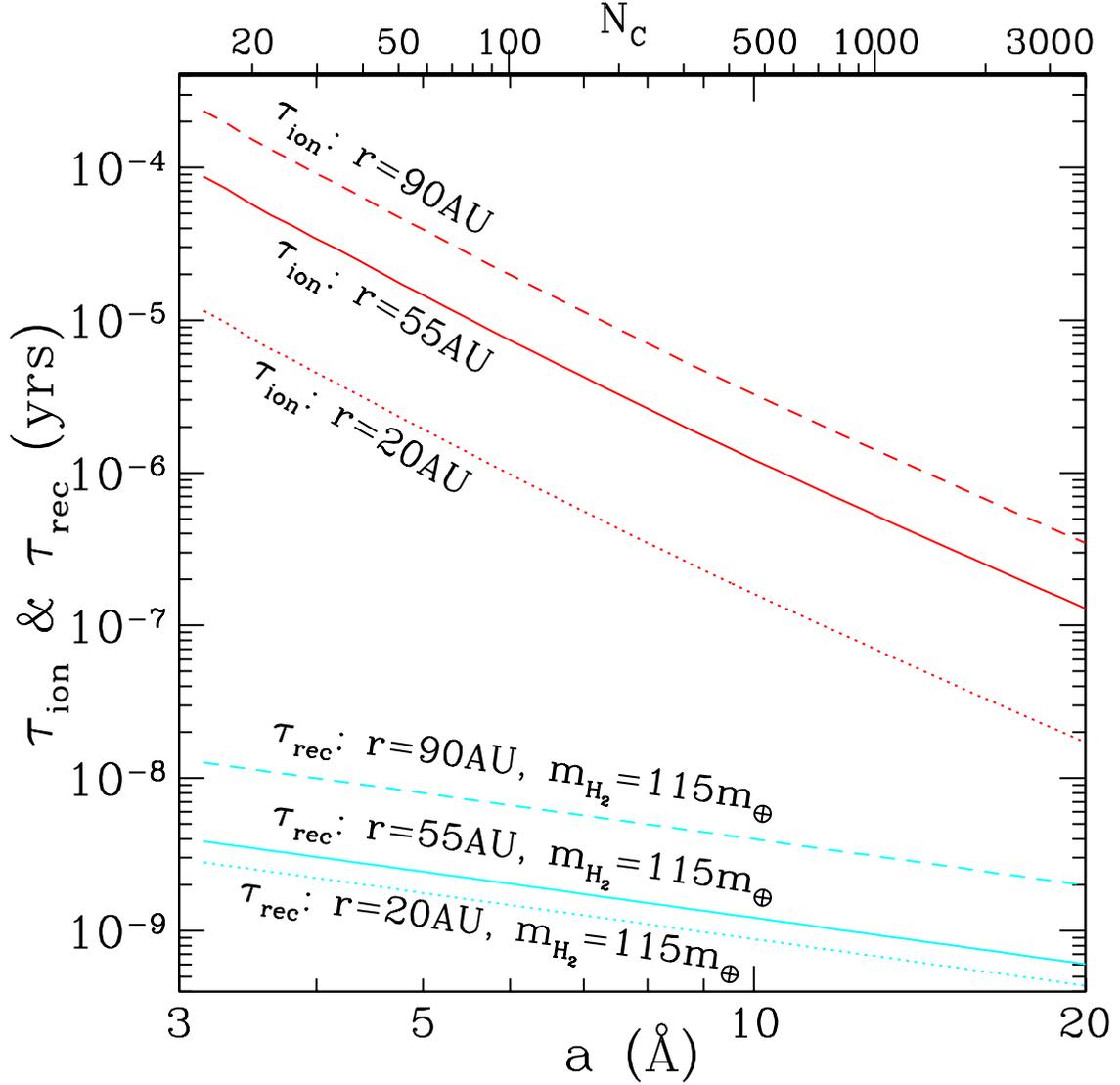}
\end{center}\vspace*{-1em}
\caption{
        \label{fig:ionrec}
        \footnotesize
        Photoionization ($\tau_{\rm ion}$; 
        upper part of this figure; see \S\ref{sec:photophys})
        and electron recombination ($\tau_{\rm rec}$; lower part;
        see \S\ref{sec:photophys} and \S\ref{sec:ne}) time scales
        for PAHs in the size ranges of $3\Angstrom<a<20\Angstrom$ 
        at $r=20\AU$ (dotted lines), 55$\AU$ (solid lines)
        and 90$\AU$ (dashed lines).
        In calculating the electron recombination time scales,
        the electron density $\nelc$ is estimated from
        the molecular hydrogen mass determined by
        Zuckerman et al.\ (1995),
        assuming that the cosmic ray ionization of $\molH$ is
        the dominant contributor to the electrons in the
        $\hda$ disk. The upper axis indicates
        the number of C atoms $N_{\rm C}$ in a PAH molecule.
        }
\end{figure}

In Figure \ref{fig:ionrec} we present the photoionization 
time scales $\tau_{\rm ion}$ ($\equiv 1/\kion$) for PAHs 
of various sizes at 3 characteristic distances $r=20,55,90\AU$
(corresponding to the maximum and half-maximums of 
the PAH distribution; see Eq.[\ref{eq:dndr1}]), 
calculated from the method described in \S\ref{sec:photophys}. 
It is seen in Figure \ref{fig:ionrec} that $\tau_{\rm ion}$ 
decreases as PAHs become larger and as PAHs are closer to the central star.
The former is because when PAHs become larger, their ionization thresholds
$E_{\rm IP}$ decrease (see Eq.[\ref{eq:Eip}]) and their UV/visible
absorption cross sections increase so that their photoionization 
rates increase (see Eq.[\ref{eq:kion}]).
The latter is because, closer to the illuminating star, 
the UV radiation field is more intense 
($\tau_{\rm ion} \propto r^{2}$; see Eq.[\ref{eq:kion}]).

Assuming that the source of the electrons in the $\hda$ disk 
is dominated by the cosmic ray ionization of $\molH$, 
and adopting a total $\molH$ mass of $m_{\molH}\approx 115\mearth$ 
inferred from submillimeter measurements of CO 
(Zuckerman, Forveille, \& Kastner 1995),
we calculate the electron recombination
time scales $\tau_{\rm rec}$ ($\equiv 1/\krec$;
see \S\ref{sec:ne} for details).
As shown in Figure \ref{fig:ionrec}, the electron recombination
time scales are by $\sim 3$ orders of magnitude smaller than 
$\tau_{\rm ion}$. Therefore, PAHs in the $\hda$ disk 
will be negatively charged.

\subsubsection{Model Parameters}\label{sec:para}
We have two sets of parameters to be specified or constrained:
\begin{enumerate}
\item The parameters $\mpah$, $\fion$, $a_0$, and $\sigma$
      determining the abundance, ionization fraction and
      size distribution of the PAH ``log-normal'' population
      (see Eqs.[\ref{eq:j_pah},\ref{eq:dnda_pah},\ref{eq:Flambda}]).
\item The parameters $\sigmap$, $P$, and $\alpha$
      determining the abundance, the fluffiness and the size 
      distribution of the porous dust population
      (see Eqs.[\ref{eq:dnda_pd},\ref{eq:brgemt},\ref{eq:Flambda},\ref{eq:dndr1},\ref{eq:dndr2},\ref{eq:dndr3}]).
\end{enumerate}

For a given set of $a_0$ and $\sigma$, 
the total amount of PAH dust $m_{\rm PAH}$ is determined 
by the absolute level of the observed PAH flux density
(see Eqs.[\ref{eq:j_pah},\ref{eq:dnda_pah},\ref{eq:Flambda}]).
Since little is known regarding the electron density $n_e$ and
its distribution in the disk, it is impossible to perform accurate 
calculations on $\fion$. 
But we have seen in \S\ref{sec:ionrec} that
PAHs in the $\hda$ disk will be negatively charged
by capturing electrons from the cosmic ray ionization of $\molH$.
Since we do not distinguish the IR properties of PAH$^-$ anions 
from PAH$^+$ cations (Li \& Draine 2001a),\footnote{%
 It has been shown that the IR properties of PAH anions closely 
 resemble those of PAH cations (e.g., see Szczepanski, Wehlburg, 
 \& Vala 1995; Langhoff 1996; Hudgins et al.\ 2000) except for 
 the very strong 3.3$\mum$ C-H stretch enhancement in the anion 
 (Szczepanski et al.\ 1995; Hudgins et al.\ 2000).
 However, Bauschlicher \& Bakes (2000) predict that
 PAH$^-$ anions have band strengths intermediate between 
 those of neutrals (strong 3.3$\mum$ C-H stretching
 and 11.3, 11.9, 12.7$\mum$ out-of-plane C-H bending modes)
 and PAH$^+$ cations (strong 6.2, 7.7$\mum$ C-C stretching 
 and 8.6$\mum$ C-H in-plane bending mode).
 In any case, the detailed differences in the band strengths
 between PAH$^-$ anions and PAH$^+$ cations would not alter
 our conclusions since, as shown in \S\ref{sec:robust}, 
 models containing a fraction ($< 50\%$) of neutral PAHs
 are also capable of fitting the observed PAH mid-IR spectrum
 (see Figures \ref{fig:ionneupah}b,c). 
 }
it is appropriate to take $\fion=1$.
But we will also consider models with $\fion=0.5,0$
(see \S\ref{sec:robust} and Figures \ref{fig:ionneupah}b,c).  

As shown in Li \& Greenberg (1998) and Li \& Lunine (2003),
cometary-type porous dust with $P\simeq 0.90$ is able to 
reproduce the observed SEDs from the IR to millimeter for 
the disks around the MS star $\beta$ Pictoris 
and the pre-MS star $\hra$; therefore, we will start with
$P=0.90$ in modeling the $\hda$ SED; but models with
larger $P$ or smaller $P$ will also be discussed 
(see \S\ref{sec:robust} and Figure \ref{fig:coldvarpara}).

It is interesting to note that, as early as 1965, 
based on a Monte Carlo simulation, 
Cameron \& Schneck (1965) showed that 
a very open structure (``fairy castle'')
with a porosity in the range of $0.83\simlt P\simlt 0.89$
(Cameron \& Schneck [1965] used the term ``{\it underdense}'') 
is expected for dust aggregates
assembled from randomly incident particles,
depending on their shapes, sizes and incidence angles
(also see Cameron 1995).
A porosity of $P\approx 0.90$ for the ``cold-coagulation'' dust 
(which implies a mass density of 
$\approx 1.7\times \left[1-\Pice\right]\g\cm^{-3}
\approx 0.46\g\cm^{-3}$ for this dust assuming
ice mantles coated on the individual subgrains 
[$\Pice\approx 0.73$]; see Appendix B in Li \& Lunine 2003)
is consistent with the mean mass density $\approx 0.5\g\cm^{-3}$
of cometary nuclei determined from studies of comet splitting 
and of non-gravitational forces (Rickman 2003; Whipple 1999).
Comets are likely to have been formed initially through
the cold accumulation of interstellar dust 
(e.g. see Greenberg \& Li 1999).
Hence, the ice-coated ``cold-coagulation'' dust aggregates 
we consider here are plausible building blocks of 
cometesimals and, hierarchically, comets.

Similar to the PAH mass parameter $\mpah$, for a given set of
$P$ and $\alpha$, the total amount of porous dust 
(characterized by $\sigmap$; 
see Eqs.[\ref{eq:pdmass},\ref{eq:dndr1},\ref{eq:dndr2},\ref{eq:dndr3}])
is determined by the absolute level of the observed flux density
at wavelengths $\lambda \simgt 10\mum$.

Therefore, we are left with $\Npar=3$ adjustable parameters: 
$a_0$, $\sigma$, and $\alpha$.

For a model with $N_{\rm par}$ adjustable parameters, 
the goodness of fit is measured by
\begin{equation}
\frac{\chi^2}{\dof} = \frac{\sum_{i=1}^{105}
                      \left(\left[\lambda F_\lambda\right]_{\rm mod} -
                      \left[\lambda F_\lambda\right]_{\rm obs}\right)^2}
                      {\Npar + 105}
\end{equation}
where 
$[\lambda F_{\lambda}]_{\rm mod}$ is the model spectrum
(see Eq.[\ref{eq:Flambda}]), and
$[\lambda F_{\lambda}]_{\rm obs}$ is the (dereddened, stellar 
photospheric contributions subtracted) observational spectrum
which includes the 10 IRAS, OSCIR, MIRLIN, and SCUBA photometric 
data points and the 95 CGS3 spectroscopic data points
(see Table \ref{tab:fluxobs}). 

\subsection{Model Results}\label{sec:results}
\begin{figure}[h]
\begin{center}
\epsfig{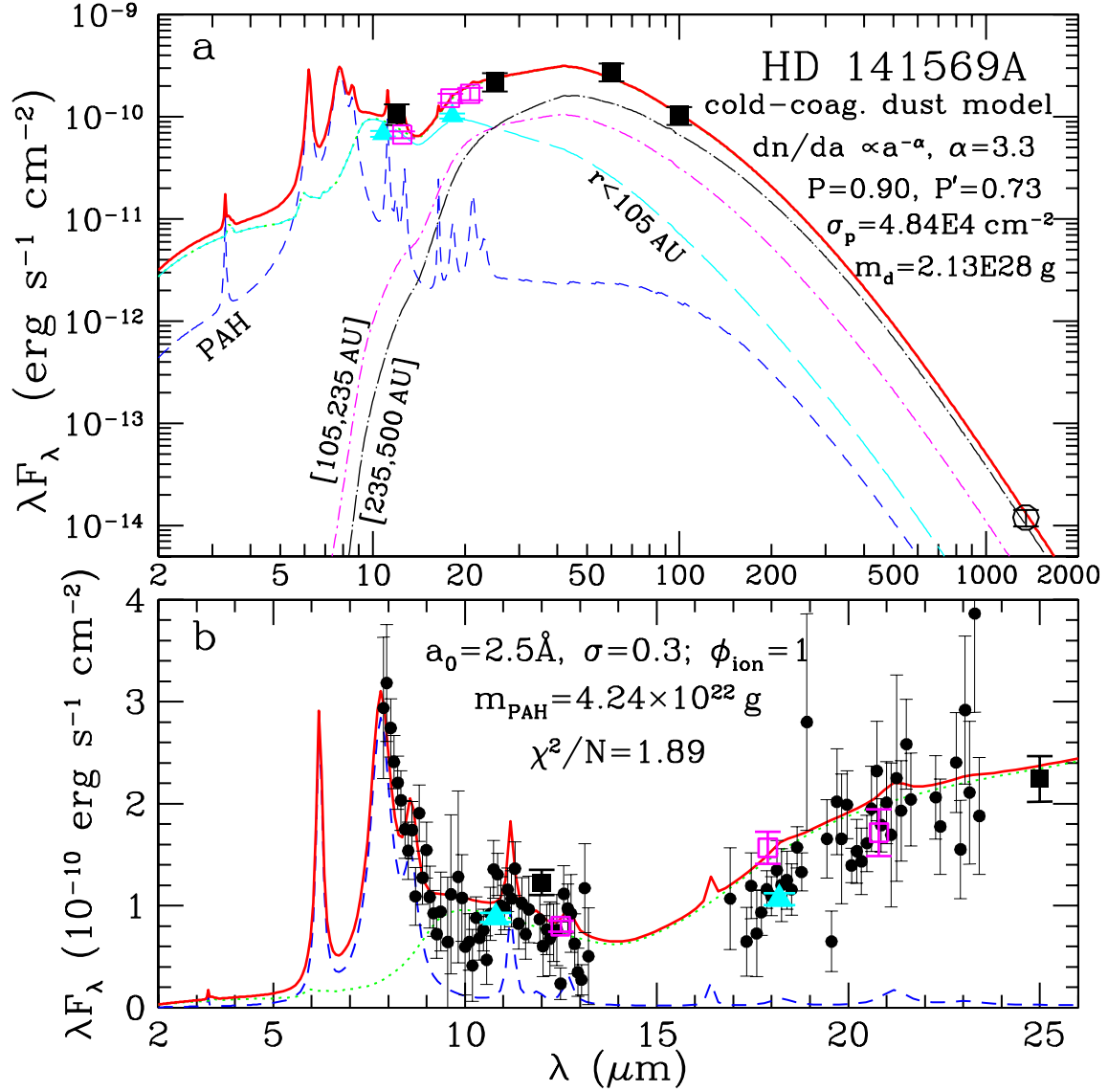}
\end{center}\vspace*{-1em}
\caption{
        \label{fig:cold90}
        \footnotesize
        Comparison of the ``cold-coagulation'' dust model
        spectrum (red solid line; model no.\,1) 
        with the observed spectral energy distribution.
        The model consists of two dust components:
        (1) the {\bf PAH cation component} ($\fion=1$; blue dashed line) 
        with a log-normal size distribution 
        (and a lower-cutoff of $\apahmin=3.5\Angstrom$)
        characterized by a peak parameter of $a_0=2.5\Angstrom$, 
        a width parameter of $\sigma=0.3$,
        and a total mass of $\mpah\approx 7.10\times 10^{-6}\mearth$,
        and (2) the {\bf porous dust component} (green dotted line) 
        characterized by a porosity of $P=0.90$ ($\Pice\approx 0.73$), 
        a power-law index of $\alpha\approx 3.3$ 
        for the dust size distribution ($a\in [1\mum,1\cm]$), 
        a dust surface density of
        $\sigmap\approx 4.84\times 10^4\cm^{-2}$, 
        and a total mass of $\md\approx 3.56\mearth$.
        We adopt the dust spatial distribution ($0.2<r<500\AU$) 
        derived from imaging observations of scattered light 
        and dust thermal emission
        (see Eqs.[\ref{eq:dndr1}-\ref{eq:dndr3}] 
        and \S\ref{sec:iremmethod}).
        Upper panel ({\bf a}) also illustrates 
        the regional contributions of the porous dust 
        in the inner warm region ($r<105\AU$; cyan long dashed line), 
        the inner ring ($105<r<235\AU$; magenta dot-dashed line),
        and the outer ring ($235<r<500\AU$; black dot-long dashed line).
        The observational SED includes 
        the 10.8 and 18.2$\mum$ OSCIR/KeckII photometry 
        (cyan filled triangles; Fisher et al.\ 2000), 
        the 12, 25, 60 and 100$\mum$ IRAS photometry 
        (black filled squares; Sylvester et al.\ 1996),
        the 12.5, 17.9, and 20.8$\mum$ MIRLIN/KeckII photometry 
        (magenta open squares; Marsh et al.\ 2002),
        and the CGS3 mid-IR spectrum 
        (black filled circles in lower panel [{\bf b}]
         which illustrates the PAH emission bands; 
         Sylvester et al.\ 1996).
        All observational data in ({\bf a}) and ({\bf b})
        have been dereddened, with the stellar photospheric 
        contribution subtracted.
        }
\end{figure}

We first consider the ``cold-coagulation'' dust model.
As shown in Figure \ref{fig:cold90}, our best-fit model
(no.\,1) provides an excellent fit to the entire spectrum 
(including both the broadband photometry and the PAH emission features)
with emission from a mixture of PAH cations ($\fion=1$) 
and porous dust with a porosity of $P=0.90$, a power-law
size distribution index of $\alpha \approx 3.3$, 
and a total dust mass of $\md\approx 3.56\mearth$.
The PAH component is characterized by $a_0\approx 2.5\Angstrom$,
$\sigma\approx0.3$, and $\mpah\approx 7.10\times 10^{-6}\mearth$. 
See Table \ref{tab:para} for a full set of model parameters.
For the sake of convenience, hereafter we call this model 
the ``canonical cold-coagulation model''.

To illustrate the contributions to the observed IR emission
from dust in different regions,
we show in Figure \ref{fig:cold90}a the IR emission
from warm dust in the inner region of $r < 105\AU$,
and from cool dust in the inner ($105 < r < 235\AU$)
and outer rings ($235 < r < 500\AU$).
While the cool dust in the two rings dominates the emission
at $\lambda \simgt 60\mum$ and accounts for $\approx 80\%$ of
the $25\mum$ IRAS flux (it is the dust in the outer ring that
is responsible for the $1350\mum$ SCUBA emission),
the mid-IR 10.8 and 18.2$\mum$ radiation detected by the OSCIR/KeckII
instrument (Fisher et al.\ 2000; plotted as filled-triangles
in Figure \ref{fig:cold90}) are exclusively emitted
by warm dust at $r < 105\AU$. This confirms earlier work
of Fisher et al.\ (2000) and Augereau et al.\ (1999a)
who suggested the existence of two separate populations of dust
grains in the $\hda$ disk: inner, warmer grains that emit 
the mid-IR (10 and 18$\mum$) radiation, and more distant grains 
that are responsible for the scattered near-IR flux seen in 
the NICMOS image and the far-IR (60 and 100$\mum$) emission 
detected by IRAS. It is worth noting that in our model, 
the $r<105\AU$ warm dust fully accounts for both the 10.8$\mum$
and the 18.2$\mum$ emission, with very little contribution from
the PAH component (also see Figure \ref{fig:cold90}b). 
This is consistent with the discovery of Fisher et al.\ (2000) 
that the emitting regions at both wavelengths are of comparable size 
and therefore the same population of dust may emit 
both the 10.8 and 18.2$\mum$ radiation. 

Our model yields a total dust IR flux of
$\int_{912\Angstrom}^{\infty} F_\lambda d\lambda \approx
5.89\times 10^{-10}\erg\s^{-1}\cm^{-2}$,
corresponding to an IR luminosity of $\sim 0.18\,L_\odot$,
$\sim 8.1\times 10^{-3}$ of the total stellar luminosity
($L_\star \approx 22.4\,L_\odot$). 
The fractional contributions of the PAH component,
the inner warm dust component, the inner ring, 
and the outer ring (see Figure \ref{fig:cold90}a), 
are approximately 12\%, 26\%, 25\%, and 37\%, respectively.

Our model predicts a vertical optical depth at visible wavelengths 
of $\tau(V)\approx 0.024$ at $\rp=203\AU$ (see Table \ref{tab:para}).
The vertical optical depths at other radial distances are
smaller than this as can be seen from Eqs.(\ref{eq:opdp}-\ref{eq:dndr3}).
The in-plane optical depth is also much smaller than one.\footnote{%
 The in-plane optical depth can be calculated from
 $\tau_\lambda \equiv \int_{\rin}^{\rout}
 \sigmar/\left[\sqrt{2\pi}H(r)\right]\,dr
 \int^{\amax}_{\amin} C^{\rm PD}_{\rm ext}(a,\lambda)
 \left(dn_{\rm PD}/da\right)\,da$ 
 where $H$ is the vertical scale height (see Eq.[\ref{eq:sclhgt}]).
 Our best-fit model leads to $\tau(V) \approx 0.097$. 
 }
This justifies the optically-thin treatment employed 
in the entire paper.
The PAH component causes little
extinction: its contribution to the visual vertical optical depth 
at $r=55\AU$ where the distribution of PAHs in the $\hda$ disk 
peaks (see \S\ref{sec:iremmethod}) is only 
$\tau(V)\approx 1.54\times 10^{-4}$.\footnote{%
 At a first glance, one may argue that the PAH optical depth
 at visual wavelengths would also be about 12\% of 
 the total since the PAH component
 emits $\sim 12\%$ of the total flux so that it must have absorbed
 $\sim 12\%$ of the total stellar radiation.  
 We remind the reader that in our model the PAH dust lies closer 
 to the star ($r<105\AU$) so that it is exposed to a radiation
 field $\approx 1.24\times 10^6/4.03\times 10^4\approx 31$ 
 times stronger than that of the porous dust component
 (see Eq.[\ref{eq:umean}] and \S\ref{sec:iremmethod}).
 Another reason lies in the fact that PAHs absorb more
 efficiently in the UV compared to the visible 
 than do micron-sized silicate or carbon grains
 (see Figure 2 in Li \& Draine 2001a, Figure 1 in Li \& Draine 2002a,
  Figures 4a and 5a in Draine \& Lee 1984).
  } 

\begin{figure}[h]
\begin{center}
\epsfig{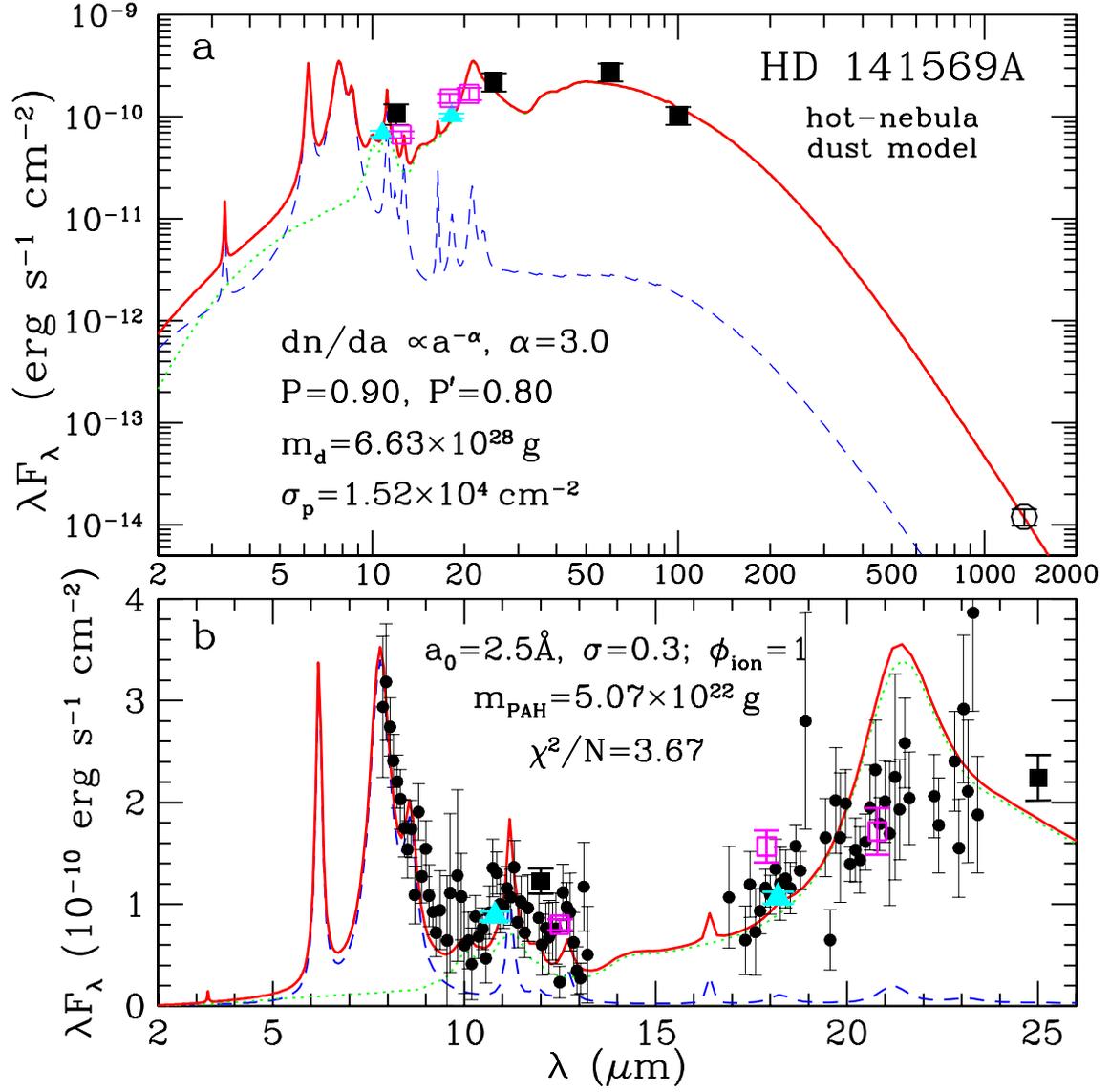}
\end{center}\vspace*{-1em}
\caption{
        \label{fig:hot90}
        \footnotesize
        Same as Figure \ref{fig:cold90} but for
        the $P=0.90$ ($\Pice\approx 0.80$) 
        ``hot-nebula'' dust model (no.\,2).  
        This model predicts a broad prominent feature at $21\mum$ 
        due to crystalline silicates which is not seen
        in the CGS3 spectrum (Sylvester et al.\ 1996).
        }
\end{figure}

Similar results are obtained for the ``hot-nebula'' dust model.
We show in Figure \ref{fig:hot90} the best-fit $P=0.90$
($\Pice\approx 0.80$) ``hot-nebula'' dust model spectrum
(model no.\,2; see Table \ref{tab:para} for model parameters).
In comparison with the observed SED,
the overall fit is acceptable except that the model predicts
a strong crystalline silicate emission band at $21\mum$
which is not seen in the CGS3 spectrum (Sylvester et al.\ 1996),
suggesting that only a small fraction of the hot silicate dust is 
in crystalline form.

\begin{figure}[h]
\begin{center}
\epsfig{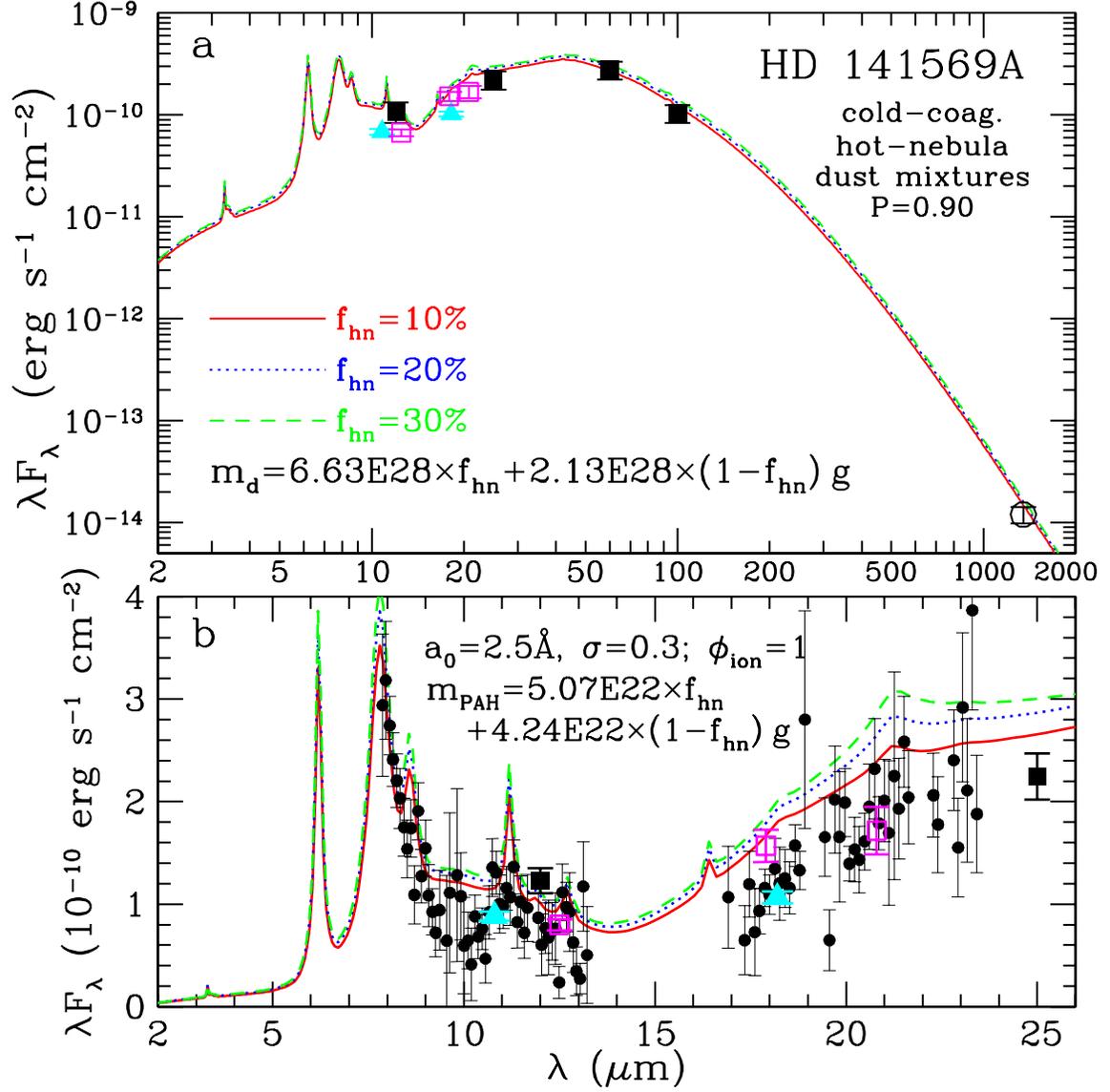}
\end{center}\vspace*{-1em}
\caption{
        \label{fig:coldhot90}
        \footnotesize
        Model spectra expected from a mixture of
        the ``cold-coagulation'' dust 
        (model no.\,1; Figure \ref{fig:cold90})
        with a mass fraction of $\left(1-\fhn\right)$
        and the ``hot-nebula'' dust 
        (model no.\,2; Figure \ref{fig:hot90})
        with a mass fraction of $\fhn$
        (red solid line -- $\fhn =10\%$;
         blue dotted line -- $\fhn =20\%$; 
         green dashed line -- $\fhn =30\%$).
        The non-detection of the 21$\mum$ crystalline silicate feature 
        limits the fraction of ``hot-nebula'' dust (i.e. crystalline
        silicates) to $\fhn\approx 10\%$.
        }
\end{figure}

One should keep in mind that 
the ``cold-coagulation'' model and the ``hot-nebula'' model 
represent two extremes; the actual dust in the $\hda$ disk 
is likely to be somewhere intermediate between these two types. 
Therefore, we consider a dust model consisting of a mixture 
of these two dust types with various mass mixing ratios.\footnote{%
  For simplicity, we assume that they are separate populations.
  In reality, it is more likely that the highly processed
  ``hot-nebula'' dust and the ``unaltered'' protostellar interstellar 
  dust are mixed to some extent and hence form 
  porous heterogeneous aggregates. 
  }
Let $\fhn$ be the mass fraction of the ``hot-nebula'' dust
in the mixture of highly-processed dust and ``unaltered'' dust.
Let $\left(\lambda F_\lambda/\md\right)_{\rm cc}$
and $\left(\lambda F_\lambda/\md\right)_{\rm hn}$
respectively be the IR flux for per gram 
``cold-coagulation'' dust (model no.\,1)
and ``hot-nebula'' dust (model no.\,2).
Let $\md^{\rm cc}$ and $\md^{\rm hn}$ respectively
be the dust mass required for 
the ``cold-coagulation'' model (no.\,1) 
and the ``hot-nebula'' model (no.\,2) 
to reproduce the observed SED.
The model spectrum expected from a mixture of these two kinds
of dust with a mixing mass fraction of $\fhn$
is simply
\begin{equation}\label{eq:mix}
\lambda F_\lambda = \left[
              \left(1-\fhn\right)
              \left(\frac{\lambda F_\lambda}{\md}\right)_{\rm cc}
        + \fhn \left(\frac{\lambda F_\lambda}{\md}\right)_{\rm hn}
               \right] 
           \left[\left(1-\fhn\right) \md^{\rm cc} 
           +  \fhn \md^{\rm hn}\right] ~~.    
\end{equation}
In Figure \ref{fig:coldhot90} we plot the model spectra 
with $\fhn=10\%$, 20\% and 30\%. 
As can be seen in Figure \ref{fig:coldhot90}b, the non-detection
of the 21$\mum$ crystalline silicate feature places an upper limit
of $\fhn\approx 10\%$. 

\begin{figure}[h]
\begin{center}
\epsfig{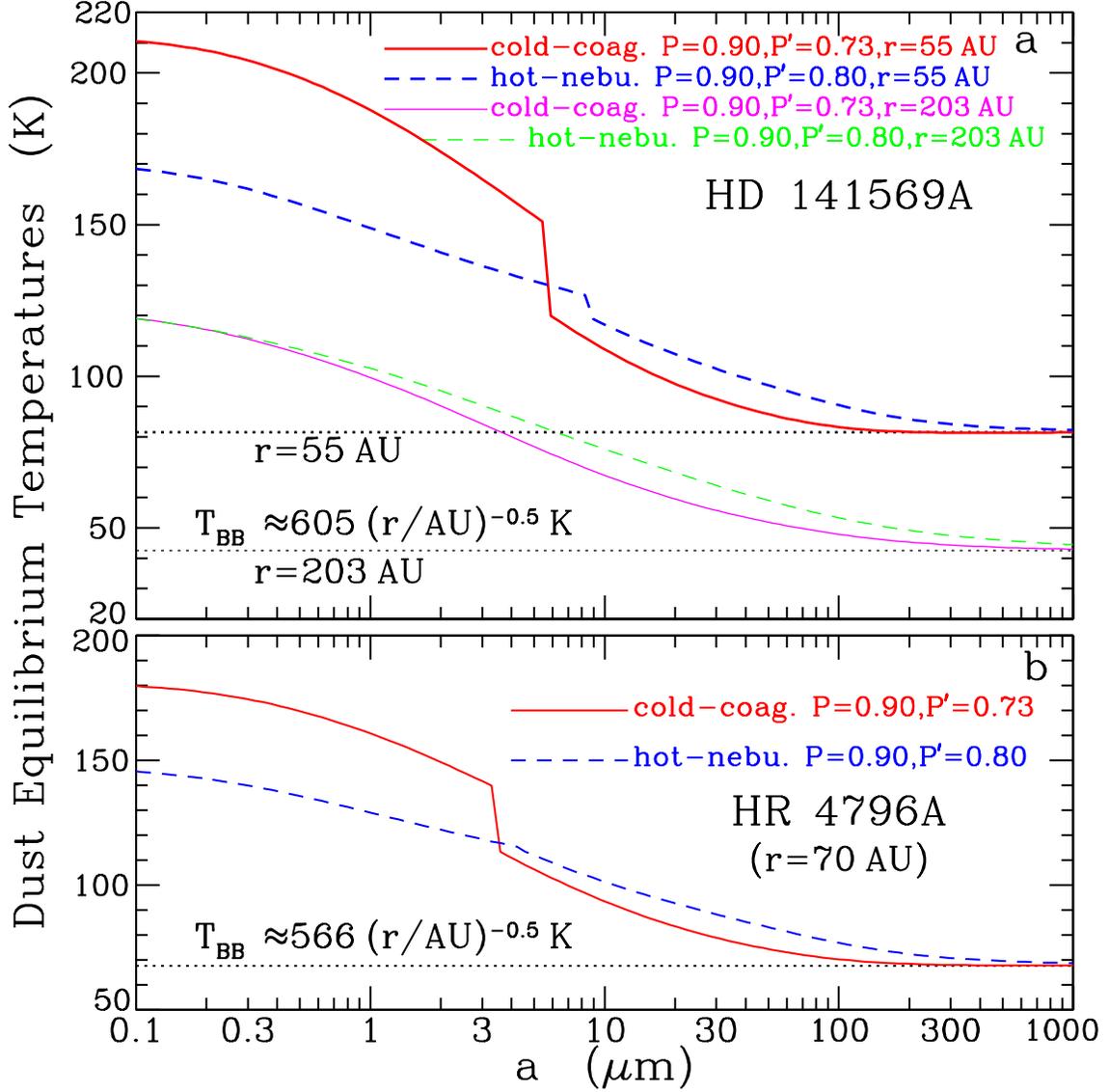}
\end{center}\vspace*{-1em}
\caption{
        \label{fig:Td}
        \footnotesize
        Equilibrium temperatures for ``cold-coagulation'' dust
        of $P=0.90$ ($\Pice\approx 0.73$) at a distance of
        $r=55\AU$ (solid line), and $r=203\AU$ (thin solid line)
        from $\hda$ (upper panel {\bf [a]}), 
        $r=70\AU$ (solid line) from $\hra$ (lower panel {\bf [b]}),
        and for ``hot-nebula'' dust
        of $P=0.90$ ($\Pice\approx 0.80$) at
        $r=55\AU$ (dashed line) and $r=203\AU$ (thin dashed line)
        from $\hda$ (upper panel {\bf [a]}), 
        and $r=70\AU$ (dashed line) from $\hra$ (lower panel {\bf [b]}).
        The horizontal dotted lines plot the black-body temperatures
        of $T_{\rm BB} \approx 81.6, 42.5, 67.6\K$ respectively
        at $r=55$, $203\AU$ from $\hda$ (upper panel {\bf [a]}), 
        and $r=70\AU$ from $\hra$ (lower panel {\bf [b]}). 
        The discontinuity at $T\approx 120\K$ in the temperature 
        profiles (as a function of grain size) results from 
        ice sublimation: grains hotter than $\sim 120\K$ 
        do not maintain icy mantles and therefore attain higher
        temperatures than their icy counterpart since silicate
        and carbon materials have a much higher UV/visual absorptivity
        than H$_2$O-dominated ice.
        }
\end{figure}

Finally, we present in Figure \ref{fig:Td} the equilibrium
temperatures for the best-fit ``cold-coagulation'' dust
($P=0.90$, $\Pice\approx 0.73$) 
and ``hot-nebula'' dust ($P=0.90$, $\Pice\approx 0.80$)  
as a function of grain size 
at a distance of $r=55$, $203\AU$ from $\hda$,
and $r=70\AU$ from $\hra$.

\begin{table}[h,t]
\begin{center}
{\tiny
%{\scriptsize
\caption[]{Models for the $\hda$ dust disk IR emission.\tablenotemark{a}\label{tab:para}}
\begin{tabular}{cccccccccccccc}
\hline \hline
	model 
        &dust 
	&$P$
        &$\Pice$\tablenotemark{b}
      	& $\alpha$\tablenotemark{c}

      	& $\langle a\rangle$\tablenotemark{d}
	& $\langle a^2\rangle$\tablenotemark{d}
	& $\langle a^3\rangle$\tablenotemark{d}
        & $m_{\rm d}$
	& $\sigmap$\tablenotemark{e}
        & $m_{\rm PAH}$\tablenotemark{f}
        &$\tau_{\rm p}(V)$\tablenotemark{g}
      	& $\chi^2/N$ 
        & note
	\\
        no.
        &type
      	&
        & 
        & 

      	&($\mu$m) %<a>
	&(${\rm \mu m}^2$) %<a>^2
	&($10^{-8}\cm^3$) %<a>^3
        &($10^{28}\g$) %m_dust;
	&($10^{4}\cm^{-2}$) %sigma_p;
	&($10^{22}\g$) %m_pah 
        &($10^{-2}$) %optical_depth_5500A
        &%chi^2/N
	&\\%note 
\hline
1	& cold-coag. & 0.90 & 0.73 % dust type & P & P'
        & 3.3 % dn/da ~ a^alpha;
        & 1.77 & 7.18 & 0.21 % <a> & <a^2> & <a^3>
        & 2.13 & 4.84 %dust mass & surface density 
        & 4.24 %m_pah;
        & 2.41 & 1.89 & preferred\\ %tau_V %chi^2/N & comments
2	& hot-nebu. & 0.90 & 0.80 % dust type & P & P'
        & 3.0 % dn/da ~ a^alpha;
        & 2.00 & 18.4 & 2.00 % <a> & <a^2> & <a^3>
        & 6.63 & 1.52 %dust mass & surface density 
        & 5.07 %m_pah;
        & 1.85 & 3.67 & \\ %tau_V %chi^2/N & comments
3	& cold-coag. & 0.90 & 0.73 % dust type & P & P'
        & 3.2 % dn/da ~ a^alpha;
        & 0.18 & 0.099 & 0.0027 % <a> & <a^2> & <a^3>
        & 2.52 & 430.5 %dust mass & surface density 
        & 4.29 %m_pah;
        & 2.18 & 2.27 & $\amax=0.1\mum$\\ %tau_V %chi^2/N & comments
4	& cold-coag. & 0.90 & 0.73 % dust type & P & P'
        & 3.6 % dn/da ~ a^alpha;
        & 16.2 & 426.5 & 9.65 % <a> & <a^2> & <a^3>
        & 1.66 & 0.081 %dust mass & surface density 
        & 5.51 %m_pah;
        & 2.21 & 1.72 & $\amax=10\mum$\\ %tau_V %chi^2/N & comments
5	& cold-coag. & 0.90 & 0.73 % dust type & P & P'
        & 3.3 % dn/da ~ a^alpha;
        & 1.77 & 7.18 & 0.21 % <a> & <a^2> & <a^3>
        & 3.01 & 4.66 %dust mass & surface density 
        & 4.32 %m_pah;
        & 2.32 & 1.70 & $\rout=1000\AU$\\ %tau_V %chi^2/N & comments
6	& cold-coag. & 0.95 & 0.87 % dust type & P & P'
        & 3.2 % dn/da ~ a^alpha;
        & 1.83 & 9.26 & 0.44 % <a> & <a^2> & <a^3>
        & 1.73 & 3.89 %dust mass & surface density 
        & 4.28 %m_pah;
        & 2.38 & 2.00 & \\ %tau_V %chi^2/N & comments
7	& cold-coag. & 0.80 & 0.46 % dust type & P & P'
        & 3.5 % dn/da ~ a^alpha;
        & 1.67 & 4.95 & 0.049 % <a> & <a^2> & <a^3>
        & 1.38 & 6.58 %dust mass & surface density 
        & 4.22 %m_pah;
        & 2.21 & 2.13 & \\ %tau_V %chi^2/N & comments
8	& cold-coag. & 0.90 & 0.73 % dust type & P & P'
        & 3.3 % dn/da ~ a^alpha;
        & 1.77 & 7.18 & 0.21 % <a> & <a^2> & <a^3>
        & 2.13 & 4.84 %dust mass & surface density 
        & 4.45 %m_pah;
        & 2.41 & 2.34 & $a_0=3.5\Angstrom, \sigma=0.4$\\ %tau_V/chi/com.
9	& cold-coag. & 0.90 & 0.73 % dust type & P & P'
        & 3.3 % dn/da ~ a^alpha;
        & 1.77 & 7.18 & 0.21 % <a> & <a^2> & <a^3>
        & 2.13 & 4.84 %dust mass & surface density 
        & 6.19 %m_pah;
        & 2.41 & 4.97 & $a_0=3\Angstrom, \sigma=0.2, \fion=0.5$\\
          %tau_V/chi/com.
10	& cold-coag. & 0.90 & 0.73 % dust type & P & P'
        & 3.3 % dn/da ~ a^alpha;
        & 1.77 & 7.18 & 0.21 % <a> & <a^2> & <a^3>
        & 2.13 & 4.84 %dust mass & surface density 
        & 9.95 %m_pah;
        & 2.41 & 24.4 & $a_0=3\Angstrom, \sigma=0.2, \fion=0$\\
          %tau_V/chi/com.
11	& cold-coag. & 0.90 & 0.73 % dust type & P & P'
        & 3.3 % dn/da ~ a^alpha;
        & 1.77 & 7.18 & 0.21 % <a> & <a^2> & <a^3>
        & 2.13 & 4.84 %dust mass & surface density 
        & 4.47 %m_pah;
        & 2.41 & 2.02 & $a_{\rm min}^{\rm PAH} =4.6\Angstrom$\\%tau_V/chi/com.
\hline
\end{tabular}
\tablenotetext{a}{Unless otherwise stated, all models assume 
                  (1) $\amin=1\mum$ and $\amax=1\cm$ for 
                      the porous dust component;
                  (2) $a_0=2.5\Angstrom$, $\sigma=0.3$, 
                      $\apahmin=3.5\Angstrom$, and $\fion=1$ 
                      for the PAH component;
                   (3) $r\in [0.2,500\AU]$ 
                       for the dust spatial distribution
                       given in Eqs.(\ref{eq:dndr1}-\ref{eq:dndr3}).}
\tablenotetext{b}{The porosity of the ice-coated porous aggregate
                  (see Appendix B in Li \& Lunine [2003]).}
\tablenotetext{c}{The index of the power-law size distribution
                  for the porous dust component:
                  $dn(a)/da \sim a^{-\alpha}$ for $a\in [\amin,\amax]$
                  (see Eq.[\ref{eq:dnda_pd}]).}
\tablenotetext{d}{$\langle a^{\gamma}\rangle = \int_{\amin}^{\amax} da\
                   a^{\gamma}\ dn(a)/da/\int_{\amin}^{\amax} da\ dn(a)/da$.
                   The dust surface mass density can be written as
                   $\sigma(r;m)=\langle m\rangle\,\sigmar$ 
                   with the mean dust mass 
                   $\langle m \rangle =(4\pi/3)\langle a^3\rangle
                   \langle\rho\rangle$ 
                   where $\langle\rho\rangle \approx 2.5\,(1-P)\g\cm^{-3}$,
                   $\langle\rho^\prime\rangle \approx 1.7\,(1-\Pice)\g\cm^{-3}$
                   for the ``cold-coagulation'' model;
                   and $\langle\rho\rangle \approx 3.5\,(1-P)\g\cm^{-3}$,
                   $\langle\rho^\prime\rangle \approx 2.4\,(1-\Pice)\g\cm^{-3}$
                   for the ``hot-nebula'' model 
                   (see Appendix B in Li \& Lunine [2003]).}
\tablenotetext{e}{The mid-plane surface density ($z=0$) at $\rp=203\AU$
                  for the porous dust component (see Eq.[\ref{eq:dndr2}]).}
\tablenotetext{f}{We assume that the spatial distribution of the PAH 
                  component follows that of the inner warm porous dust
                  (which is responsible for the 10.8 and 18.2$\mum$
                   mid-IR emission observed by Fisher et al.\ [2000];
                   see Figure \ref{fig:cold90}a and Eq.[\ref{eq:dndr1}])
                  at $r\simlt 105\AU$ (see \S\ref{sec:iremmethod}).}
\tablenotetext{g}{The vertical optical depth at $\lambda=0.55\mum$
                  and $\rp=203\AU$. The contribution to $\tau(V)$ 
                  from the PAH component
                  is negligible (see \S\ref{sec:results}).}
}
\end{center}
\end{table}

\section{Discussion\label{sec:discussion}}
The narrow, ring-like structure was originally suggested by 
Sylvester \& Skinner (1996) for the $\hda$ disk in modeling 
the IRAS data (plus the JCMT upper limit at $\lambda = 1.1\mm$). 
However, their best-fitting models required a huge inner cavity 
with a radius of $\simgt 670\AU$. 
On the other hand, it was shown by Malfait, Bogaret, \& Waelkens
(1998) that a flat, conventional disk extending from $\sim 7\AU$
to $\sim 1400\AU$ with a power-law index of $-0.6$ for the dust 
spatial distribution was also able to fit the IRAS data.

Malfait et al.\ (1998) modelled the $\hda$ IRAS data
in terms of dust with a hypothetical opacity law of 
$\kabs \propto \lambda^{-1}$ and a hypothetical radial-dependent
temperature of $T_{\rm d}(r) = 6180\left(r/R_\star\right)^{-2/5}\K$.
Their inferred dust spatial distribution of $dn(r)/dr \propto r^{-0.6}$
with $r\in [7,1400\AU]$ is inconsistent with the current view
of a central void at $r\simgt 20\AU$ 
(e.g. Marsh et al.\ 2002; Mouillet et al.\ 2001).
In the work of Sylvester \& Skinner (1996), the dust in 
the $\hda$ disk was modelled either as a mixture of two
separate components (amorphous silicate and amorphous carbon) 
or as a single pure amorphous carbon component. 
The dust was taken to be compact spheres with 
a power-law size distribution of $dn/da\sim a^{-3.5}$ 
in the size range of $50\Angstrom < a < 100\mum$.
The variation of dust density within the disk with
distance from the star was also treated as a power-law.
Their model did not treat the PAH emission bands.
It is therefore not surprising that their model 
spectrum was deficient at $\lambda \simlt 15\mum$. 

Using the OSCIR/KeckII 10.8 and 18.2$\mum$ flux ratio,
Fisher et al.\ (2000) placed an upper limit of $\amax \simeq 1\mum$
on the radius of the dust grains responsible for 
the mid-IR emission at {\it both} wavelengths 
if they are made of compact silicate spheres.
Since this conclusion was drawn from the assumption 
that the dust is located at a distance of 20$\AU$ 
from the star, a wide range of grain sizes may be 
allowed if the mid-IR emitting dust is wide spread 
within $r<105\AU$ (see Eq.[\ref{eq:dndr1}]).

In contrast, a lower limit of 
$\amin \simeq \left(0.6\pm0.2\right)\mum/\left(1-P\right)$
was derived by Boccaletti et al.\ (2003) for grains
with a porosity of $P$, assuming that the slight surface
brightness difference between 
their ground-based 2.2$\mum$ imaging data
and the HST NICMOS data at 1.1$\mum$ (Weinberger et al.\ 1999) 
and 1.6$\mum$ (Augereau et al.\ 1999a)
is caused by the color effect of the grains.
A grain size distribution steeper than $dn/da \propto a^{-3}$
was also inferred from such an analysis. 
However, these conclusions are very sensitive to 
the scaling factor between the target star and 
the calibrator star which is uncertain. 
Once the scaling factor uncertainty is included
in their analysis, a much wider range of grain sizes
is allowed.

In comparison with the narrow, sharply bounded $\hra$ disk
(Schneider et al.\ 1999), the dust annulus around $\hda$ is 
about 9--10 times wider and displays a more complex
ring-gap-ring morphology (Augereau et al.\ 1999a; 
Weinberger et al.\ 1999; Clampin et al.\ 2003).
In addition, for the $\hda$ disk, there is clear evidence 
for the existence of a warm dust population
at $r<105\AU$ inward of the inner ring (Marsh et al.\ 2002).
While a warm ``zodiacal'' dust component at a radial distance of
a few AU from $\hra$ was invoked by Koerner et al.\ (1998) 
and Augereau et al.\ (1999b) to account for the emission at 
$\lambda \simlt 12\mum$, it was shown later by Li \& Lunine (2003) 
that the porous dust model naturally explains the entire dust
emission including that at $\lambda \simlt 12\mum$ and therefore,
there does not seem to be any strong evidences for a ``zodiacal''
dust component in the inner disk of $\hra$.
Very recently, Marsh et al.\ (2002) found that the radial 
optical depth profile of the $\hda$ disk at $\lambda=12.5\mum$ 
appears to increase inside of $r=20\AU$ at a $2\sigma$ level.
Does this indicate the existence of a ``zodiacal'' dust component
in the $\hda$ disk? SIRTF imaging at the 3.6, 4.5, 
and 5.8$\mum$ IRAC bands (see Table \ref{tab:sirtf} 
and \S\ref{sec:sirtf}) may be able to address this question
since the ``zodiacal'' dust may also reveal its existence 
by emitting at the IRAC bands. It is interesting to note that 
the M band (4.77$\mum$) photometry does appear to show a small 
excess over the stellar photosphere (see Figure \ref{fig:starobs}a).
But the uncertainties both in the photometric flux determination
and in the Kurucz model atmosphere representation prevent 
a definite conclusion.

The time for complete sublimation of the ice mantles 
coated on the individual tenth-micron subgrains of 
a porous aggregate of temperature $T$ 
(see Figure \ref{fig:Td}) is 
$\tausubl \approx 1.5\times 10^{-12}
\left(\Delta a/{\rm \mu m}\right)
\left(\rhoice/{\rm g}\cm^{-3}\right)
10^{\left(2480\K/T\right)}\left(T/{\rm K}\right)^{-3.5}\yr$
(Backman \& Paresce 1993)
where $\Delta a \approx 0.04\mum$ is the ice mantle thickness.
By integrating $\tausubl$ over the dust size distribution 
$dn_{\rm PD}/da$, we estimate the mean sublimation time scales
to be $\langle \tausubl\rangle \approx 7.6\times 10^5, 
3.3\times 10^{32}, 2.8\times 10^{10}\yr$
respectively for the best-fit ``cold-coagulation'' dust at 
a distance of $r=55\AU$, $203\AU$ from $\hda$, and $70\AU$ from $\hra$.
If PAHs in dust disks indeed originate from the sublimation
of icy mantles, we expect to see PAHs in the inner warm region
of the $\hda$ disk ($r<100\AU$) where $\tausubl < \tau_{\rm age}$, 
but not in the ring regions of the $\hda$
($r>100\AU$) and $\hra$ disks ($r\gtsim 70\AU$)
where $\tausubl \gg \tau_{\rm age}$.

The 3.3, 6.2, 7.7, 8.6 and 11.3$\mum$ vibrational bands 
diagnostic of PAH molecules are not seen in the $\hra$ disk.
As mentioned above, this is mainly because in the $r=70\AU$ ring 
region, the sublimation of icy mantles which ejects PAHs
is too slow ($\tausubl \gg \tau_{\rm age}\approx 8\pm 3\Myr$).
Even if PAHs are produced by mechanisms
other than sublimation (e.g. shattering of carbonaceous grains),
the recondensation of PAHs in the icy mantles of the porous dust 
component occurs on a time scale of a few thousand years,
much shorter than the age of the $\hra$ system:
$\tau_{\rm cond} \approx \left(v_{\rm PAH} n_{\rm d}
\pi \langle a^2\rangle\right)^{-1} 
\approx \left(\pi\langle m_{\rm PAH}\rangle/
8 k T_{\rm gas}\right)^{1/2} \left(\sqrt{2\pi}H/\sigmap\right)
\left(\pi \langle a^2\rangle\right)^{-1}
\approx 1570\yr$
where $T_{\rm gas}(r) \approx 566 \left(r/{\rm AU}\right)^{-1/2}\K 
\approx 67\K$ is the gas temperature; 
$\langle m_{\rm PAH}\rangle \approx 7.99\times 10^{-22}\g$
is the average mass of the PAH molecules with a size distribution 
like that of the $\hda$ disk 
(i.e., $a_0=2.5\Angstrom$ and $\sigma=0.3$; see \S\ref{sec:results}); 
$H \approx 0.5\AU$ is the vertical scale height (Kenyon et al.\ 1999); 
$\sigmap \approx 4.90\times 10^4\cm^{-2}$ (Li \& Lunine 2003) 
and $n_{\rm d}\approx \sigmap/\sqrt{2\pi}H$ are respectively
the mid-plane surface density and volume density of 
the porous dust component at $\rp=70\AU$;
$\pi \langle a^2\rangle \approx 9.02\times 10^{-7}\cm^2$ 
(Li \& Lunine 2003) is the total grain surface areas
integrated over the $dn/da\propto a^{-2.9}$ size distribution
with $a\in [1\mum,1\cm]$ 
(see Table 1 [model no.\,2] in Li \& Lunine 2003).
Similarly, the time scale for the accretion of PAHs on
grains at $r=203\AU$ from the $\hda$ disk is about $9\times 10^5\yr$,
shorter than the age of the $\hda$ system.
Therefore, free-flying PAH molecules are unlikely to exist
in the ring regions of the $\hda$ and $\hra$ disks.

To summarize, the respective presence and absence of free-flying PAHs 
in the $\hda$ and $\hra$ disks can be attributed to the structural
differences between these two disks.
It is worth noting that the $\beta$ Pictoris disk,
modelled best as a wedge-shaped disk with a constant opening angle
(Artymowicz, Burrows, \& Paresce 1989, 
Backman, Gillett, \& Witteborn 1992), 
is more like the $\hda$ disk than the $\hra$ disk does in the sense
that the $\beta$ Pictoris disk also extends to the inner warm
region where the sublimation of icy mantles occurs.
However, the existing mid-IR spectra of the $\beta$ Pictoris disk
do not appear to show any evidence of PAHs 
(Telesco \& Knacke 1991, Knacke et al.\ 1993, 
Pantin, Waelkens, \& Malfait 1999, 
Weinberger, Becklin, \& Zuckerman 2003).
It is not clear whether the absence of PAHs 
in the $\beta$ Pictoris disk is due to its evolutionary status 
or just because the PAH emission features are 
swamped by the strong 9.7$\mum$ and 11.3$\mum$ silicate bands.
With an age of $\approx 12^{+8}_{-4}\Myr$
(Barrado y Navascu\'{e}s et al.\ 1999, Zuckerman et al.\ 2001),
the $\beta$ Pictoris disk is more evolved than the $\hda$ disk.
SIRTF spectroscopy of $\beta$ Pictoris at $\lambda \simlt 8\mum$ 
will allow us to draw a more definite conclusion regarding
the presence or absence of PAHs in the inner disk of
$\beta$ Pictoris.

The presence of transiently heated PAH molecules in the dusty 
environments of Herbig Ae/Be stars has been observationally 
well established (see e.g. Whittet et al.\ 1983;
Brooke, Tokunaga, \& Strom 1993; Sylvester et al.\ 1996; 
Siebenmorgen et al.\ 2000). 
Natta \&  Kr\"{u}gel (1995) have calculated the mid-IR emission 
spectra expected for PAHs around Herbig Ae/Be stars, 
assuming a spherical shell geometry for the dust distribution. 
Sylvester, Skinner, \& Barlow (1997) have modelled the SEDs of 
8 Vega-like dust disks with temperature-fluctuating 
small silicate and amorphous carbon grains of $a=3\Angstrom$ 
and $5\Angstrom$ included. So far, however, little has been done in 
modeling the PAH component in the $\hda$ disk.
This work actually represents the first successful modeling of 
its entire SED, including the PAH emission features.

\subsection{Robustness\label{sec:robust}}
The dust model presented here is fully described by
10 parameters: 
(1) $\apahmin$, $a_0$, $\sigma$, and $\fion$ 
for the PAH component;
(2) $\amin$, $\amax$, $\alpha$, and $P$ 
for the porous dust component; 
(3) $\rin$ and $\rout$ for the dust spatial distribution $dn(r)/dr$.
But, as discussed in \S\ref{sec:para}, we actually have
only 3 free parameters: $a_0$, $\sigma$, and $\alpha$.

Other parameters have been constrained 
by rather general considerations:
(1) we take $\apahmin = 3.5\Angstrom$, the radius of 
the smallest survival interstellar PAH molecule (\S\ref{sec:pah});
(2) we take $\fion=1$ since PAHs in the $\hda$ disk 
are expected to be negatively charged (\S\ref{sec:ionrec});
(3) we take $\amin=1\mum$ since this is the smallest size 
that makes sense in the framework of interstellar 
dust aggregation (see Footnote-\ref{fnt:pd});
(4) we take $\amax=1\cm$; this size is not well constrained
but neither is it crucial, since grains larger 
than $\sim 100\mum$ emit like blackbodies and 
their IR emission spectra are size-insensitive
(see Figure \ref{fig:Td});
(5) we take $P=0.90$ since dust of such a high porosity
reproduces the SEDs of the $\beta$ Pictoris disk and 
the $\hra$ disk; a porosity of $P\approx 0.90$ is expected 
for fluffy aggregates formed by the accumulation of interstellar 
dust and is also consistent with the low density nature of 
cometary nuclei (\S\ref{sec:para});
(6) we take $\rin$ to be the radial distance where the refractory
dust starts to evaporate (\S\ref{sec:iremmethod});
(7) we take $\rout=500\AU$ which is expected from disk truncation 
caused by the tidal effects of $\hdb$ (Footnote-\ref{fnt:rout}).
The dust spatial distribution is taken to be that derived from
the near-IR imaging of scattered starlight 
and mid-IR imaging of dust thermal emission 
(Eqs.[\ref{eq:dndr1}-\ref{eq:dndr3}] and \S\ref{sec:iremmethod}).

\begin{figure}[h]
\begin{center}
\epsfig{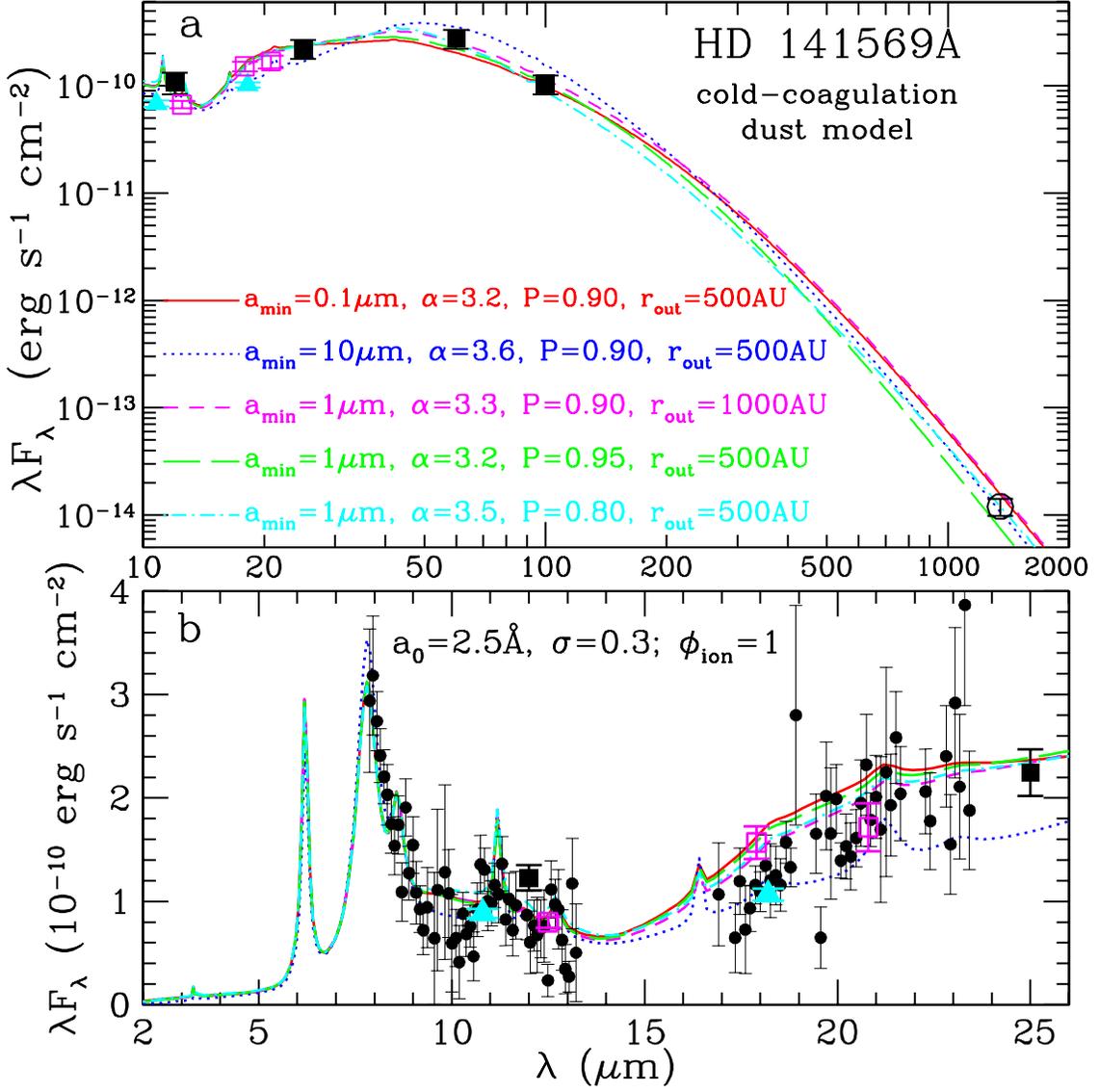}
\end{center}\vspace*{-1em}
\caption{
        \label{fig:coldvarpara}
        \footnotesize
        Comparison of the observed SED with the theoretical
        spectra calculated from the ``cold-coagulation'' dust
        models with $\amin=0.1\mum$ (model no.\,3; $\alpha=3.2$; 
        red solid line),
        $\amin=10\mum$ (model no.\,4; $\alpha=3.6$; blue dotted line),
        $\rout=1000\AU$ (model no.\,5; $\alpha=3.3$; magenta dashed line),
        $P=0.95$ (model no.\,6; $\alpha=3.2$; green long dashed line),
        and $P=0.80$ (model no.\,7; $\alpha=3.5$; cyan dot-dashed line).
        Unless otherwise specified,
        the remaining parameters of these models (see Table \ref{tab:para})
        are the same as those of the canonical ``cold-coagulation'' 
        model (no.\,1) described in \S\ref{sec:results}
        (see Figure \ref{fig:cold90}).
        }
\end{figure}

To be complete, we have also considered models with 
parameters differing from the nominal ones described above.
In the following, except the specifically stated parameters,
all other parameters remain the same as those of the
canonical ``cold-coagulation'' model derived in 
\S\ref{sec:results} (model no.\,1; see Figure \ref{fig:cold90}).
See Table \ref{tab:para} for all the key parameters.
\begin{itemize}
\item The $\amin=0.1\mum$ model 
      (no.\,3; $\alpha=3.2$; Figure \ref{fig:coldvarpara}; 
      red solid line) --- 
      is almost identical to the $\amin=1\mum$ model 
      (no.\,1; Figure \ref{fig:cold90}) 
      except its slight deficiency at the IRAS 60$\mum$ band.

\item The $\amin=10\mum$ model 
      (no.\,4; $\alpha=3.6$; Figure \ref{fig:coldvarpara}; 
      blue dotted line) --- 
      emits too much at the IRAS 60 and 100$\mum$ bands,
      and too little at the MIRLIN/KeckII 17.9, 
      20.8$\mum$ bands and the IRAS 25$\mum$ band.
\item The $\rout=1000\AU$ model 
      (no.\,5; $\alpha=3.3$; Figure \ref{fig:coldvarpara}; 
      magenta dashed line) --- 
      is almost identical to the $\rout=500\AU$ model 
      (no.\,1; Figure \ref{fig:cold90}) since the cool dust at
      $500<r<1000\AU$ emits very little in the wavelength 
      range of interest here. It emits only $\sim 10\%$ more 
      at $\lambda >200\mum$ than the $\rout=500\AU$ model.
\item The $P=0.95$ model 
      (no.\,6; $\alpha=3.2$; Figure \ref{fig:coldvarpara}; 
      green long dashed line) --- 
      is almost identical to the $P=0.90$ model 
      (no.\,1; $\alpha=3.3$; Figure \ref{fig:cold90}) 
      except its slight deficiency at the SCUBA 1350$\mum$ band.
      Its emission at 1350$\mum$ is $\sim 50\%$ lower than that 
      of the $P=0.90$ model. 
\item The $P=0.80$ model 
      (no.\,7; $\alpha=3.5$; Figure \ref{fig:coldvarpara}; 
      cyan dot-dashed line) --- 
      is almost identical to the $P=0.90$ model 
      (no.\,1; $\alpha=3.3$; Figure \ref{fig:cold90}) 
      except that it emits $\sim 20\%$ less at $100<\lambda<600\mum$.
\end{itemize}

In summary, although the canonical ``cold-coagulation'' dust model
presented in \S\ref{sec:results} is preferred, some flexibilities
are allowed in modeling the $\hda$ SED; 
i.e., {\it no fine tuning is needed} in order for our models
to achieve an excellent fit to the observed entire SED.
Therefore, the porous dust model made of interstellar materials, 
previously shown successful in reproducing the SEDs of the dust
disks around $\beta$ Pictoris and $\hra$, is also robust
in modeling the IR emission from the $\hda$ disk.

\begin{figure}[h]
\begin{center}
\epsfig{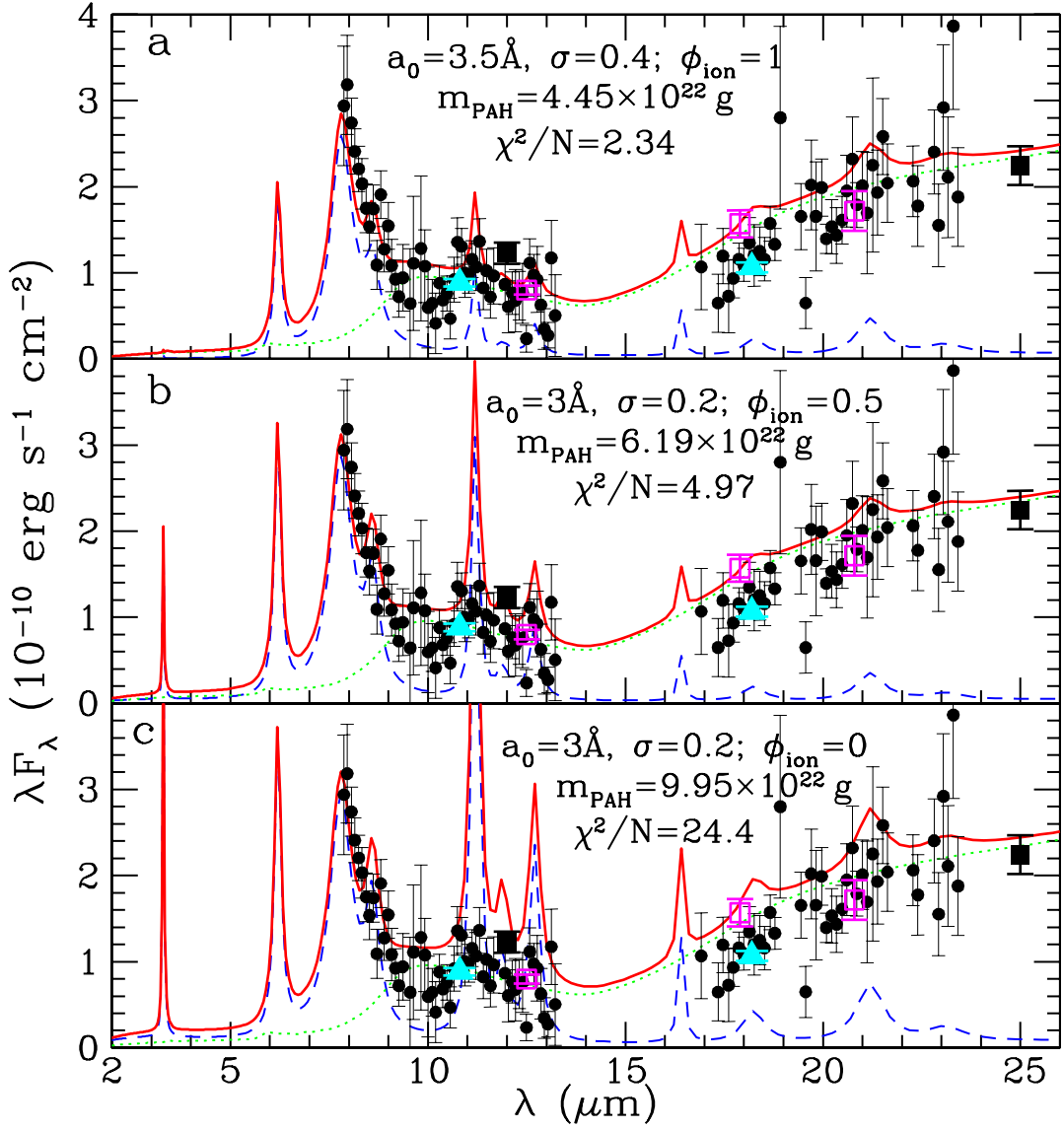}
\end{center}\vspace*{-1em}
\caption{
        \label{fig:ionneupah}
        \footnotesize
        Best-fits to the CGS3 mid-IR spectrum provided by
        (a) charged PAHs ($\fion=1.0$) of interstellar size
        distribution ($a_0=3.5\Angstrom$ and $\sigma=0.4$; model no.\,8);
        (b) a mixture of neutral and charged PAHs 
        ($\fion=0.5$, $a_0=3.0\Angstrom$ and $\sigma=0.2$; model no.\,9); 
        and (c) neutral PAHs 
        ($\fion=0$, $a_0=3.0\Angstrom$ and $\sigma=0.2$; model no.\,10).
        Other dust parameters (see Table \ref{tab:para}) are
        the same as those of the canonical ``cold-coagulation''
        model (no.\,1; Figure \ref{fig:cold90}).
        }
\end{figure}

\begin{figure}[h]
\begin{center}
\epsfig{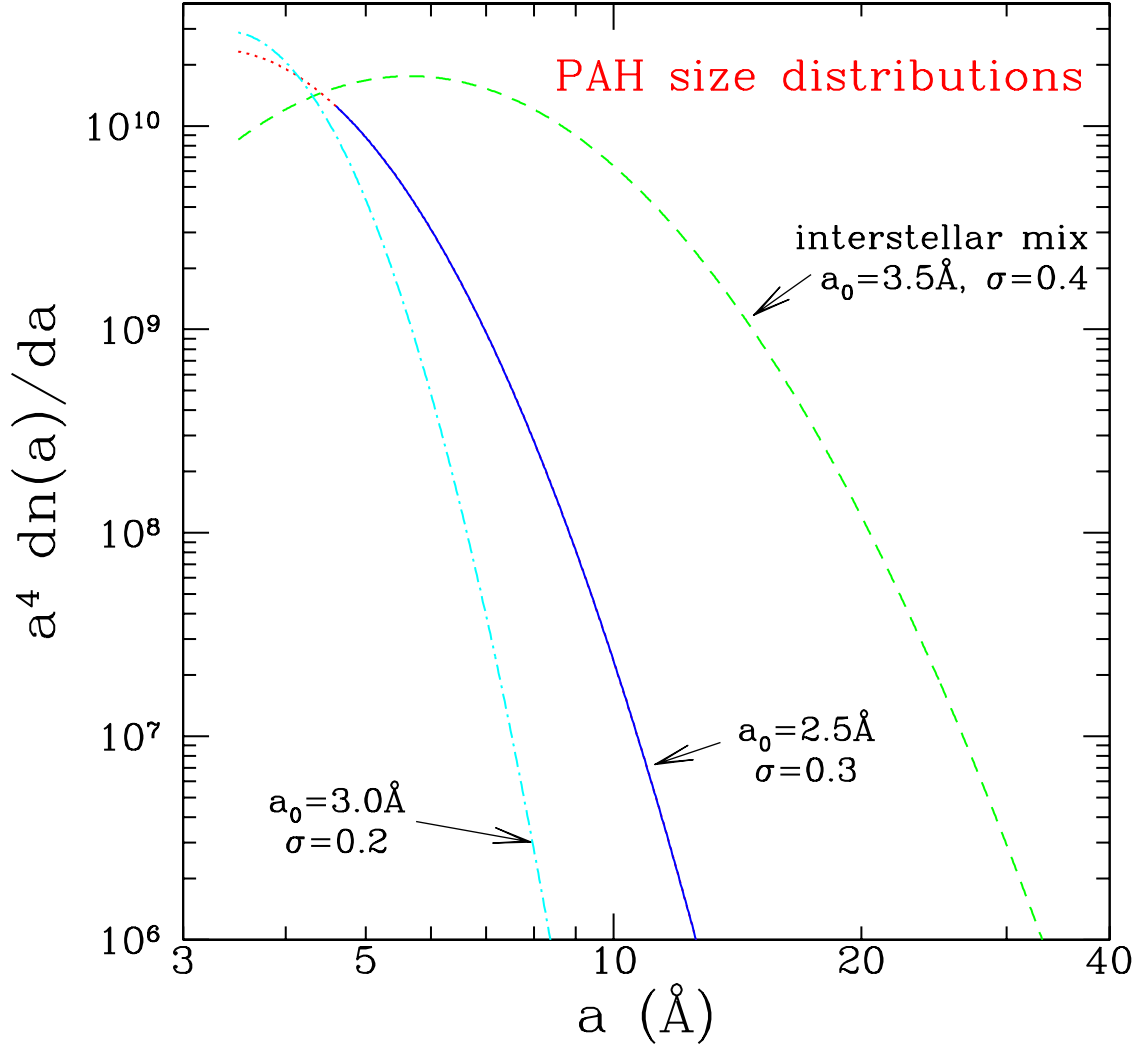}
\end{center}\vspace*{-1em}
\caption{
        \label{fig:dndapah}
        \footnotesize
        Size distributions for the PAH population.
        A log-normal functional form (see Eq.[\ref{eq:dnda_pah}]) 
        is assumed. The distribution is determined by
        a peak-parameter $a_0$ and a width-parameter $\sigma$:
        $a_0 = 2.5\Angstrom$ 
        and $\sigma = 0.3$ (solid and dotted lines) for
        the canonical best-fit model (no.\,1; see Figure \ref{fig:cold90}
        and \S\ref{sec:results});
        $a_0 = 3.5\Angstrom$ and $\sigma = 0.4$ 
        (dashed line) for the interstellar PAH model 
        (see Li \& Draine 2001a; see Figure \ref{fig:ionneupah}a
        and \S\ref{sec:robust} in this paper);
        $a_0 = 3.0\Angstrom$ and $\sigma = 0.2$ 
        (dot-dashed line) for the $\fion=0.5$ and $\fion=0$ model
        (see Figures \ref{fig:ionneupah}b,c and 
        \S\ref{sec:robust} in this paper).
        The PAH lower-cutoff sizes are all set 
        at $\apahmin=3.5\Angstrom$ except for one case 
        (dotted line: $\apahmin=4.6\Angstrom$ [model no.\,11]; 
        see \S\ref{sec:pahdes} and Figure \ref{fig:cold90paha46}).
        }
\end{figure}

We have also tried to fit the observed SED using 
the interstellar PAH size distribution, characterized 
by $a_0=3.5\Angstrom$ and $\sigma=0.4$ (Li \& Draine 2001a).
As shown in Figure \ref{fig:ionneupah}a, the model spectrum
predicted from the interstellar mixture of PAH cations (model no.\,8)
is also in an excellent agreement with the observed mid-IR spectrum.
In comparison with the canonical best-fit mixture 
($a_0=2.5\Angstrom$ and $\sigma=0.3$; 
model no.\,1; see Figure \ref{fig:cold90}),
the interstellar mixture, rich in relatively large PAHs
(see Figure \ref{fig:dndapah}), emits a little bit more (less) 
at wavelengths longward (shortward) of $\sim 10\mum$. 
This is because large PAHs do not reach temperatures as high
as do small PAHs when heated by energetic photons. 

So far, all models assume $\fion=1$. 
We now consider models containing neutral PAHs. 
For illustration, two models are considered: 
a model with a mixture of both neutral and ionized PAHs 
($\fion=0.5$; model no.\,9), and a model only consisting of 
pure neutral PAHs ($\fion=0$; model no.\,10).
In Figure \ref{fig:ionneupah}b and \ref{fig:ionneupah}c
we respectively show the best-fit model spectra calculated from
the $\fion=0.5$ model and from the neutral PAH model.
The PAH component of both dust models has a size distribution of 
$a_0\approx3.0\Angstrom$ and $\sigma\approx 0.2$
(see Figure \ref{fig:dndapah}). Again, other parameters 
(see Table \ref{tab:para}) are the same as those of 
the canonical ``cold-coagulation'' model (no.\,1).
An inspection of Figure \ref{fig:ionneupah}c would 
immediately lead us to rule out the neutral PAH model since
this model produces too strong an 11.3$\mum$ feature 
(as well as a 12.7$\mum$ feature) to reconcile
with the CGS3 spectrum of Sylvester et al.\ (1996).
The fit to the CGS3 spectrum by the $\fion=0.5$ model 
is acceptable except that it also appears to emit a bit 
too much in the 11.3$\mum$ band, although the quality of 
the CGS3 spectrum is inadequate to disprove this model. 
Note that the model-predicted 11.3$\mum$ feature is much sharper 
than the observed one since we adopt the mean interstellar 
PAH bandwidths (see Table 7 of Li \& Draine 2001a).
As a matter of fact, the integrated flux in the calculated 
11.3$\mum$ feature is comparable to the observed. 
The determination of the band strength of the 11.3$\mum$ feature 
is also complicated by its adjacent continuum. 
In our dust model, the warm porous dust in the inner region 
($r<105\AU$) dominates the continuum underneath the 11.3$\mum$ feature.
Future observational studies (e.g. SIRTF) will provide 
more information on $\fion$. 
But at this moment, it is safe to state that a large fraction 
of the PAH molecules responsible for
the observed mid-IR spectrum in the $\hda$ disk must be ionized.

\subsection{PAH Destruction}\label{sec:pahdes}
\begin{figure}[h]
\begin{center}
\epsfig{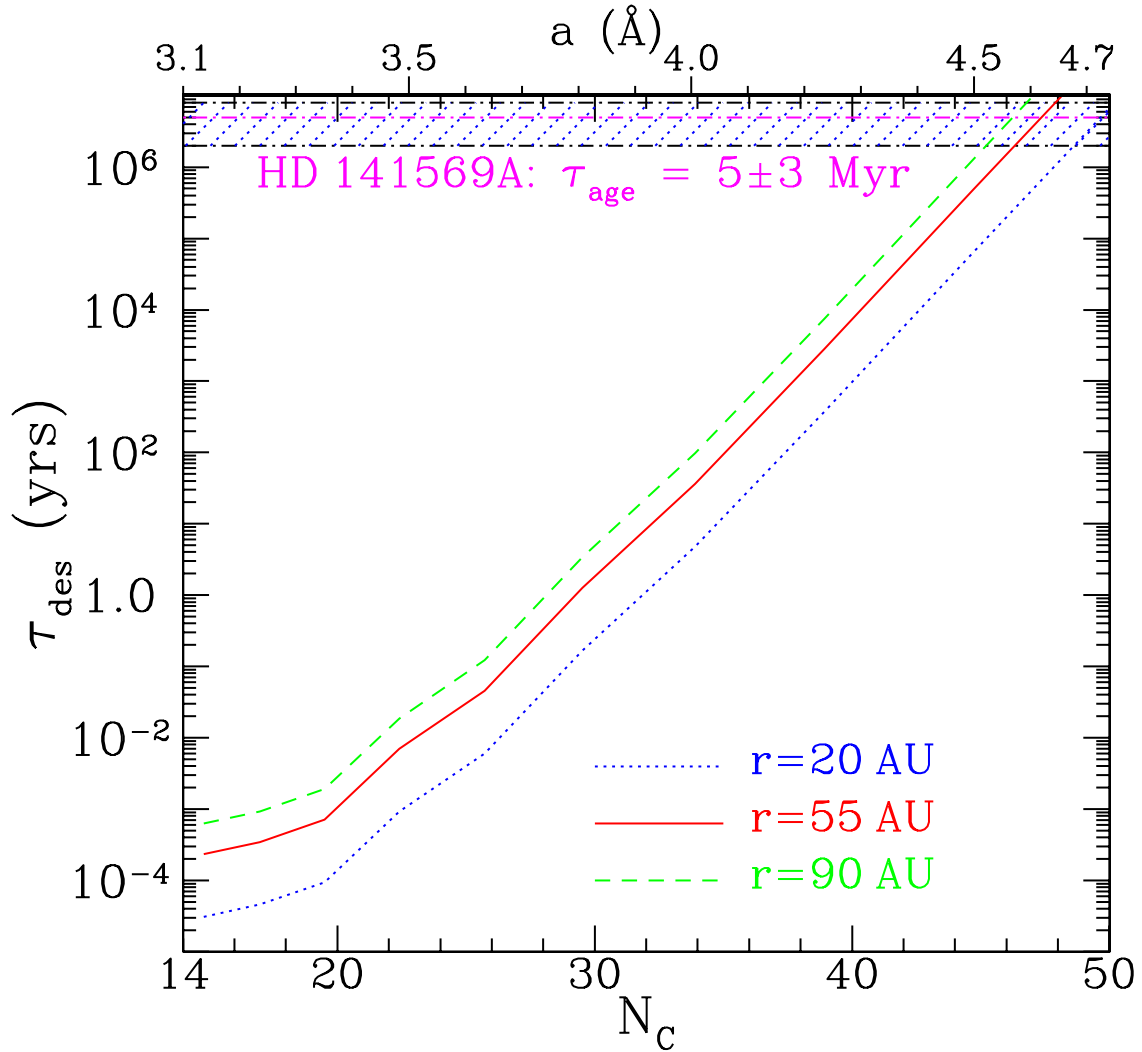}
\end{center}\vspace*{-1em}
\caption{
        \label{fig:pahdes}
        \footnotesize
        Photodestruction time scales for PAHs at
        $r = 20, 55, 90\AU$ as a function of size (upper axis) 
        or $N_{\rm C}$ (the number of carbon atoms in a PAH molecule;
        lower axis). Shaded region depicts the age of the $\hda$ system.
        It is seen that the smallest size of surviving PAHs is
        $\sim 4.6\Angstrom$. But a stable distribution of a population
        of {\it smaller} PAHs is not impossible as long as 
        they are continuously replenished, presumably by sublimation 
        of icy mantles of large grains (in which interstellar PAHs 
        have condensed during the dense molecular cloud phase)
        produced by collisions of large bodies such as 
        planetesimals and comets.
        }
\end{figure}

Upon absorption of an energetic photon, small PAHs,
with an insufficient number of internal vibrational
modes in which to distribute this photon energy, 
may be dissociated by ejection of a hydrogen atom, 
a hydrogen molecule, and/or an acetylene molecule (C$_2$H$_2$).
See \S\ref{sec:photophys} for a detailed discussion 
of this photoprocess. 
Following the method described in \S\ref{sec:photophys},
we have calculated the photodestruction rates ($\kdes$)
for small PAHs exposed to stellar UV photons in the $\hda$ disk, 
represented by the photoejection rates of an acetylene molecule.
In Figure \ref{fig:pahdes} we show the PAH destruction time scales 
($\taudes\equiv 1/\kdes \propto r^{-2}$) for PAHs 
at $r=20,55,90\AU$ as a function of PAH size. 
Apparently, PAHs smaller than $a\simeq 4.6\Angstrom$ 
at $r<105\AU$ are photolytically unstable: 
they are expected to be photodestroyed in a timescale 
shorter than the lifetime of the $\hda$ system. 
Larger PAHs are more stable since they have a larger number
of vibrational modes so that they can easily accommodate 
the absorbed photon energy and it is therefore less likely for 
photodissociation to occur.

In order to maintain a stable distribution of small PAHs
in the inner region ($r<105\AU$) of the $\hda$ disk,
there must exist a source continuously replenishing 
the small PAHs at a rate of
\begin{equation}\label{eq:mpahdot}
\mpahdot = \int_{\rin}^{105\AU} \sigma_{\rm PAH}(r) 2\pi r dr
           \int_{\apahmin}^{\infty} da \frac{dn_{\rm PAH}}{da}
           \frac{\left(4\pi/3\right) a^3 \rho_{\rm PAH}}{\taudes(a,r)}
\end{equation} 
where the PAH surface density distribution, $\sigma_{\rm PAH}(r)$,
is taken to be the same as that of the inner warm porous dust
($r<105\AU$; see \S\ref{sec:iremmethod} and Eq.[\ref{eq:dndr1}])
\begin{equation}\label{eq:dndrpah}
\sigma_{\rm PAH}(r) = \sigma_{\rm PAH}^{\rm p}
\exp\left\{-4\ln2\left[\left(r/{\rm AU}-55\right)/70\right]^2\right\}
\end{equation}
where $\sigma_{\rm PAH}^{\rm p}$ 
is the PAH mid-plane surface density at $r=55\AU$.
For the canonical ``cold-coagulation'' dust model 
(no.\,1; \S\ref{sec:results}; Figure \ref{fig:cold90})
with $a_0=2.5\Angstrom$, $\sigma=0.3$, $\apahmin=3.5\Angstrom$,
and $\sigma_{\rm PAH}^{\rm p}\approx 1.17\times 10^{13}\cm^{-2}$,
we arrive at $\mpahdot \approx 1.11\times 10^{-3}\mearth/{\rm yr}$! 
However, the PAH photodestruction time scales $\taudes$ derived above 
may have been substantially underestimated for two reasons: 
(1) the photoejection of an acetylene molecule is not equivalent to 
the complete destruction of a PAH molecule; 
(2) PAHs can be repaired by accreting carbon atoms and/or ions 
from the gas (e.g. see Allain et al.\ 1996);
the presence of carbon ions in the $\hda$ disk 
was revealed by the detection of the [CII] 158$\mum$ emission 
(see Lorenzetti et al.\ 2002).
Therefore, it is very likely that 
the required PAH replenishment rate estimated above 
has been considerably overestimated.

\begin{figure}[h]
\begin{center}
\epsfig{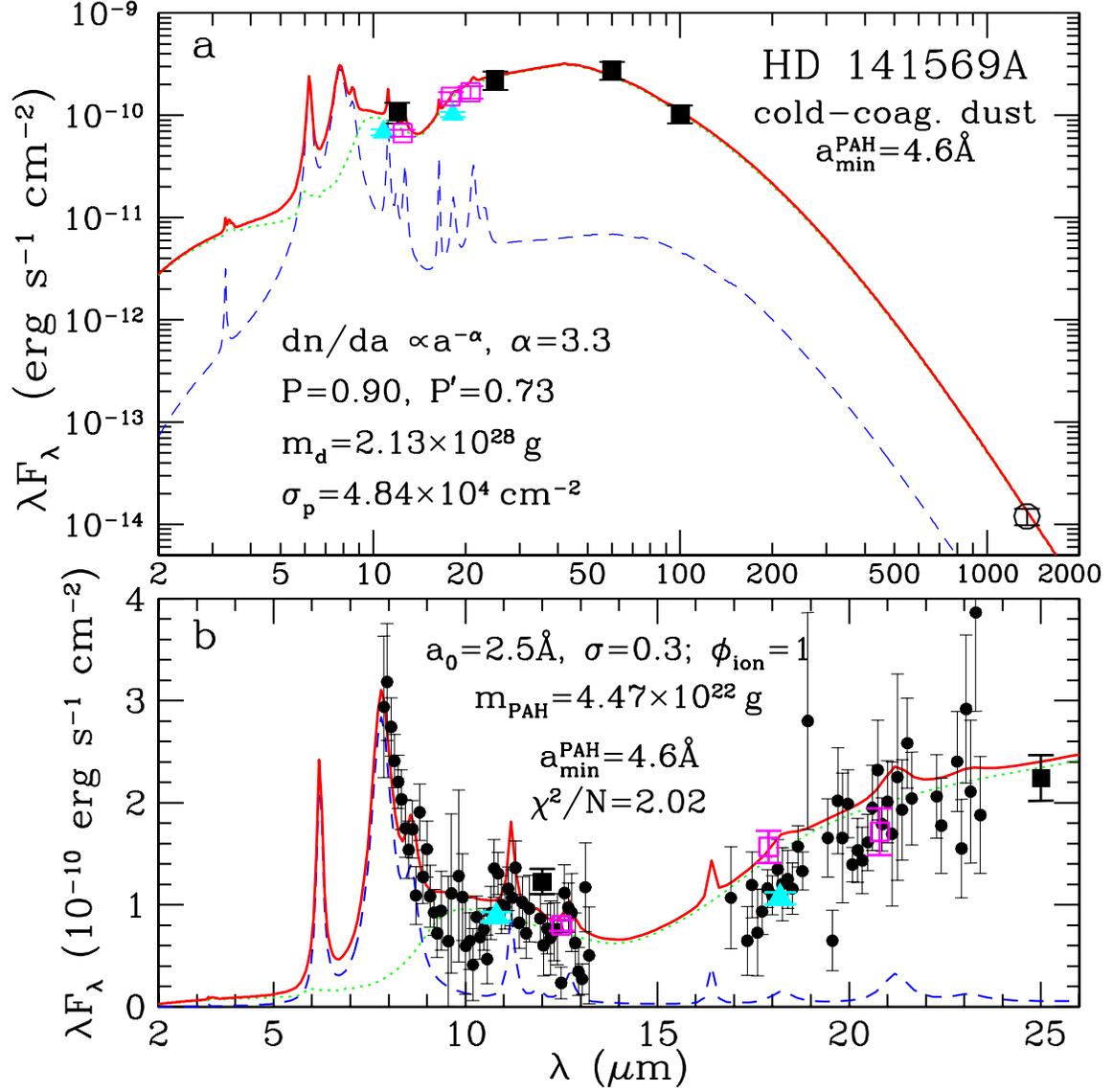}
\end{center}\vspace*{-1em}
\caption{
        \label{fig:cold90paha46}
        \footnotesize
        Same as Figure \ref{fig:cold90} (model no.\,1) 
        except that the PAH component has a larger 
        lower-cutoff size $\apahmin = 4.6\Angstrom$ (model no.\,11).
        In comparison with the canonical ``cold-coagulation'' model
        with $\apahmin = 3.5\Angstrom$ 
        (model no.\,1; see Figure \ref{fig:cold90}), 
        the $\apahmin = 4.6\Angstrom$ model 
        predicts weaker 3.3 and 6.2$\mum$ features since this model
        is relatively lacking in small (and hence hot) PAHs
        (see Figure \ref{fig:dndapah}). 
        }
\end{figure}

But even if we assume that the above derived PAH mass loss rate
due to photodestruction is valid {\it and} there is insufficient
replenishment implying that there will be no PAHs smaller than 
$\sim 4.6\Angstrom$, our model with $\apahmin = 4.6\Angstrom$
is still able to reproduce the observed SED.    
This is demonstrated in Figure \ref{fig:cold90paha46}
for the ``cold-coagulation'' dust model with
$\apahmin = 4.6\Angstrom$ (model no.\,11). 
Therefore, the exact knowledge of the PAH lower-cutoff size 
$\apahmin$ is not critical in modeling the $\hda$ SED.

\subsection{Radiation Pressure and Poynting-Robertson Drag\label{sec:rppr}}
In addition to the gravitational attraction from the central star,
grains in the $\hda$ disk are subject to 
(1) radiative outward propulsion
as a consequence of momentum transfer from stellar photons, 
and (2) Poynting-Robertson drag which
causes them to spiral toward the gravitational force center 
as a consequence of angular momentum loss through radiation
(Burns, Lamy, \& Soter 1979; Backman \& Paresce 1993; Kr\"ugel 2003).

The ratio of the radiative force to the gravitational force is 
\begin{equation}\label{eq:betarp}
\beta_{\rm RP}(a) = \frac{R_\star^2 \int_{912\Angstrom}^{\infty} 
                 F_\lambda^{\star} 
                \left[ C_{\rm abs}(a,\lambda) 
                + \left(1-\langle g \rangle\right)
                   C_{\rm sca}(a,\lambda)\right] d\lambda}
                 {4 c G M_\star \left(4\pi/3\right) a^3 \rho_{\rm PAH}}
\end{equation}
where $G$ is the gravitational constant, 
$M_\star$ ($\approx 2.3\,m_\odot$) is the stellar mass,
$C_{\rm sca}(a,\lambda)$ is the scattering cross section
for a grain of size $a$ at wavelength $\lambda$,
$\langle g \rangle$ (``asymmetry factor'') is the average 
value of the cosine of the scattering angle.

\begin{figure}[h]
\begin{center}
\epsfig{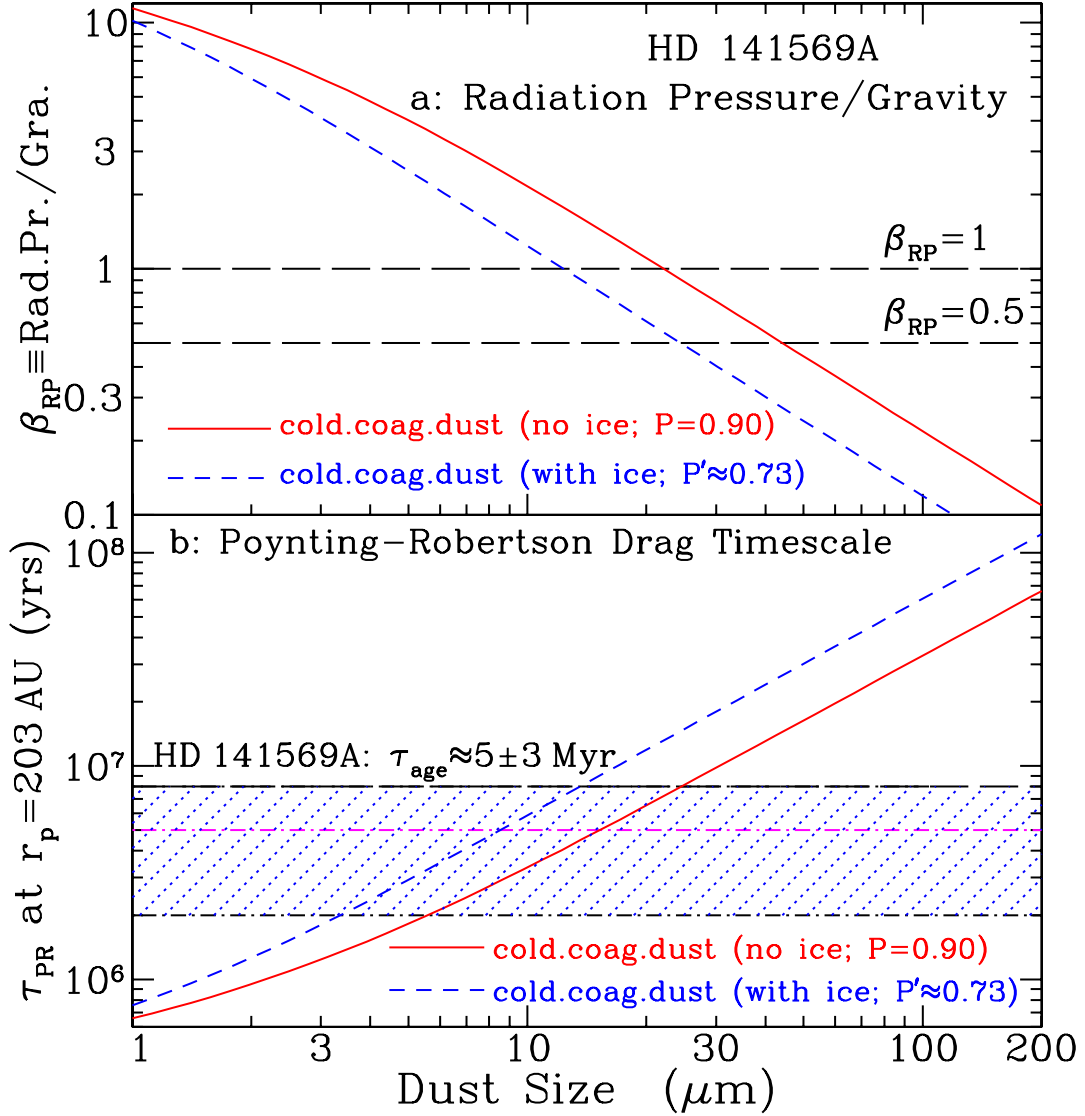}
\end{center}\vspace*{-1em}
\caption{
        \label{fig:rppr}
        \footnotesize
        Upper panel (a): ratio of the radiative repulsion to 
        the gravitational attraction ($\beta_{\rm RP}$) for 
        the best-fit canonical ``cold-coagulation'' dust 
        (model no.\,1; without ice $P=0.90$ [solid line] 
        or with ice $\Pice=0.73$ [dashed line]).
        The long-dashed horizontal lines plot 
        $\beta_{\rm RP}=1,0.5$
        (the smaller value accounting for the fact that
        particles released at periastron from parent bodies 
        moving on circular orbits only need to have 
        $\beta_{\rm RP}=0.5$ for ejection [Burns et al.\ 1979], 
        assuming that there is no significant gas drag).
        Lower panel (b): the orbit decay timescales $\tau_{\rm PR}$
        due to the Poynting-Robertson drag for 
        the best-fit canonical ``cold-coagulation'' dust
        (model no.\,1; without ice $P=0.90$ [solid line] 
        or with ice $\Pice=0.73$ [dashed line])
        at a radial distance of $\rp=203\AU$ from the central star
        (note $\tau_{\rm PR}\propto r^{2}$; see Eq.[\ref{eq:taupr}]).
        The shaded region depicts the $\hda$ age ($\approx 5\pm3 \Myr$).     
        }
\end{figure}

We have calculated $\beta_{\rm RP}$ as a function of dust size
for the best-fit canonical ``cold-coagulation'' dust model 
(no.\,1; see Figure \ref{fig:cold90} and \S\ref{sec:results}).
As shown in Figure \ref{fig:rppr}a, for grains smaller than 
$\sim 10-20\mum$, the radiation pressure (RP) overcomes the
gravitational attraction (i.e. $\beta_{\rm RP}\simgt 1$) 
and, therefore, these grains will be blown out from the $\hda$ disk.
But we note that a stable distribution of small grains 
($\simlt 10-20\mum$) in the disk is possible provided that
they are continuously replenished by collisions of
larger bodies and/or larger grains so that at any time 
the disk contains a substantial population of small particles 
(see Krivov, Mann, \& Krivova 2000).

The dust removal rate due to the radiation pressure (RP) expulsion
can be estimated from
\begin{equation}\label{eq:mpahdot}
\mrpdot = \int_{\rin}^{\rout} \sigmar 2\pi r dr
           \int_{\amin}^{10\mum} da \frac{dn_{\rm PD}}{da}
           \frac{\left(4\pi/3\right) a^3 \rho_{\rm PD}}{\taurp(r)} ~~;
\end{equation} 
\begin{equation}\label{eq:taurp}
\taurp(r) \approx 1900\,\left(\frac{r}{203\AU}\right)^{3/2}\yr  
\end{equation} 
where $\taurp$, the RP timescale, is assumed to be comparable to 
the local dynamical timescale $\tau_{\rm dyn} = 2\pi/\Omega(r)$ 
where $\Omega(r)\equiv \left(GM_{\star}/r^3\right)^{1/2}$ 
is the Keplerian frequency. 
For the best-fit canonical ``cold-coagulation'' model 
(no.\,1; see Figure \ref{fig:cold90} and \S\ref{sec:results}),
the RP dust mass loss rate is 
$\approx 7.87\times 10^{-6}\,m_\oplus/{\rm yr}$.

We have also calculated the Poynting-Robertson (PR) drag timescale 
$\taupr(a,r)$ (the time it takes for a grain of size $a$ at a distance 
of $r$ to fall into the central star) from
\begin{equation}\label{eq:taupr}
\taupr(a,r) = \frac{c^2\,r^2\,\left(4\pi/3\right)a^3\rho_{\rm PD}}
              {R_\star^2 \int_{912\Angstrom}^{\infty} 
               F_\lambda^{\star} 
               C_{\rm abs}(a,\lambda) d\lambda} ~~.
\end{equation}
In Figure \ref{fig:rppr}b we show that,
for grains smaller than $\sim 20\mum$ at a radial distance
of $\rp=203\AU$, the lifetime associated with the PR drag is
shorter than the $\hda$ age and hence, these grains will fall 
into the central star even if they are not ejected 
by radiation pressure. 
For the inner warm dust component at $r<105\AU$, 
grains in the range of $10-20\mum \simlt a\simlt 100-200\mum$, 
although stable against radiation pressure ejection,
will be removed from the disk by the PR drag.
Since the best-fit ``cold-coagulation'' model (no.\,1) has 
$\sim 80\%$ of the total surface areas in grains smaller than $100\mum$, 
these grains must be efficiently replenished by cascade collisions 
of planetesimals and larger grains. 
By integrating the PR dust removal rate
$\left(4\pi/3\right) a^3\rho_{\rm PD}/\tau_{\rm PR}(a,r)$ 
over the whole size range and over the entire disk, 
we estimate the PR dust mass loss rate to be 
$\approx 1.38\times 10^{-8}\,m_\oplus/{\rm yr}$ 
for the best-fit canonical ``cold-coagulation'' model (no.\,1).
Therefore, over the life span of $\hda$, 
roughly $39\,m_\oplus$ of dust is lost by radiation pressure 
and Poynting-Robertson drag.

For the PAH component, the ratio of radiation pressure to
gravity $\beta_{\rm RP}\approx 87.4$ is almost independent of size.
This is because PAHs are in the Rayleigh limit at UV/visible
wavelengths where the absorption of stellar photons is such 
that $C_{\rm abs}(a,\lambda) \propto a^3/\lambda$
and $\langle g \rangle \simeq 0$ (Bohren \& Huffman 1983; Kr\"ugel 2003).
The PR time scale for PAHs at $r=55\AU$ is about 6000$\yr$,
also independent of size for the same reason.
The PAH mass loss rates due to radiation pressure and
Poynting-Robertson drag are 
$\approx 3.09\times 10^{-8}\,m_\oplus/{\rm yr}$ and 
$\approx 1.95\times 10^{-9}\,m_\oplus/{\rm yr}$, respectively.
This implies that a total amount of $\sim 0.16\mearth$
of PAH molecules needs to be replenished over the life span of $\hda$.

\subsection{Predictions for {\it SIRTF}\label{sec:sirtf}}
The {\it Space Infrared Telescope Facility} (SIRTF) will be
capable of sensitive imaging using the {\it Infrared Array Camera} 
(IRAC) at 3.6, 4.5, 5.8, and 8.0$\mum$, and using the 
{\it Multiband Imaging Photometer} (MIPS) at 24, 70, and 160$\mum$.
In Table \ref{tab:sirtf} we show the band-averaged intensities
for our preferred dust models.

SIRTF will also be able to perform low-resolution 5--40$\mum$
and high-resolution 10--37$\mum$ spectroscopic observations 
using the {\it Infrared Spectrograph} (IRS) instrument.
IR spectroscopy and imaging will provide powerful constraints on 
the $\hda$ dust spatial distribution and its chemical composition.

Our best-fit canonical ``cold-coagulation'' dust model (no.\,1;
Figure \ref{fig:cold90}) predicts strong PAH C-C stretching features 
at 6.6$\mum$ and 7.7$\mum$, a weak in-plane C-H bending feature
at 8.6$\mum$, a weak out-of-plane C-H bending feature at 11.3$\mum$,
a very weak C-H stretching feature at 3.3$\mum$, 
very weak out-of-plane C-H bending features at 11.9$\mum$ and 12.7$\mum$,
and very weak C-C bending features at 16.4, 18.3, 21.2 and 23.1$\mum$
(see Li \& Draine 2001a for details).
With neutral PAHs included (e.g., see Figure \ref{fig:ionneupah}b
for the $\fion=0.5$ model [no.\,9] which is not ruled out in the 
present study), stronger 3.3$\mum$ and 11.3$\mum$ features are expected.
SIRTF spectroscopy at $\lambda \simlt 8\mum$ will provide
an important test of the PAH model with respect to 
the PAH size distribution, charging, and survival in the $\hda$ disk.    

The ``cold-coagulation'' dust model (no.\,1) produces a broad, 
smooth amorphous silicate feature at 9.7$\mum$ which forms 
a plateau underneath the 11.3$\mum$ PAH feature 
(see Figure \ref{fig:cold90} and \S\ref{sec:results}). 
In contrast, the ``hot-nebula'' dust model 
(no.\,2; see Figure \ref{fig:hot90} and \S\ref{sec:results})
predicts two crystalline silicate features at 11.3 and 23$\mum$.
While the 11.3$\mum$ feature is swamped by 
the PAH C-H out-of-plane bending mode which is also at 11.3$\mum$,
the 21$\mum$ feature is prominent and the CGS3 mid-IR spectrum readily
places an upper limit of about 10\% on crystalline silicate dust
(see Figure \ref{fig:coldhot90} and \S\ref{sec:results}).
With high resolution spectroscopy,
SIRTF will allow us to perform more detailed studies of 
the silicate emission features so as to infer 
the degree of processing which the $\hda$ dust has experienced. 

\begin{table}[h]
\begin{center}
\caption[]{Dust IR emission (Jy) integrated over SIRTF bands 
           predicted for our preferred models\label{tab:sirtf}.}
\begin{tabular}{cccccccc}
\hline
model & IRAC & IRAC & IRAC & IRAC & MIPS & MIPS &MIPS \\
no.   & 3.6$\mum$ 
      & 4.5$\mum$ 
      & 5.8$\mum$ 
      & 8.0$\mum$ 
      & 24$\mum$
      & 70$\mum$
      & 160$\mum$ \\
\hline
1	&0.011
	&0.018
	&0.12
	&0.41
	&1.86
	&4.45
	&2.27\\
8	&0.0098
        &0.015
	&0.090
	&0.38
	&1.94
	&4.57
	&2.37\\

9	&0.034
	&0.023
	&0.14
	&0.43
	&1.91
	&4.52
	&2.32\\
11	&0.0098
	&0.016
	&0.10
	&0.40
	&1.91
	&4.55
	&2.35\\
\hline
\end{tabular}
\end{center}
\end{table}

\section{Conclusion}\label{sec:conclusion}
We have modelled the mid-infrared to submillimeter spectral 
energy distribution of the double-ring disk around $\hda$
in terms of the porous dust model previously shown successful
in reproducing the SEDs observed for the disks around 
$\beta$ Pictoris (Li \& Greenberg 1998) and $\hra$ (Li \& Lunine 2003).
The dust is modelled as porous aggregates of either unaltered
or heavily processed interstellar materials.
While the former (``cold-coagulation'' dust) 
is made of amorphous silicate dust, carbonaceous dust, 
H$_2$O-dominated ices, and vacuum,
the latter (``hot-nebula'' dust) consists of a mixture of 
crystalline silicates, ices, and vacuum (\S\ref{sec:model}).
For both dust types, a vacuum volume fraction of 90\% 
is assumed (\S\ref{sec:para}).
The dust spatial distribution is taken to be that 
derived from imaging observations of scattered light 
and dust thermal emission (Eqs.[\ref{eq:dndr1}-\ref{eq:dndr3}]; 
\S\ref{sec:iremmethod}).
We take a simple power-law for the dust size distribution 
($dn/da\propto a^{-\alpha}$) in the size range 
of $1\mum \simlt a \simlt 1\cm$.
In addition, a population of PAH molecules with a log-normal
size distribution characterized by a peak-parameter $a_0$ 
and a width-parameter of $\sigma$ is invoked to
account for the observed 7.7$\mum$ and 11.3$\mum$ 
``unidentified infrared'' emission features that
are generally attributed to PAHs.
Our principal results are:
\begin{enumerate}
\item The ``cold-coagulation'' dust model with a porosity of
      $P=0.90$, a size-distribution power index of $\alpha\approx 3.3$,
      and a total dust mass of $\md \approx 3.56\mearth$,
      together with a population of charged PAHs with
      $a_0\approx 2.5\Angstrom$, $\sigma\approx 0.3$
      and a total mass of $\mpah\approx 7.10\times 10^{-6}\mearth$ 
      provides an excellent fit to the entire SED, 
      including the PAH mid-IR emission features
      (\S\ref{sec:results}; Figure \ref{fig:cold90}).
\item Although the $P=0.90$ ``hot-nebula'' dust model provides
      an overall good fit to the observed SED, it predicts a strong
      crystalline silicate emission feature at $21\mum$ which is
      not seen in the currently available observational spectrum
      (\S\ref{sec:results}; Figure \ref{fig:hot90}).
      This places an upper limit of $\sim 10\%$ on 
      the fraction of ``hot-nebula'' dust 
      (i.e. crystalline silicates;
      \S\ref{sec:results}; Figure \ref{fig:coldhot90}).
\item The PAH molecules in the $\hda$ disk are expected to
      be negatively charged (Figure \ref{fig:ionrec}; 
      \S\ref{sec:ionrec}; \S\ref{sec:photophys}), assuming that 
      the cosmic ray ionization of H$_2$ is the dominant 
      source of electrons (\S\ref{sec:ne}).
      But it has also been shown that models consisting of 
      a mixture of both neutral and charged PAHs are also 
      capable of reproducing the observed PAH IR emission features, 
      provided that the fraction of neutral PAHs is smaller than 50\% 
      (Figure \ref{fig:ionneupah}; \S\ref{sec:robust}).
\item Approximating the photodestruction rate of a PAH molecule
      by the photoejection rate of an acetylene molecule,  
      PAHs smaller than $\sim 4.6\Angstrom$ will be photodestroyed
      in a time scale shorter than the age of the $\hda$ system
      (Figure \ref{fig:pahdes}; \S\ref{sec:pahdes}; \S\ref{sec:photophys}).
      In order to have a stable distribution of small PAHs
      ($3.5<a<4.6\Angstrom$) in the $\hda$ disk, we require 
      a continuous replenishment of this component 
      at a rate of $\sim 1.11\times 10^{-3}\mearth/{\rm yr}$.
      Such a high replenishment rate may not be essential, 
      since (1) the photoejection of a C$_2$H$_2$ molecule 
      does not necessarily lead to the complete destruction 
      of a PAH molecule, and 
      (2) the repair of a PAH molecule with a broken C-C bond
      through reacting with gas-phase C atoms and/ions is ignored
      in deriving the PAH destruction time scale.
      Furthermore, we have shown in \S\ref{sec:pahdes} 
      (Figure \ref{fig:cold90paha46})
      that a model exclusively consisting of PAHs larger than 
      $\sim 4.6\Angstrom$ also achieves a good fit to 
      the observed SED.
\item Grains smaller than $\sim 10-20\mum$ will be radiatively 
      expelled from the disk; grains at $r=55\AU$ in the size range 
      of $\sim 10-20\mum \simlt a \simlt 100-200\mum$ will also 
      be removed from the disk due to the Poynting-Robertson 
      inward spiralling drag (at a {\it closer} [{\it further}] 
      distance from the star, {\it larger} [{\it smaller}] grains 
      will be removed). Collisions of planetesimals must continuously 
      replenish the dust in the disk at a rate of 
      $\approx 7.88\times 10^{-6}\,m_\oplus/{\rm yr}$ (\S\ref{sec:rppr}).
      PAHs will also be rapidly removed from the disk by radiation
      pressure and, to a less degree, by Poynting-Robertson drag. 
      A replenishment rate of 
      $\approx 3.28\times 10^{-8}\,m_\oplus/{\rm yr}$
      is needed to maintain the PAH disk (\S\ref{sec:rppr}).
\item Spectroscopic and broadband photometric predictions
      are made for SIRTF observations (\S\ref{sec:sirtf}).
\end{enumerate}

\acknowledgments
We thank M.J. Barlow for providing us with 
the $\hda$ CGS3 mid-IR spectrum.  
We thank D.E. Backman, A.G.W. Cameron, C. Kulesa, V. La Page, 
S. Leach, T.W. Rettig, and M.D. Silverstone for helpful 
discussions, suggestions, and/or clarifications.
%We thank S. Leach and V. La Page for clarification
%on PAH photophysics. 
A. Li thanks the University of Arizona for the ``Arizona 
Prize Postdoctoral Fellowship in Theoretical Astrophysics''.
This research was supported in part by a grant from the NASA 
origins research and analysis program.
%The most obvious way to form a gap in a disk is with 
%a planet. The planet does not have to be in the gap.
%It could either be sweeping up the dust and rocks from
%the disk as it travels in its orbit around the star, 
%or the gravity of the planet could knock the dust out 
%of one part of the disk.
%%%%%%%%%%%%%%%%%%%%%%%%%%% Appendix %%%%%%%%%%%%%%%%%%%%%%%%%%%%
\appendix
\section{Photophysics of Polycyclic Aromatic Hydrocarbon Molecules
in the HD\,141569A Disk\label{sec:photophys}}  

Following the absorption of an energetic photon, a PAH molecule
has 3 major competing decay channels to relax its energy: 
emission;\footnote{%
  The emission process is dominated by fluorescence 
  (transitions between vibrational states of same 
  multiplicity) in the IR and part in the visible.
  Phosphorescence (transitions between vibrational states 
  of different multiplicity) is less important.
  In this work we will thus only consider IR emission
  via fluorescence.
  }
ionization; and photodissociation.
The photodissociation process, critical for small PAHs, 
has 3 major loss channels: the loss of a hydrogen atom, 
a hydrogen molecule, and an acetylene molecule (C$_2$H$_2$).

We use the Rice, Ramsperger, and Kassel (RKK) theory
(Forst 1973) for the photodissociation rate for
a PAH molecule of internal energy $E$ after excitation
\begin{eqnarray}\label{eq:rkk}
\nonumber
k^{\rm x}(E) &=&
\nud^{\rm x}\left(1-\frac{\Ed^{\rm x}}{E}\right)^{3N-7} ~, 
~~ E > \Ed^{\rm x} ~;
\\ 
&=& 0 ~, ~~ E \leq \Ed^{\rm x} ~;
\end{eqnarray}
where $N$ is the number of atoms in the molecule;
$\Ed^{\rm x}$ is the activation energy for the loss of 
the species-${\rm x}$ (${\rm \equiv H, H_2, C_2H_2}$) 
related to bond dissociation (i.e., the minimum energy 
required for the dissociation to occur);
$\nud^{\rm x}$ is the frequency factor for the ${\rm x}$-loss 
channel which reflects the efficiency of 
intramolecular redistribution of the energy over the vibrational 
levels of the molecule's ground electronic state
after internal conversion.
In Table \ref{tab:rkk} we tabulate the $\Ed$ and $\nud$ parameters
for the loss of H, H$_2$ and C$_2$H$_2$ experimentally obtained
by Jochims et al.\ (1994) for small PAHs.

\begin{table}[h,t]
\begin{center}
%{\tiny
%{\scriptsize
{\footnotesize
\caption[]{Activation energy $\Ed$ and frequency factor $\nud$
(Jochims et al.\ 1994) \label{tab:rkk}.}
\begin{tabular}{lccl}
\hline \hline
	loss channel
        &$\Ed^{\rm x}$
        &$\nud^{\rm x}$\\
        x
        &(eV)
        &(s$^{-1}$)\\
\hline
H       &2.8
        &$1\times 10^{16}$\\
H$_2$   &2.9
        &$1\times 10^{16}$\\
C$_2$H$_2$  &2.9
        &$1\times 10^{15}$\\
\hline
\end{tabular}
}
\end{center}
\end{table}

The IR photon emission rate for a PAH molecule of
internal energy $E$ is 
\begin{equation}\label{eq:kir}
\kir(E) = \int^{\infty}_{912\Angstrom} \frac{C^{\rm PAH}_{\rm abs}(\lambda) 
4\pi B_{\lambda}(T[E])}{hc/\lambda} d\lambda
%\times \frac{E}{\int^{\infty}_{hc/E} C^{\rm PAH}_{\rm abs}(\lambda)
%4\pi B_{\lambda}(T[E]) d\lambda}
\end{equation}
where the vibrational temperature $T$ is obtained 
from the vibrational energy $E$ using the Debye-model
approximation for the PAH enthalpy (see Eq.[15] in Li \& Draine 2001).

Assuming that IR emission, ionization, and dissociation 
are the only relaxation processes for a PAH molecule excited
by an energetic photon, the photodestruction rate 
(approximated by the ejection of an acetylene molecule) $\kdes$, 
and the dehydrogenation rate $\kdhy$
for PAHs at a distance of $r$ from the central star
can be obtained from following equations\footnote{%
 For simplicity, we assume that the absorbed photon will
 {\it either} ionize the neutral PAH molecule {\it or} 
 participate in the photodissociation and vibrational relaxation.
 }
\begin{equation}\label{eq:kdes}
\kdes(r) = \left(\frac{\Rstar}{2r}\right)^2
\int^{912\Angstrom}_{hc/\Ed^{\rm C_2H_2}} \frac{k_{\rm C_2H_2}}
{k_{\rm H} + k_{\rm H_2} + k_{\rm C_2H_2} + \kir}
\frac{C^{\rm PAH}_{\rm abs}(\lambda) \Fstar}{hc/\lambda} d\lambda ~~;
\end{equation}

\begin{equation}\label{eq:kdhy}
\kdhy(r) = \left(\frac{\Rstar}{2r}\right)^2
\int^{912\Angstrom}_{hc/\Ed^{\rm H}} \frac{k_{\rm H}}
{k_{\rm H} + k_{\rm H_2} + k_{\rm C_2H_2} + \kir}
\frac{C^{\rm PAH}_{\rm abs}(\lambda) \Fstar}{hc/\lambda} d\lambda ~~.
\end{equation}

Following Allain, Leach, \& Sedlmayr (1996),
we adopt an analytical formula for the ionization yield 
$\Yion$ which was based on the experimental data of
pyrene and coronene (Verstraete et al.\ 1990)
\begin{eqnarray}\label{eq:Yion}
\nonumber
\Yion &=& 0 ~, ~~ E < \EIP ~;
\\ 
\nonumber
&=& \frac{0.1\,\left(E-\EIP\right)}{8.90\eV-\EIP} ~, ~~ \EIP \leq E < 8.9\eV ~;
\\
\nonumber
&=& 0.25\,\left(E/{\rm eV}\right) - 2.125 ~, ~~ 8.9 \leq E < 11\eV ~;
\\
&=& 0.0596\,\left(E/{\rm eV}\right) - 0.03 ~, ~~ 11 \leq E \leq 13.6\eV ~;
\end{eqnarray}
where $E$ and $\EIP$ (ionization threshold) are in units of $\eV$.
We adopt a simple expression for the ionization threshold
(see Eq.[2] in Weingartner \& Draine 2001 for $Z=0$),
\begin{equation}\label{eq:Eip}
\left(\frac{\EIP}{{\rm eV}}\right) = 
4.4 + \left(\frac{7.19\Angstrom}{a}\right)
+\left(\frac{2.94\Angstrom}{a}\right)^2
\end{equation}
where the PAH radius $a$ is in unit of angstrom.
The photoionization rate for a PAH molecule located at a distance of $r$
from the central star is
\begin{equation}\label{eq:kion}
\kion(r) = \left(\frac{\Rstar}{2r}\right)^2
\int^{912\Angstrom}_{hc/\EIP} 
\frac{\Yion C^{\rm PAH}_{\rm abs}(\lambda) \Fstar} 
{hc/\lambda} d\lambda ~~.
\end{equation}

The electronic recombination rate for a PAH cation is
\begin{equation}\label{eq:krec}
\krec(r) = \int_{0}^{\infty}d\ve
           \nelc \ve \pi a^2 \left(1+\frac{2e \psi}{m\ve^2}\right)
           4\pi\ve^2 \left(\frac{\me}{2\pi k\Tg}\right)^{3/2}
           \exp\left(-\frac{\me \ve^2}{2 k \Tg}\right) 
         = \nelc \left(\frac{8 k \Tg}{\pi \me}\right)^{1/2}
           \pi a^2 \left(1+\frac{e^2}{a k\Tg}\right)
\end{equation}
where $\nelc$ is the electron density; 
$\ve = \left(8 k \Tg/\pi \me\right)^{1/2}$ is the electron
thermal velocity; $\me$ is the electron mass;
$e$ is the electron charge;
$\psi=e/a$ is the PAH electrostatic potential; 
$k$ is the Boltzmann constant;
$\Tg$ is the gas temperature. We will take 
$\Tg(r) = \left(R_\star/2r\right)^{1/2} \Teff
\approx 605\,\left(r/{\rm AU}\right)^{-1/2}\K$
for the $\hda$ disk.
In Eq.(\ref{eq:krec}) we have assumed that the sticking
probability for recombination of an electron on the
PAH cation equals one. The term $2e \psi/m\ve^2$ accounts
for the ``Coulomb focussing'' effect (Spitzer 1941).

\section{Electron Density in the HD\,141569A Disk\label{sec:ne}}  
Molecular CO gas in the $\hda$ disk has been detected in emission 
through its rotational transitions at submillimeter wavelengths
(Zuckerman, Forveille, \& Kastner 1995) and through its ro-vibrational
transitions at near-IR (Brittain \& Rettig 2002; Brittain et al.\ 2003). 
Brittain \& Rettig (2002) also reported the detection of
$\triHi$ emission in the $\hda$ disk. 
In principle, one can estimate the electron density from the
$\triHi$ and CO measurements based on the following assumptions -- 
(1) $\triHi$ is produced mainly through the cosmic ray (CR)
ionization of $\molH$ to $\molHi$: 
$\molH + {\rm CR} \rightarrow \molHi + e^{-}$ 
and the ion-neutral reaction:
$\molHi + \molH \rightarrow \triHi + H$;
(2) $\triHi$ is destroyed primarily by its recombination
with an electron: $\triHi + e^{-} \rightarrow \molH + {\rm H}$
and by the ion-neutral reaction with CO: 
$\triHi + {\rm CO} \rightarrow {\rm H_2CO}^{+} + {\rm H}$
(McCall et al.\ 1998).
Let $\nelc$, $\nmolH$, $\nco$, and $\ntriHi$ be the number densities
of electrons, $\molH$, CO, and $\triHi$, respectively;
$\rcr$ be the cosmic ionization rate;
$\ke$ and $\kco$ be the rate constants for 
the $\triHi$-electron recombination
and the $\triHi$-CO reaction, respectively. 
One can derive the electron density from
\begin{equation}\label{eq:ne}
\nelc = \frac{\rcr \nmolH - \kco \nco\ntriHi}
             {\ke \ntriHi} ~~.
\end{equation}

Brittain \& Rettig (2002) estimated the column densities
of CO and $\NtriHi$ respectively to be $\Nco \approx 10^{11}\cm^{-2}$ 
and $\NtriHi \approx 7\times 10^{10}\cm^{-2}$
from the CO and $\triHi$ emission spectra obtained
for the $\hda$ disk with the NASA {\it Infrared Telescope} (IRTF)
using the {\it Cryogenic Echelle Spectrograph} (CSHELL).\footnote{%
 This assumes that both CO and $\triHi$ are contained in the
 $17<r<50\AU$ region around $\hda$.
 }
With $\rcr = 3\times 10^{-17}\s^{-1}$,
$\kco = 2.0\times 10^{-9}\cm^3\s^{-1}$,    
$\ke = 1.8\times 10^{-7}\cm^3\s^{-1}$ (see McCall et al.\ 1998 
and references therein), and the nominal $\NmolH/\Nco \approx 10^4$
(Thi et al.\ 2001), one finds $\nelc <0$ from Eq.(\ref{eq:ne})!
This suggests that CO and $\triHi$ may be distributed in different 
regions. Brittain \& Rettig (2002) suggest that $\triHi$ may originate 
from the extended envelope of a gas giant protoplanet at $\sim 7\AU$.

Therefore, it is not appropriate to derive $\nelc$ from the $\triHi$ data;
instead, we will place an upper limit on $\nelc$ from $\nmolH$ 
assuming that the cosmic ray ionization of $\molH$ 
is the major source of electrons: $\nelc \approx \taustar\rcr\nmolH$ 
where $\taustar\approx 5\Myr$ is the stellar age.\footnote{%
 Atomic gas has also been seen in the $\hda$ disk.
 A double-peaked H$\alpha$ emission was detected 
 by Andrillat, Jaschek, \& Jaschek (1990) 
 and Dunkin, Barlow, \& Ryan (1997), indicating a rotating gas disk
 close to the star. Dunkin et al.\ (1997) also detected circumstellar
 NaI and CaII absorption lines. Recently, Lorenzetti et al.\ (2002) 
 reported the detection of C$^{+}$ gas.
 However, little is known regarding 
 the atomic gas column density and its contribution to the electrons
 in the disk. Since it is shown in \S\ref{sec:ionrec} that 
 the amount of electrons provided by $\molH$ alone is sufficient to
 (negatively) charge the PAH molecules in the $\hda$ disk,
 inclusion of electrons from the cosmic ray ionization 
 of atomic gas would further support our conclusion
 (i.e., a large fraction of the PAH molecules {\it are} 
 [see \S\ref{sec:ionrec} and Figure \ref{fig:ionrec}]
 and {\it must be} [see \S\ref{sec:robust} 
 and Figure \ref{fig:ionneupah}] charged in the $\hda$ disk).
 Furthermore, an additional amount of electrons would 
 be provided by the X-ray ionization of $\molH$.
 }
Below we will discuss how the number density of 
molecular hydrogen $\nmolH$ is obtained.

Assuming that the disk is in vertical hydrostatic equilibrium,
the spatial distribution of molecular hydrogen is
\begin{equation}\label{eq:nH}
\nmolH(r,z) = \frac{\sigmaH(r)}{\sqrt{2\pi} H(r)}
              \exp\left[-\frac{1}{2}\left(\frac{z}{H}\right)^2\right] ~~,
\end{equation}
\begin{equation}\label{eq:sigmaH}
\sigmaH(r) = \int_{-\infty}^{+\infty} \nmolH(r,z) dz
\end{equation}
where $\sigmaH$ is the surface density of $\molH$;
the vertical scale height $H$ is
\begin{equation}\label{eq:sclhgt}
H(r) = \left(\frac{2k\Tg}{G\Mstar\mu \mH}\right)^{1/2} r^{3/2}
\end{equation}
where $\Mstar (\approx 2.3\msun)$ is the stellar mass, 
$\mH$ is the mass of a hydrogen atom,
$\mu$ is the mean molecular weight ($\mu \approx 2.34$ for 
a mixture of $\molH$ and He).
The total molecular hydrogen mass in the $r< 130\AU$ region
around $\hda$ was estimated to be $\mHtot \approx 5-115\mearth$
based on the IRAM ({\it Institut de Radio-Astronomie Millimetrique})
observations of the $J=2-1$ and $J=1-0$ rotational transitions of CO
(Zuckerman et al.\ 1995).\footnote{%
  Zuckerman et al.\ (1995) originally gave
  $\mHtot \approx 20-460\mearth$. But this was derived 
  from the assumption that the distance from $\hda$ to
  the Earth is $d\approx 200\pc$. Adopting $d\approx 99\pc$
  (van den Ancker et al.\ 1998), $m_{\rm H_2}$ should be reduced
  by a factor of $\approx 4$.
  }   
More recently, Brittain \& Rettig (2002) found that the CO gas,
pumped by strong stellar UV radiation,
is confined to the $17<r<50\AU$ region with 
a total mass of $\approx 10^{19}\g$.
Assuming the standard ratio of $\molH$ to CO (about $10^4$),
this leads to a total $\molH$ mass of $\approx 2\times 10^{-5}\mearth$.

The $\molH$ mass inferred by Brittain \& Rettig (2002)
differs by a factor of $\approx 7\times 10^6$ from that 
of Zuckerman, Forveille, \& Kastner (1995).
This is because the UV-pumped CO detected by Brittain \& Rettig (2002)
only represents material near the star/disk interface
and is not indicative of the total gas mass. 
The UV radiation is quickly scattered by dust
and may be influential for only a small annulus of the disk.
Submillimeter measurements of CO can measure the cold 
regions inward of $\sim 130\AU$ (Zuckerman et al.\ 1995),
where most of the cold CO molecules
are in the ground vibrational state 
to which the near-IR observations are not sensitive
(Brittain et al.\ 2003; T.W. Rettig, private communication).
Therefore, we will adopt the Zuckerman et al.\ (1995) results
for the $\molH$ mass. 

We assume that in the $r< 130\AU$ region,
the spatial distribution of $\molH$ follows 
that of the dust (see Eq.[\ref{eq:dndr1}]), i.e.,
\begin{equation}\label{eq:sigH}
\sigmaH(r) = \sigmaH^{\rm p}  
\exp\left\{-4\ln2\left[\left(r/{\rm AU}-55\right)/70\right]^2\right\}
~, ~~ \rin < r < 130\AU ~;
\end{equation}
where $\sigmaH^{\rm p}$ is the mid-plane surface density of $\molH$ 
at $r=55\AU$ which can be derived from 
$\mHtot$ ($\approx 5-115\mearth$; Zuckerman et al.\ 1995)
\begin{equation}\label{eq:mHtot}
\mHtot = \int_{\rin}^{130\AU} 2\,\mH\,\sigmaH(r) 2\pi rdr ~~.
\end{equation}


\begin{thebibliography}{}
\bibitem[]{}Allain, T., Leach, S., \& Sedlmayr, E. 1996,
            A\&A, 305, 602
\bibitem[]{}Allamandola, L.J., Tielens, A.G.G.M., \& Barker, J.R. 
            1985, ApJ, 290, L25
\bibitem[]{}Andrillat, Y., Jaschek, M., \& Jaschek, C. 
            1990, A\&A, 233, 474
%\bibitem[]{}Artymowicz, P. 2000, Space Sci. Rev., 92, 69
\bibitem[]{}Artymowicz, P., \& Lubow, S.H. 1994, ApJ, 421, 651
\bibitem[]{}Artymowicz, P., Burrows, C., \& Paresce, F.
            1989, ApJ, 337, 494
\bibitem[]{}Augereau, J.C., Lagrange, A.M., Mouillet, D.,
            \& M\'{e}nard, F. 1999a, A\&A, 350, L51
\bibitem[]{}Augereau, J.C., Lagrange, A.M., Mouillet, D., 
            Papaloizou, J.C.B., \& Gorod, P.A. 1999b, A\&A, 348, 557
\bibitem[]{}Backman, D.E., \& Paresce, F. 1993, in Protostars 
            and Planets III, ed. E.H. Levy \& J.I. Lunine 
            (Tucson: Univ. Arizona Press), 1253 
\bibitem[]{}Backman, D.E., Gillett, F.C., \& Witteborn, F.C.
            1992, ApJ, 385, 680
\bibitem[]{}Barrado y Navascu\'{e}s, D., Stauffer, J.R., Hartmann, L.,
            \& Balachandran, S.C. 1997, ApJ, 475, 313
\bibitem[]{}Barrado y Navascu\'{e}s, D., Stauffer, J.R., Song, I., 
            \& Caillault, J.-P. 1999, ApJ, 520, L123
\bibitem[]{}Bauschlicher, C., \& Bakes, E.L.O. 2000, Chem. Phys., 262, 285
\bibitem[]{}Beckwith, S.V.W., Henning, Th., \& Nakagawa, Y. 2000, 
            in Protostars and Planets IV, ed. V. Mannings, A.P. Boss, 
            \& S.S. Russell (Tucson: Univ. Arizona Press), 533
\bibitem[]{}Boccaletti, A., Augereau, J.C., Marchis, F.,
            \& Hahn, J. 2003, ApJ, 585, 494 
\bibitem[]{}Bohren, C.F., \& Huffman, D.R. 1983, Absorption and 
            Scattering of Light by Small Particles (New York: Wiley)
\bibitem[]{}Brittain, S.D., \& Rettig, T.W. 2002, Nature, 418, 57
\bibitem[]{}Brittain, S.D., Rettig, T.W., Simon, T., Kulesa, C.,
            DiSanti, M.A., \& Dello Russo, N. 2003, ApJ, 588, 535
\bibitem[]{}Brooke, T.Y., Tokunaga, A.T., \& Strom, S.E. 
            1993, AJ, 106, 656
\bibitem[]{}Burns, J.A., Lamy, P.L., \& Soter, S. 1979, Icarus, 40, 1 
\bibitem[]{}Cameron, A.G.W. 1975, Icarus, 24, 128
\bibitem[]{}Cameron, A.G.W. 1995, Meteoritics, 30, 133
\bibitem[]{}Cameron, A.G.W., \& Schneck, P.B. 1965, Icarus, 4, 396 
\bibitem[]{}Clampin, M., et al.\ 2003, AJ, in press
\bibitem[]{}Draine, B.T., \& Lee, H.M. 1984, ApJ, 285, 89
\bibitem[]{}Draine, B.T., \& Li, A. 2001, ApJ, 551, 807
\bibitem[]{}Dunkin, S.K, Barlow, M.J., \& Ryan, S.G. 
            1997, MNRAS, 286, 604
\bibitem[]{}Fisher, R.S., Telesco, C.M., Pi\~{n}a, R.K.,
            Knacke, R.F., \& Wyatt, M.C. 2000, ApJ, 532, L141
\bibitem[]{}Forst, W. 1973, Theory of Unimolecular Reactions
            (New York: Academic)
\bibitem[]{}Greaves, J.S., et al. 1998, ApJ, 506, L133
\bibitem[]{}Greenberg, J.M. 1968, in Stars and Stellar Systems, Vol. VII, 
            ed. B.M. Middlehurst \& L.H. Aller (Chicago: Univ. of Chicago 
            Press), 221
\bibitem[]{}Greenberg, J.M., \& Li, A. 1999, Space Sci. Rev., 90, 149
\bibitem[]{}de Grijp, M.H.K., Miley, G.K., \& Lub, J.
            1987, A\&AS, 70, 95
\bibitem[]{}Habing, H.J., et al.\ 2001, A\&A, 365, 545
\bibitem[]{}Henry, T.J., Soderblom, D.R., Donahue, R.A.,
            \& Baliunas, S.L. 1996, AJ, 111, 439
\bibitem[]{}Holland, W.S., et al. 1998, Nature, 392, 788
\bibitem[]{}Holland, W.S., et al. 2003, ApJ, 582, 1141
\bibitem[]{}Hudgins, D.M., Bauschlicher, C.W., Allamandola, L.J., 
            \& Fetzer, J.C. 2000, J. Phys. Chem., 104, 3655 
\bibitem[]{}Jaschek, M., Jaschek, C., \& Egret, D. 
            1986, A\&A, 158, 325
\bibitem[]{}Jayawardhana, R., Fisher, R.S., Hartmann, L., Telesco, C.M., 
            Pi\~{n}a, R.K., \& Fazio, G. 1998, ApJ, 503, L79
\bibitem[]{}Jochims, H.W., R\"{u}hl, E., Baumg\"{a}rtel, H., 
            Tobita, S., \& Leach, S. 1994, ApJ, 420, 307
\bibitem[]{}Jura, M., Malkan, M., White, R., Telesco, C.M., Pi\~{n}a, R.K., 
            \& Fisher, R.S. 1998, ApJ, 505, 897
\bibitem[]{}Kenyon, S.J., Wood, K., Whitney, B.A., \& Wolff, M.J. 
            1999, ApJ, 524, L119
\bibitem[]{}Knacke, R.F., Fajardo-Acosta, S.B., Telesco, C.M.,
            Hackwell, J.A., Lynch, D.K., \& Russell, R.W.
            1993, ApJ, 418, 440
\bibitem[]{}Koerner, D.W., Ressler, M.E., Werner, M.W., \& Backman, D.E.
            1998, ApJ, 503, L83
\bibitem[]{}Krivov, A.V., Mann, I., \& Krivova, N.A. 2000, A\&A, 362, 1127
\bibitem[]{}Kr\"ugel, E. 2003, Physics of Interstellar Dust
            (Bristol: IoP)
\bibitem[]{}Kurucz, R.L. 1979, ApJS, 40, 1
\bibitem[]{}Lagrange, A.-M., Backman, D.E., \& Artymowicz, P.
            2000, in Protostars and Planets IV, ed. V. Mannings, 
            A.P. Boss, \& S.S. Russell (Tucson: Univ. Arizona Press), 639
\bibitem[]{}Langhoff, S.R. 1996, J. Phys. Chem., 100, 2819
\bibitem[]{}L\'eger, A., \& Puget, J.L. 1984, A\&A, 137, L5
%\bibitem[]{}Le Page, V., Snow, T.P., \& Bierbaum, V.M. 
%            2001, ApJS, 132, 233 
\bibitem[]{}Li, A., \& Draine, B.T. 2001a, ApJ, 554, 778
\bibitem[]{}Li, A., \& Draine, B.T. 2001b, ApJ, 550, L213
\bibitem[]{}Li, A., \& Draine, B.T. 2002a, ApJ, 572, 232
\bibitem[]{}Li, A., \& Draine, B.T. 2002b, ApJ, 576, 762
\bibitem[]{}Li, A., \& Greenberg, J.M. 1997, A\&A, 323, 566
\bibitem[]{}Li, A., \& Greenberg, J.M. 1998, A\&A, 331, 291
\bibitem[]{}Li, A., \& Lunine, J.I. 2003, ApJ, 590, 368
\bibitem[]{}Lorenzetti, D., Giannini, T., Nisini, B., 
            Benedettini, M., Elia, D., Campeggio, L., 
            \& Strafella, F. 2002, A\&A, 395, 637
\bibitem[]{}Malfait, K., Bogaret, E., \& Waelkens, C. 
            1998, A\&A, 331, 211
\bibitem[]{}Marsh, K.A., Silverstone, M.D., Becklin, E.E.,
            Koerner, D.W., Werner, M.W., Weinberger, A.J.,
            \& Ressler, M.E. 2002 ApJ, 573, 425
\bibitem[]{}Mathis, J.S., Mezger, P.G., \& Panagia, N. 
            1983, A\&A, 128, 212
\bibitem[]{}McCall, B.J., Geballe, T.R., Hinkle, K.H.,
            \& Oka, T. 1998, Science, 279, 1910
\bibitem[]{}Mouillet, D., Lagrange, A.M., Augereau, J.C., 
            \& M\'{e}nard, F. 2001, A\&A, 372, L61 
\bibitem[]{}Natta, A., \& Kr\"{u}gel, E. 1995, A\&A, 302, 849
%\bibitem[]{}Natta, A., Prusti, T., \& Kr\"{u}gel, E. 
%            1993, A\&A, 275, 527
\bibitem[]{}Pantin, E., Waelkens, C., \& Malfait, K. 
            1999, in The Universe as Seen by ISO, 
            ed. P. Cox \& M.F. Kessler
            (ESA SP-427; Noordwijk: ESA/ESTEC), 38
\bibitem[]{}Rickman, H. 2003, in ASP Conf. Ser.,
            Cometary Nuclei in Space and Time,
            ed. M.F. A'Hearn (San Francisco: ASP), in press
\bibitem[]{}Schneider, G., et al. 1999, ApJ, 513, L127
%\bibitem[]{}Siebenmorgen, R., Kr\"ugel, E., \& Mathis, J.S.
%	    1992, A\&A, 266, 501
\bibitem[]{}Siebenmorgen, R., Prusti, T., Natta, A., 
            \& M\"ueller, T.G. 2000, A\&A, 361, 258
\bibitem[]{}Spitzer, L., Jr. 1941, ApJ, 93, 369
\bibitem[]{}Sylvester, R.J., \& Skinner, C.J. 1996, MNRAS, 283, 457
\bibitem[]{}Sylvester, R.J., Skinner, C.J., \& Barlow, M.J 
            1997, MNRAS, 289, 831
\bibitem[]{}Sylvester, R.J., Dunkin, S.J., \& Barlow, M.J 
            2001, MNRAS, 327, 133
\bibitem[]{}Sylvester, R.J., Skinner, C.J., Barlow, M.J, 
            \& Mannings, V. 1996, MNRAS, 279, 915
\bibitem[]{}Szczepanski, J., Wehlburg, C., \& Vala, M. 1995, 
            Chem. Phys. Lett., 232, 221
\bibitem[]{}Takeuchi, T., \& Artymowicz, P. 2001, ApJ, 557, 990
\bibitem[]{}Telesco, C.M., \& Knacke, R.F. 1991, ApJ, 372, L29
\bibitem[]{}Telesco, C.M., et al.\ 2000, ApJ, 530, 329
\bibitem[]{}Thi, W.F., et al.\ 2001, ApJ, 561, 1074
\bibitem[]{}van den Ancker, M.E., de Winter, D., \& 
            Tjin A Djie, H.R.E. 1998, A\&A, 330, 145
\bibitem[]{}Verstraete, L., L\'{e}ger, A., d'Hendecourt, L.,
            Defourneau, D., \& Dutuit, O. 1990, A\&A, 237, 436
\bibitem[]{}Walker, H.J., \& Wolstencroft, R.D. 1988, PASP, 100, 150
\bibitem[]{}Weidenschilling, S.J., \& Cuzzi, J.N. 1993, in Protostars 
            and Planets III, ed. E.H. Levy \& J.I. Lunine 
            (Tucson: Univ. Arizona Press), 1031 
\bibitem[]{}Weinberger, A.J., Becklin, E.E., 
            \& Zuckerman, B. 2003, ApJ, 584, L33
\bibitem[]{}Weinberger, A.J., Rich, R.M., Becklin, E.E.,
            Zuckerman, B., \& Matthews, K. 2000, ApJ, 544, 937
\bibitem[]{}Weinberger, A.J., Becklin, E.E., Schneider, G., 
            Smith, B.A., Lowrance, P.J., Silverstone, M.D.,
            Zuckerman, B., \& Terrile, R.J. 1999, ApJ, 525, L53
\bibitem[]{}Weingartner, J.C., \& Draine, B.T. 2001, ApJS, 134, 263
\bibitem[]{}Whipple, F.L. 1999, Planet. Space Sci., 47, 301
\bibitem[]{}Whittet, D.C.B., Williams, P.M., Zealy, W.J., Bode, M.F.,
            \& Davies, J.K. 1983, A\&A, 123, 301
\bibitem[]{}Wyatt, M.C., Dermott, S.F., Telesco, C.M., Fisher, R.S.,
            Grogan, K., Holmes, E.K., \& Pi\~{n}a, R.K. 1999, ApJ, 527, 918
\bibitem[]{}Zuckerman, B. 2001, ARA\&A, 39, 549
\bibitem[]{}Zuckerman, B., Forveille, T., \& Kastner, J.H.
            1995, Nature, 373, 494
\bibitem[]{}Zuckerman, B., Song, I., Bessell, M.S., 
            \& Webb, R.A. 2001, ApJ, 562, L87
\end{thebibliography}
\end{document}